\documentclass[11pt,a4paper]{article}
\usepackage[utf8x]{inputenc}
\usepackage{epsfig}
\usepackage[T1]{fontenc}    
\usepackage{graphics}
\usepackage{graphicx}
\usepackage{pstricks,pst-coil,pst-fill,pst-plot}
\usepackage[fleqn]{amsmath}    
\usepackage{amssymb}    
\usepackage{amsfonts}   
\usepackage{verbatim}   
\usepackage{mathrsfs}   
\usepackage{dsfont}
\usepackage{euscript}
\usepackage{yfonts}
\usepackage{amsthm}         
\usepackage{txfonts}
\usepackage{marvosym}
\usepackage{vmargin}        
\usepackage{wasysym}		

\usepackage{color}
\usepackage[normalem]{ulem}
\usepackage{cite}
\usepackage{microtype}

\setmarginsrb{1.8cm}{2cm}{1.8cm}{2cm}{1cm}{1cm}{1cm}{1.6cm}
 \makeatletter
 \@addtoreset{equation}{section}
 \makeatother

\providecommand{\bysame}{\leavevmode\hbox to3em{\hrulefill}\thinspace}
\providecommand{\MR}{\relax\ifhmode\unskip\space\fi MR }

\providecommand{\href}[2]{#2}

\let\tend=\rightarrow

\long\def\symbolfootnote[#1]#2{\begingroup%
\def\thefootnote{\fnsymbol{footnote}}\footnote[#1]{#2}\endgroup}

\newtheorem{theorem}{Theorem}[section]
\newtheorem{prop}[theorem]{Proposition}
\newtheorem*{theorem*}{Theorem}
\newtheorem{cor}[theorem]{Corollary}
\newtheorem{defin}[theorem]{Definition}

\newtheorem{lemme}{Lemma}[section]

\def\Proof{\medskip\noindent {\it Proof --- \ }}

\def\qed{\hfill\rule{2mm}{2mm}}

\newcommand\beq{\begin{equation}}
\newcommand\enq{\end{equation}}
\newcommand\bem{\begin{multline}}
\newcommand\enm{\end{multline}}

\def\beqa{\begin{eqnarray}}
\def\eeqa{\end{eqnarray}}
\def\ba{\begin{array}}
\def\ea{\end{array}}

\newcommand{\f}[2]{{\ensuremath{%
    \mathchoice%
    {\dfrac{#1}{#2}}
    {\dfrac{#1}{#2}}
    {\frac{#1}{#2}}
    {\frac{#1}{#2}}
}}}
\newcommand{\tf}[2]{\ensuremath{#1/#2}}

\def\a{\alpha}

\def\be{\beta}
\def\ga{\gamma}

\def\de{\delta}

\def\De{\Delta}
\def\eps{\epsilon}
\def\veps{\varepsilon}
\def\la{\lambda}
\def\La{\Lambda}

\def\sg{\sigma}

\def\ups{\upsilon}
\def\th{\theta}

\def\Om{\Omega}
\def\om{\omega}

\newcommand{\mc}[1]{\ensuremath{\mathcal{#1}}}
\newcommand{\mf}[1]{\ensuremath{\mathfrak{#1}}}
\newcommand{\msc}[1]{\ensuremath{\mathscr{#1}}}

\newcommand{\bs}[1]{\ensuremath{\boldsymbol{#1}}}

\DeclareFontFamily{OT1}{pzc}{}
\DeclareFontShape{OT1}{pzc}{m}{it}{<-> s * [1.10] pzcmi7t}{}
\DeclareMathAlphabet{\mathpzc}{OT1}{pzc}{m}{it}

\def \i{ \mathrm i}

\newcommand{\ov}[1]{\ensuremath{\overline{#1}}}
\newcommand{\wt}[1]{\ensuremath{\widetilde{#1}}}
\newcommand{\wh}[1]{\ensuremath{\widehat{#1}}}

\newcommand{\Int}[2]{\ensuremath{\int\limits_{#1}^{#2}}}
\newcommand{\Oint}[2]{\ensuremath{\oint\limits_{#1}^{#2}}}

\newcommand{\sul}[2]{\ensuremath{\sum\limits_{#1}^{#2}}}
\newcommand{\pl}[2]{\ensuremath{\prod\limits_{#1}^{#2}}}

\newcommand{\R}{\ensuremath{\mathbb{R}}}
\newcommand{\Cx}{\ensuremath{\mathbb{C}}}

\newcommand{\Dp}[1]{\ensuremath{\partial_{#1}}}

\newcommand{\limit}[2]{\ensuremath{\underset{#1 \tend #2}{\longrightarrow} }}

\newcommand{\ex}[1]{\ensuremath{\e{e}^{#1}}}

\newcommand{\op}[1]{ \boldsymbol{ \texttt{#1} } }

\newcommand{\norm}[1]{\ensuremath{  || #1 || }}

\newcommand{\dd}{\mathrm{d}}
\newcommand{\e}[1]{\ensuremath{\mathrm{#1}}}

\newcommand{\intff}[2]{\ensuremath{ [  #1 \,; #2 ] }}
\newcommand{\intfo}[2]{\ensuremath{ [  #1 \,; #2 [ }}

\newcommand{\intoo}[2]{\ensuremath{ ]  #1 \,; #2 [ }}

\newcommand{\intn}[2]{\ensuremath{[\![ \, #1 \,;\, #2 \,]\!]}}

\begin{document}

\begin{center}
\begin{LARGE}
{\bf \boldmath Thermodynamics of the spin-$1/2$ Heisenberg-Ising chain at high temperatures: a rigorous approach}
\end{LARGE}

\vspace{1cm}

{\large Frank G\"{o}hmann\footnote{e-mail: goehmann@uni-wuppertal.de}}
\\[1ex]
Fakult\"at f\"ur Mathematik und Naturwissenschaften, Bergische Universit\"at Wuppertal, 42097 Wuppertal, Germany.\\[2.5ex]

{\large Salvish  Goomanee\footnote{e-mail: salvish.goomanee@ens-lyon.fr},  Karol K. Kozlowski\footnote{e-mail: karol.kozlowski@ens-lyon.fr} }
\\[1ex]
Univ Lyon, ENS de Lyon, Univ Claude Bernard Lyon 1, CNRS, Laboratoire de Physique, F-69342 Lyon, France \\[2.5ex]

{\large Junji Suzuki\footnote{e-mail: suzuki.junji@shizuoka.ac.jp}}%
\\[1ex]
Department of Physics, Shizuoka University, Ohya 836, Shizuoka,  Japan.

\par

\end{center}

\vspace{40pt}

\centerline{\bf Abstract} \vspace{1cm}
\parbox{16cm}{\small
This work develops a rigorous setting allowing one to prove several features related to the behaviour of the Heisenberg-Ising (or XXZ) spin-$1/2$ chain at finite temperature $T$. 
Within the quantum inverse scattering method  the physically pertinent observables at finite $T$, such as the \textit{per}-site free energy or the 
correlation length, have been argued to admit integral representations  whose integrands are expressed in terms of solutions to auxiliary non-linear integral equations. The 
derivation of such representations was based on numerous conjectures: the possibility to exchange the infinite volume and the infinite Trotter number limits, 
the existence of a real, non-degenerate, maximal in modulus Eigenvalue of the quantum transfer matrix, the existence and uniqueness of solutions to the auxiliary
non-linear integral equations, as well as the possibility to take the infinite Trotter number limit on their level. 
We rigorously prove all these  conjectures for temperatures large enough. As a by product of our analysis, we obtain the large-$T$ asymptotic expansion for 
a subset of sub-dominant Eigenvalues of the quantum transfer matrix and thus of the associated correlation lengths. This result was never obtained previously, not even on heuristic grounds. }

\section{Introduction and main results}

Quantum statistical mechanics deals with the description of the large volume behaviour of interacting particle systems   coupled to a heat bath
of temperature $T$. 
In many situations of interest to   physics, the particle degrees of freedom are attached to a finite lattice $\La \subset \mathbb{Z}^d$.
The interactions between these degrees of freedom are captured by the Hamiltonian $\op{H}$,   an operator on the Hilbert space $\mf{h}$ of the model which   for a finite lattice $\La$ has 
a tensor-product structure $\mf{h}=\bigotimes_{a \in \La} \mf{h}_a$. Here $\mf{h}_a$ plays the role of a local quantum space where the degrees of freedom associated with the lattice site $a\in \La$
evolve. The most basic physically interesting quantity characterising such models is the \textit{per}-site free energy  
\beq
   - T \lim_{ \La \tend \mathbb{Z}^d } \f{1}{|\La|} \ln  \e{tr}_{\mf{h}}\Big[ \ex{ - \f{ \op{H} }{ T }  } \Big] \;. 
\enq
In this expression $|\La|$ stands for the number of points in $\La$.
It turns out that for quite general Hilbert spaces and Hamiltonians, one may rigorously prove the existence of the above limit, 
provided that $ \La \tend \mathbb{Z}^d $ in some appropriate sense \cite{RuelleRigorousResultsForStatisticalMechanics}. 
However, obtaining an explicit and thorough characterisation of the limit turned out to be a notoriously hard problem. 

So far, rigorous and explicit descriptions of the \textit{per}-site free energy could only be obtained for one dimensional, \textit{viz}. when $d=1$, 
quantum systems and, even in such a simplified setting, only for an extremely limited number of cases. Basically the only rigorous results concern models 
equivalent to free fermions \cite{LewisPuleZagrebnovLDPpourMesuresKacEtHardBosons,SuzukiInoueMoreDvpmtInterchangeabilityTrotterAndVolumeLimitInPartFcton}, \textit{viz}. non-interacting models, and the non-linear 
Schrödinger model \cite{DorlasLewisPuleRigorousProofYangYangThermoEqnNLSE}. In both cases the models belong to the class of quantum integrable models
for which one may rely on powerful algebraic tools, originating from the theory of quantum groups, to obtain a certain amount of information on the spectra, Eigenvectors and
many other observables.  Still, the models for which the free energy could  be computed rigorously are very specific, 
even among the quantum integrable models. This implicates, in particular,  that the techniques of proofs developed for these models have low chances to be
applicable  to more complex quantum integrable models. It is the purpose of this paper to develop
new strategies and techniques of proofs which will have a large scope of applicability. In this work, we will focus on the  XXZ spin-$1/2$ chain,
but more generally, we expect our approach to work for any quantum integrable model associated with a fundamental $R$-matrix.

The XXZ spin-$1/2$ chain is an archetypical example of a quantum integrable model. It refers to the Hamiltonian operator
\beq
\op{H} \, = \, J \sum_{a=1}^{L} \Big\{ \sigma_a^x \,\sigma_{a+1}^{x} + \sigma_a^y \,\sigma_{a+1}^{y} +  \De \, (\sigma_a^z \,\sigma_{a+1}^{z} + 1)  \Big\} \, - \, \f{h}{2} \sul{a=1}{L} \sigma_a^z  \; . 
\label{ecriture hamiltonien XXZ}
\enq
Here  $J>0$ represents the so-called exchange interaction, $\De \in {\mathbb R}$ is the anisotropy parameter,
$h>0$ the external magnetic field and $L \in 2\mathbb{N}$ corresponds to the number of sites. $\op{H}$ acts on the Hilbert space $\mf{h}_{XXZ}=\bigotimes_{a=1}^{L}\mf{h}_a$ with $\mf{h}_a \simeq \Cx^2$,
 $\sg^{\a}$, $\a\in \{x, y, z \}$, are the Pauli matrices, and the operator $\sg_{a}^{\a}$ acts as the Pauli matrix $\sg^{\a}$
on $\mf{h}_a$ and as the identity on all the other spaces:
\beq
\sg_{a}^{\a} \, = \, \underbrace{ \e{id} \otimes \cdots \otimes \e{id} }_{a-1} \otimes \, \sg^{\a} \otimes \underbrace{ \e{id} \otimes \cdots \otimes \e{id} }_{L-a}  \;. 
\enq
Finally, the model is subject to periodic boundary conditions, \textit{viz.} $\sg_{L+1}^{\a}=\sg_{1}^{\a}$.

The \textit{per}-site free energy of the XXZ chain is defined as
\beq
f\, = \; -T \lim_{L\tend + \infty} \Big\{ \f{1}{L} \ln \e{tr}_{\mf{h}_{XXZ}} \Big[ \ex{-\frac{1}{T} \op{H}  } \Big] \Big\} \;. 
\label{definition energie libre}
\enq
As already mentioned, the existence of the limit in \eqref{definition energie libre} follows from well-known results in the theory of thermodynamic limits, see \textit{e.g.} \cite{RuelleRigorousResultsForStatisticalMechanics}. 
However, such existence results do not provide one with any direct characterisation of the value of this limit. 
First approaches based on the integrability of the XXZ chain model and aiming at a characterisation of this limit go back to the works of 
Gaudin \cite{GaudinTBAXXZMassiveInfiniteSetNLIE} and Takahashi \cite{TakahashiTBAforXXZFiniteTinfiniteNbrNLIE}. Those works 
provided a formal computation of the assumedly dominant contribution, in the large-$L$ limit, to the partition function $\e{tr}_{\mf{h}_{XXZ}} \Big[ \ex{-\frac{1}{T} \op{H}  } \Big] $
by generalising the arguments first invoked by Yang and Yang \cite{YangYangNLSEThermodynamics} in the context of studying the thermodynamics of another quantum integrable model, the one-dimensional
non-linear Schrödinger model \cite{BrezinPohilFinkelbergFirstIntroBoseGas,LiebLinigerCBAForDeltaBoseGas}.  While the approach of Yang and Yang could, \textit{in fine}, 
be put on rigorous grounds by Dorlas, Lewis and Pul\'{e} \cite{DorlasLewisPuleRigorousProofYangYangThermoEqnNLSE}
for the non-linear Schrödinger model, it appears that doing so in the case of the XXZ spin-chain is hardly possible. 
Indeed, the proof of \cite{DorlasLewisPuleRigorousProofYangYangThermoEqnNLSE} invokes, at some stage, the completeness of the Bethe states, a set of eigenstates of an integrable model's Hamiltonian 
which is parameterised by roots of a coupled system of algebraic equations in many variables, the Bethe Ansatz equations \cite{BetheSolutionToXXX}. Whilst completeness of Bethe states could be proven to hold for the 
non-linear Schrödinger model \cite{DorlasOrthogonalityAndCompletenessNLSE}, it turns out that completeness issues are, by far more, subtle for the XXZ chain \cite{MukhinTarasovVarchenkoCompletenessandSimplicitySpectBA}. 
Gaudin and Takahashi based their calculations on the so-called string hypothesis, assuming that in the large-$L$ limit all solutions of the Bethe Ansatz
equations can be classified according to certain patterns of roots in the complex plane, called "string solutions" and discovered by Bethe in his original work
\cite{BetheSolutionToXXX}. Although the result of Gaudin and Takahashi for the dominant contribution to $ \e{tr}_{\mf{h}_{XXZ}} \Big[ \ex{-\frac{1}{T} \op{H}  } \Big]$ is very likely to be correct, their derivation based on the string hypothesis
is certainly not. It is well-established for a while
\cite{CauxHagemansDeformedStringsInXXX,DestriLowensteinFirstIntroHKBAEAndArgumentForStringIsWrong,EsslerKorepinSchoutensBAE2ParticleSectorXXXAndSomeNonStandardSols,MukhinTarasovVarchenkoCompletenessandSimplicitySpectBA} 
that that the so-called "string solutions" do not provide one with a complete set of Eigenvectors for the XXZ chain and, in some cases, do not give rise to solutions of the Bethe equations.

 Thus, an alternative approach to the thermodynamics had to be devised.
 An important step forward in this direction was made by Koma \cite{KomaIntroductionQTM6VertexForThermodynamicsOfXXX,KomaIntroductionQTM6VertexForThermodynamicsOfXXZ}. 
By implementing the Trotter discretisation of the Boltzman statistic operator first proposed by Suzuki \cite{SuzukiArgumentsForInterchangeabilityTrotterAndVolumeLimitInPartFcton}, 
and further developed by Suzuki and Inoue \cite{SuzukiInoueMoreDvpmtInterchangeabilityTrotterAndVolumeLimitInPartFcton},
Koma proposed a Trotter approximant of the finite-$L$ partition function of the XXX \cite{KomaIntroductionQTM6VertexForThermodynamicsOfXXX} and then of the XXZ \cite{KomaIntroductionQTM6VertexForThermodynamicsOfXXZ} chain. 
By this one means a representation of the type 
\beq
\e{tr}_{\mf{h}_{XXZ}} \Big[ \ex{-\frac{1}{T} \op{H} } \Big] \; = \; \lim_{N\tend +\infty} \e{tr}_{ \mf{h}_{\mf{q}} } \Big[ \mc{T}_{N,T}^{L} \Big] \;. 
\label{ecriture limite Trotter a la Koma}
\enq
More precisely, Koma rewrote the original finite-$L$ partition function as resulting from taking an infinite Trotter limit $N \rightarrow + \infty$ of the trace of the $L^{\e{th}}$-power of an operator $ \mc{T}_{N,T}$
acting on an auxiliary Hilbert space $\mf{h}_{\mf{q}}$. Koma's transfer matrix $\mc{T}_{N,T}$ corresponds to the transfer matrix of a
vertex model related to the product of two Ising models and acts on an auxiliary Hilbert space $\mf{h}_{ \mf{q} }$ whose dimension
blows up exponentially with $N$. Koma built on the setting of  \cite{SuzukiArgumentsForInterchangeabilityTrotterAndVolumeLimitInPartFcton,SuzukiInoueMoreDvpmtInterchangeabilityTrotterAndVolumeLimitInPartFcton}
so as to establish the validity of the
commutativity of the $L\tend +\infty$ and $N\tend +\infty$ limits when substituting \eqref{ecriture limite Trotter a la Koma} into \eqref{definition energie libre}. 
He also proved that the transfer matrix $\mc{T}_{N,T}$ admits a non-degenerate, real,  maximal in modulus Eigenvalue
$\La_{\e{max}}(\mc{T}_{N,T})$. This led to the representation 
\beq
f\, = \; -T \lim_{N\tend + \infty} \Big\{ \ln\big[ \La_{\e{max}}(\mc{T}_{N,T}) \big] \Big\} \;. 
\label{ecriture energie libre comme limite trotter via QTM Koma}
\enq
Within Koma's approach, the Eigenvalue $\La_{\e{max}}(\mc{T}_{N,T})$ was expressed in terms of a particular solution to the Bethe Ansatz equations. 
However, Koma only managed to treat rigorously the calculation of the Trotter limit when $\De=0$, which is a trivial situation and which was dealt with, 
by much   simpler means, in \cite{SuzukiInoueMoreDvpmtInterchangeabilityTrotterAndVolumeLimitInPartFcton}. 
For general $\De$, Koma did not manage to establish a way of effectively taking the Trotter limit in \eqref{ecriture energie libre comme limite trotter via QTM Koma}. 
He could only perform a numerical  analysis \cite{KomaIntroductionQTM6VertexForThermodynamicsOfXXX}  in the case $\De=1$.  
Later, Takahashi \cite{TakahashiThermoXXZInfiniteNbrRootsFromQTM} proposed a formal scheme for taking the infinite Trotter number limit of the Bethe Ansatz equations obtained by Koma, 
but that did not lead to any closed formula for $f$, nor was it the result of rigorous handlings. 

\vspace{2mm}

The above discussion stresses the serious problems arising within Koma's transfer matrix approach relating to taking care, in an efficient way, of the 
infinite Trotter number limit. Furthermore, Koma's transfer matrix does not appear to exhibit enough algebraic structure so as to allow one for conforming it to a larger class of problems such as the 
computation of thermal correlation functions, which are the objects of main interest to the physics of the model. 
For that purpose, as observed in \cite{GohmannKlumperSeelFinieTemperatureCorrelationFunctionsXXZ}, it is more 
adapted to use a different quantum transfer matrix $\op{t}_{\mf{q}}$ directly related to a staggered six-vertex model, whose 
construction was pioneered in \cite{AkustsuSuzukiWadatiModernFormOfQTM4XXZ} and then further improved in \cite{KlumperNLIEfromQTMDescrThermoRSOSOneUnknownFcton},
where $\op{t}_{\mf{q}}$ was introduced as a member of a commuting family of `column-to-column' transfer matrices. The latter construction has proved to be most convenient in the context of integrable lattice models. Over the years
it was used to push rather far, many concrete calculations of various observables associated with the finite temperature XXZ spin-$1/2$ 
chain \cite{KozDugaveGohmannThermaxFormFactorsXXZ,KozDugaveGohmannThermaxFormFactorsXXZOffTransverseFunctions,
KozDugaveGohmannSuzukiLowTFFSeriesMassiveXXZQTMandXXX,KozGohmannKarbachSuzukiFiniteTDynamicalCorrFcts,GohmannKlumperSeelFinieTemperatureCorrelationFunctionsXXZ,GohmannKlumperSeelElementaryBlocksFiniteTXXZ}. 
Those results, which stress the wide scope of applicability of the quantum transfer matrix, constitute an important motivation for putting the handlings based on the use of $\op{t}_{\mf{q}}$ on rigorous grounds.

In the approach based on $\op{t}_{\mf{q}}$, it holds that
\beq
\e{tr}_{\mf{h}_{XXZ}} \Big[ \ex{-\frac{1}{T} \op{H}  } \Big] \; = \; \lim_{N\tend +\infty} \e{tr}_{ \mf{h}_{\mf{q}} } \Big[ \op{t}_{ \mf{q} }^{L} \Big] \;, 
\enq
see, \textit{e.g.}\ \cite{GohmannKlumperSeelFinieTemperatureCorrelationFunctionsXXZ} for an exposure thereof in a modern language. The quantum transfer matrix acts 
on the `quantum' space $\mf{h}_{\mf{q}} \, = \, \bigotimes_{a=1}^{2N} \mf{h}_a$, with $\mf{h}_a \simeq \Cx^2$. 
It is defined as the trace over the auxiliary space $\mf{h}_{0}$ of the quantum monodromy matrix 
\beq
 \op{t}_{ \mf{q} } \, = \, \e{tr}_{\mf{h}_0} \Big[ \op{T}_{\mf{q};0}(0) \Big] \;,
\label{definition QTM}
\enq
where the latter object is defined as an ordered product of $\op{R}$-matrices: 
\beq
\op{T}_{\mf{q};0} (\xi) \; = \; \op{R}_{2N,0}^{\mf{t}_{2N}}\Big( -\tfrac{ \aleph }{N}-\xi \Big) \, \op{R}_{0,2N-1}\Big( \xi -\tfrac{  \aleph  }{N}\Big) \cdots  
 \op{R}^{\mf{t}_{2}}_{2, 0}\Big( -\tfrac{  \aleph  }{N}-\xi \Big)\, 
\op{R}_{0, 1}\Big( \xi -\tfrac{ \aleph }{N}\Big) \cdot   \ex{ \f{h}{2T} \sg^z_0}   \quad \e{with} \quad  \aleph =  \f{ J }{ T } \sinh(\eta)   \;. 
\label{ecriture matrice de monodromie quantique}
\enq
Above, $\op{R}$ represents the $\op{R}$-matrix of the six-vertex model 
\beq
\op{R}(\la)\; = \;  \f{1}{ \sinh(\eta) } \left( \ba{cccc} \sinh\big( \eta + \la \big) 	& 0 	& 0	 & 0  \\
							    0   &  \sinh\big( \la \big)   &   \sinh(\eta)   &  0  \\ 
								0   & \sinh(\eta)  & \sinh\big(  \la \big)   &    0  \\ 
							      0 	& 0		 &    0  			&   \sinh\big( \eta + \la \big)  \ea \right) \; ,  
\label{ecriture matrice R six vertex}
\enq
while the notation $\op{R}_{ab}$ stands for the embedding of $\op{R}$ into an operator on $\mf{h}_0\otimes \mf{h}_{\mf{q}}$ which acts as the identity operator 
on all spaces $\mf{h}_{k}$, $k \not= a,b$ and as $\op{R}(\la)$ on the spaces $\mf{h}_{a} \otimes \mf{h}_b$. 
Finally, the superscript $\mf{t}_a$ stands for the partial transposition  with respect to the space $\mf{h}_a$.

Here and in the following, for definiteness, we choose the parameterisation 
\beq
\De = \cosh(\eta) \quad \eta=-\i\zeta \quad \e{and} \quad \zeta \in \intoo{0}{\pi}
\enq
 corresponding to the regime $-1< \De <1$. We shall always assume, without further notice, that $\zeta$ is generic. However, with minor modifications, the whole analysis developed in the following also carries through 
to the regimes $|\De|\geq 1$.

\subsection{Main results of the paper}


%
%
%
%
%
%
%
%
%

On formal grounds, the calculation of the free energy based on the quantum transfer matrix $\op{t}_{\mf{q}}$ follows the same strategy as with Koma's Trotter approximant $\mc{T}_{N,L}$. 
One first \textit{assumes} the exchangeability of the thermodynamic ($L\tend + \infty$) and the Trotter ($N\tend + \infty$) limits. 
Then, one \textit{assumes} that $\op{t}_{\mf{q}}$ admits a maximal in modulus, real non-degenerate Eigenvalue $\wh{\La}_{\e{max}}$. The two conjectures put together, allow one to write 
the \textit{per}-site free energy of the model as 
\beq
f\, = \; -T \lim_{N\tend + \infty} \Big\{ \ln\big[ \wh{\La}_{\e{max}} \big] \Big\} \;. 
\label{ecriture lien energie libre et vp dom QTM}
\enq
On the basis of Bethe Ansatz calculations and a certain amount of numerical input, one then \textit{argues} an integral representation for  $\wh{\La}_{\e{max}}$
which is given in terms of a solution $\wh{\mathfrak{A}}$ to an auxiliary non-linear integral equation. By assuming that it is licit to take the infinite Trotter limit formally on the level of this
non-linear integral equation, one obtains an integral representation for $f$. This strategy was developed in \cite{DestriDeVegaAsymptoticAnalysisCountingFunctionAndFiniteSizeCorrectionsinTBAFirstpaper,KlumperNLIEfromQTMDescrThermoXYZOneUnknownFcton}. 
While all of the above assumptions were supported  by thorough numerical calculations, no mathematically rigorous proof was ever given. The main progress achieved in this work 
consists in establishing rigorously, for $T$ large enough:
\begin{itemize}
 
 \item[i)] the exchangeability of the Trotter and infinite volume limits;
 
 \item[ii)] the existence of a maximal in modulus Eigenvalue of the quantum transfer matrix which, furthermore, is real and non-degenerate; 
 
 \item[iii)] the well-definiteness of the class of non-linear integral equations describing the Eigenvalues of $\op{t}_{\mf{q}}$, as well as the existence and uniqueness of their solutions;
 
 \item[iv)] the rigorous identification of the non-linear integral equation describing the dominant Eigenvalue of the quantum transfer matrix. 
 
\end{itemize}

Points i)-iv), all put together, allow one to establish the below theorem which crowns the efforts of this paper. 
In order to state the theorem, we shall agree that $\mc{D}_{z_0,\a} \subset \Cx$ stands for the open disc of radius $\a$ centred at $z_0$, \textit{c.f.}\ \eqref{definition bande largeur alpha et disque}. 
Also, $\mc{B}_r$ stands for the space of holomorphic functions on the strip of width   $\tfrac{1}{2}\e{min}(\zeta, \pi-\zeta)$ around $\R$
whose $L^{\infty}$-norm on this strip is bounded by $r$, \textit{c.f.}\ \eqref{defintion fct holomorphes bornees sur la bande}.

\begin{theorem}
\label{Theorem principal aticle f a T fini} 
 
 There exist $T_0>0$ and $\eps>0$ such that for any $T>T_0$, the \textit{per}-site free energy of the XXZ chain defined by \eqref{definition energie libre} admits the 
 integral representation 
\beq
- \frac{f}{T} \, = \, \   \f{h}{2T} - \f{2 J}{T}  \cos(\zeta)   \,  - \,   \Oint{ \Dp{}\mc{D}_{0,\eps}  }{ }  \f{ \sin(\zeta) \,  \mc{L}\mathrm{n} \big[ 1+  \ex{\mf{A} } \, \big](u) }{ \sinh(u-\i \zeta)\,  \sinh(u) } \cdot \f{ \dd u }{ 2\pi }   
\enq
 in which $\mf{A}$ is the unique solution to the non-linear integral equation on $\mc{B}_{\mf{c}}$
\beq
\mf{A}(\xi) \, = \, -\f{1}{T} \bigg\{  h \, - \, \f{2J \sin^2(\zeta) }{ \sinh(\xi) \sinh(\xi-\i\zeta) } \bigg\} 
\; + \; \Oint{ \Dp{} \mc{D}_{0,\eps} }{}  \f{\sin(2\zeta) \cdot   \mc{L}\mathrm{n}\big[ 1+ \ex{\mf{A}} \big](u) }{ \sinh(\xi-u+\i\zeta) \sinh(\xi-u-\i\zeta) }  \cdot \f{ \dd u  }{ 2\pi }  \;. 
\label{ecriture NLIE A}
\enq
Here, the logarithm is defined as 
\beq
 \mc{L}\mathrm{n}\big[ 1+ \ex{\mf{A}} \big](\xi) \, = \, \Int{\eps}{\xi} \f{ \mf{A}^{\prime}(u) } { 1+ \ex{-\mf{A}(u)} }  \cdot \f{\dd u}{2\i\pi}   \; + \; \ln\big[ 1+  \ex{\mf{A}(\eps) } \, \big]
%
%
\enq
 in which the integral runs from $\eps$ to $\xi \in \Dp{} \mc{D}_{0,\eps}$ in positive direction along $\Dp{} \mc{D}_{0,\eps}$ and $\ln$ stands for the principal branch of the logarithm with $\e{arg} \in \intfo{-\pi}{\pi}$. 
The claim of uniqueness and existence of solutions to \eqref{ecriture NLIE A} is part of the statement of the theorem.
 
 Finally, the \textit{per}-site free energy admits the high-$T$ asymptotic expansion:
\beq
- \frac{f}{T}\, = \, \ln 2 \, - \, \f{ J \cos(\zeta) }{ T } \, + \, \e{O}\Big( T^{-2} \Big)\;. 
\label{ecriture DA energie libre}
\enq
\end{theorem}

 The expansion \eqref{ecriture DA energie libre} was already obtained in the literature, without the rigorous justification, 
 through various methods \cite{DestriDeVegaAsymptoticAnalysisCountingFunctionAndFiniteSizeCorrectionsinTBAFiniteMagField,RojasSouzaThomaszHighTExpFormalChainModel},
 up to order $\e{O}\big( T^{-5} \big)$ for the XXZ chain in a magnetic field 
and up to order $\e{O}(T^{-100})$ for the XXX chain at zero magnetic field \cite{ShiroisihiTakahashiHighTExpansionFreeEnergyXXXzerohUpToOrder100}.  
In fact, taking for granted the conclusions of Theorem \ref{Theorem principal aticle f a T fini}, it is not a problem to  build on the non-linear integral equation \eqref{ecriture NLIE A} so as 
to push the expansion \eqref{ecriture DA energie libre} to very high orders in $T^{-1}$, this with the help of formal computer algebra.

 \vspace{3mm}

 The techniques developed in this work allow one, in fact, to obtain a much larger range of results. Indeed, while the dominant Eigenvalue  $\wh{\La}_{\e{max}}$ of the quantum transfer matrix 
provides a means to access the \textit{per}-site free energy through \eqref{ecriture lien energie libre et vp dom QTM}, the other Eigenvalues  $\wh{\La}_{\e{ex};k}$ 
provide an access to the correlation lengths $\xi_{k}$ of the model. These control the speed of the exponential decay of multi-point correlation functions seen as a function of the distances separating the local operators, 
see \textit{e.g.} \cite{KozDugaveGohmannThermaxFormFactorsXXZ,KozDugaveGohmannThermaxFormFactorsXXZOffTransverseFunctions} for more details. 
The correlation lengths are defined as
\beq
\ex{-\f{1}{\xi_k} } \; = \; \lim_{N\tend + \infty} \bigg\{ \f{ \wh{\La}_{\e{ex};k} }{ \wh{\La}_{\e{max}} } \bigg\} \;. 
\label{definition longueur de correlation via limite Trotter}
\enq
Our analysis allows us to show that, for temperatures large enough, and for a rather large subset of the Eigenvalues $\wh{\La}_{\e{ex};k}$
the limit \eqref{definition longueur de correlation via limite Trotter} exists and is characterised in terms of the unique solution to a non-linear integral equation. 
See Proposition \ref{Proposition correspondance solutions NLIE originelle et NLIE type point fixe}, Theorem \ref{Theorem existence et unicite point fixe de OTN} and 
Proposition \ref{Proposition forme solution NLIE avec part trous a gde T} for more details. 
These results allow one, in the end, to characterise rigorously the large-temperature behaviour of a certain subset of the correlation lengths.

 \begin{theorem}
\label{Theorem longueur de correlations}

Fix integers $n_x$, $n_y$ and pick some $\varrho>0$ small enough. Let $\{y_a\}_{1}^{n_y}$ correspond to a solution to the system 
\beq
(-1)^{n_x-n_y + 1}  
\pl{ b=1}{n_y} \Big\{   \sinh(\i\zeta + y_{b} - y_{a}) \Big\}  \cdot  \Big(   \sinh(\i\zeta + y_{a} ) \Big)^{n_x}  
	    \, = \,  \pl{ b=1}{n_y}\Big\{ \sinh(\i \zeta + y_{a} - y_{b}) \Big\} \cdot \Big( \sinh(\i\zeta - y_{a} )  \Big)^{n_x}  
\label{ecriture BAE XXZ spin 1}
\enq
for $a=1,\dots, n_y $, such that the following three subsidiary constraints are satisfied
\begin{enumerate}
\item $y_a\not= y_b \pm \i \zeta$, \; $y_a \not= y_b$ \;\; $\mathrm{mod} \, \i \pi \,  \mathbb{Z}$ \quad for \; \,  $a, b \in \intn{1}{n_y}$ ;
 \item  $\bigg| (-1)^{n_x-n_y }    \pl{ b=1}{n_y} \f{  \sinh(\i\zeta + y_{b} ) }{ \sinh(\i\zeta - y_{b} ) } \, + \, 1 \bigg| > \varrho $  ;
 \item   for $a =1,\dots, n_y$,    
\beq
y_a \in \Big\{ z \in \Cx \; : \; |\Im(z)| \, \leq \,  \f{\pi}{2} \; , \; z \not\in \mc{D}_{\pm \i \zeta_{\e{m}}, \varrho} \cup \{0\} \Big\} \quad  with \quad \zeta_{\e{m}} \, = \, \e{min}\big\{ \zeta, \pi - \zeta \big\}  \;. 
\nonumber
\enq
\end{enumerate}

Finally, let $h_1,\dots, h_{n_x} \in \mathbb{Z}$ be pairwise distinct. Then, there exists a correlation length $\xi_k$ whose large-$T$ asymptotics take the form 

\beq
\ex{-\f{1}{\xi}_k } \cdot \ex{-\f{f}{T}} \; = \;  \f{1}{T^{n_x} }   \pl{a=1}{n_x} \bigg\{ \f{ -2 \i J  }{ (2h_a+1+n_x-n_y)\pi-\sum_{a=1}^{n_y}\th_+(-y_a) }   \bigg\}   \pl{a=1}{n_y} \f{ \sinh(y_a-\i\zeta) }{ \sinh(y_a)}
\enq
where $-\i\th$ is a determination of the logarithm of $\la \mapsto \tf{ \sinh(\i\zeta + \la) }{ \sinh(\i\zeta - \la)  }$ and the $+$ subscript in $\th_{+}$
indicates that one should take the $+$ boundary value if $-y_a$ is located on the logarithm's cut.

\end{theorem}

This theorem thus provides a classification of a subset of the correlation lengths of the spin-$1/2$ XXZ chain in terms of 
solutions to the Bethe Ansatz equations \eqref{ecriture BAE XXZ spin 1} describing the spectrum of a spin-1 XXZ chain of length $n_x$, 
\textit{c.f.} \cite{SogoWadatRederivationLaxMatrixXYZandHigherSpinXXX,SogoLowLyingExcitationsHigherSpinXXZCalculation}. 
Also, note that the choice of the determination of the logarithm in Theorem \ref{Theorem longueur de correlations} is irrelevant in that it simply results in shifts, by a global integer, of 
the integers $h_1,\dots, h_{n_x}$.

\subsection{Notations}

\begin{itemize}

  \item Given $\a>0$, $z_0\in \Cx$, we shall denote
\beq
\mc{S}_{\a} \; = \; \Big\{ z \in \Cx \, : \, |\Im(z)|<\a \Big\}  \qquad \e{and} \qquad 
\mc{D}_{z_0,\a} \; = \; \Big\{ z \in \Cx \, : \, |z-z_0|<\a \Big\} \;. 
\label{definition bande largeur alpha et disque}
\enq

  \item Given an open subset $U\subset \Cx$, $\mc{O}(U)$ stands for the ring of holomorphic functions on $U$, and for any function $g$ on $U$,
  we denote $\norm{g}_{L^{\infty}(U)} \; = \; \e{supess}_{u \in U} |g(u)|$. 
 
 \item  Given $r>0$, set 
\beq
\mc{B}_{ r } \; = \; \Big\{ \xi \in \mc{O}(\mc{S}_{\tf{\zeta_{\e{m}} }{2}}) \, : \, \norm{ \xi }_{ L^{\infty}(\mc{S}_{\tf{ \zeta_{\e{m}} }{2}})} \, \leq \, r   \Big\}
\quad \e{with} \quad  \zeta_{\e{m}} \,  = \, \e{min}(\zeta, \pi-\zeta)\;. 
\label{defintion fct holomorphes bornees sur la bande}
\enq

\item For integers $a<b$, $\intn{a}{b}=\{a,a+1,\dots,b\}$. 


 \item Consider $n$ complex numbers $z_1,\dots, z_n \in \Cx$, distinct or not. If some roots coincide, the total number of roots equal to a given complex number $z$ is called their multiplicity and denoted $k_{z}$. 
 One then defines the set 
\beq
\{ z_a \}_{1}^{n} \; = \; \Big\{ (\la, k_{\la} ) \; : \; \la \in \{z_1,\dots, z_n \}  \Big\}
\enq
in which  $\{z_1,\dots, z_n \} $ stands for the usual set build up from the numbers  $z_1,\dots, z_n$. 
\item Given $\Om$ a finite set and a map  $n : (\Om,x)\tend (\mathbb{Z}, n_x) $, the weighted cardinality $|A|$ of the  set $A=\big\{ (x, n_x) \, : \, x \in \Om \big\}$ is defined by $|A|=\sul{ x \in \Om }{} n_x$. 

\item Given a function $f$ on the set $\Om$ and $A$ defined as above, we agree upon the shorthand notation 
\beq
\sul{\la \in A}{} f(\la) \, \equiv \,\sul{x \in \Om }{} n_x \,  f(x)  \qquad \e{and} \qquad \pl{\la \in A}{} f(\la) \, \equiv \, \pl{x \in \Om }{} \big\{ f(x)\big\}^{n_x}  \;. 
\enq

\item Given two sets $A=\big\{ (x, n_x) \, : \, x \in \Om_{A} \big\}$ and $B=\big\{ (y, n_y) \, : \, x \in \Om_{B} \big\}$, one defines their algebraic sum $\oplus$ and difference $\ominus$
as 
\beq
A \oplus B \; = \; \Big\{ ( x, n_x+m_x ) \; : \; x \in \Om_A\cup \Om_B \Big\} \;, \qquad A \ominus B \; = \; \Big\{ ( x, n_x-m_x ) \; : \; x \in \Om_A\cup \Om_B \Big\}
\label{definition difference algebrique ensemble}
\enq
in which one understands that the maps $n$ and $m$ are extended as $n_x=0$, resp. $m_x=0$, on $\Om_{B}\setminus \Om_{A}$, resp. $\Om_{A}\setminus \Om_{B}$. 
Furthermore, if $\Om_A$, $\Om_B$ are two sets, then $\Om_A \ominus \Om_B \; \equiv  \; A\ominus B$, where $A,B=\big\{ (x, 1) \, : \, x \in \Om_{A,B} \big\}$

\item These notation allow one to introduce  compact notations for sums and products. Given a function $f$ on $\Om_A\cup\Om_B$ parameterising sets $A,B$ as above,our conventions imply
\beq
\sul{\la \in A\ominus B}{}   f(\la)   \; = \; 
\sul{x \in \Om_A}{} n_x  f(x) \, - \,  \sul{y \in \Om_B}{} m_y f(y)  \;\;  ,   \quad
\pl{\la \in A\ominus B}{} f(\la) \, = \,  \f{ \pl{ x \in \Om_A }{} \big\{ f(x) \big\}^{n_x}  }{  \pl{y \in \Om_B }{}  \big\{ f(y) \big\}^{m_y}  }   \;. 
\label{defintion convention somme produit et cardinalite ensembles}
\enq

\item Given a point $x \in \Cx$, $\{x\}^{\oplus n }$ denotes the set $\{(x,n)\}$, meaning that one should understand 
\beq
\sul{t \in \{x\}^{\oplus n} }{} f(t) \equiv n f(x) \, . 
\label{definition ensemble repetee}
\enq

\item Given an operator $\op{A}$,  $|||\op{A}|||$ stands for its norm,  $\sg(\op{A})$ for its spectrum and $r_{S}(\op{A})$ for its spectral radius. 
  
\item $\ln$ stands for the principal branch of the logarithm continued to $\R^{-}\setminus \{0\}$ with the convention $\e{arg}(z) \in \intfo{-\pi}{\pi}$.   
  
\end{itemize}

\subsection{Outline of the paper}

The paper is organised as follows. In Section \ref{Section Comportement Grand T traces de QTM} we develop a setting allowing one to establish, for $T$ large enough, various 
structural properties of the spectrum of the quantum transfer matrix. In Section \ref{Section VP dominante et echange limites} we apply these results to the proof of properties 
i)-ii) stated earlier on. Then, in Section \ref{Section connection spectre qtm et NLIE}, we recall the connection, at finite Trotter number, between the Bethe Ansatz 
approach to the characterisation of the spectrum of $\op{t}_{\mf{q}}$ and non-linear integral equations. In Section \ref{Section existance et unicite sols NLIE} we develop the rigorous 
treatment, for $T$ large enough, of these non-linear integral equations. In Section \ref{Section caracterisation VP a haute tempe}, we  gather all of the previous results
so as to establish Theorem \ref{Theorem principal aticle f a T fini}. As a byproduct we also characterise the expression, at high $T$, of the model's correlation lengths. 
Quite remarkably, these appear to be governed by solutions to the Bethe Ansatz equation for the spin-1 XXZ chain.
Finally, in Section \ref{Section analyse numerique}, we illustrate our analysis by providing numerical results for the solution sets
 to the Bethe Ansatz equations describing the spectrum 
of the quantum transfer matrix at finite Trotter number.




\section{\boldmath Traces of powers of the quantum transfer matrix at high temperatures}

\label{Section Comportement Grand T traces de QTM}

It is useful to recall that given an elementary matrix $\op{E}^{ik}$ acting on $\mc{L}(\Cx^2)$,  $\op{E}_{a}^{ik}$  stands for its canonical embedding 
into an operator on $\mf{h}_0\otimes \mf{h}_{\mf{q}}$ acting non-trivially only on the $a^{\e{th}}$ factor in the tensor product:
\beq
\op{E}_{a}^{ik} \, = \, \underbrace{ \e{id}\otimes \cdots  \otimes \e{id} }_{a }  \otimes \, \op{E}^{ik} \otimes \underbrace{ \e{id} \otimes \cdots \otimes \e{id} }_{ 2N - a } \;. 
\enq
This notation allows one to write down embeddings of operators on $\Cx^2\otimes \Cx^2$ into operators on $\mf{h}_0\otimes \mf{h}_{\mf{q}}$. 
For instance, $\op{P}_{ab}$ stands for the embedding of the permutation operator  $\op{P}$  on $\Cx^2\otimes \Cx^2$ into operators on $\mf{h}_0\otimes \mf{h}_{\mf{q}}$:
$\op{P}_{ab}$ acts as $\op{P}$ on $\mf{h}_{a}\otimes \mf{h}_{b}$ and as the identity operator on the other spaces in the tensor product decomposition of $\mf{h}_0\otimes \mf{h}_{\mf{q}}$. 
This operator, along with its partial transpose on the $a^{\e{th}}$ space $\op{P}_{a\, b}^{\mf{t}_a} $, can be recast in terms of the elementary matrices introduced above as
\beq
\op{P}_{a\, b} \; = \; \sul{i,k=1}{2} \op{E}_{a}^{ik}\, \op{E}_{b}^{ki} \; , \qquad
\op{P}_{a\, b}^{\mf{t}_a} \; = \; \sul{i,k=1}{2} \op{E}_{a}^{ik} \, \op{E}_{b}^{ik} \;. 
\label{ecriture expression explicite matrice permutation et sa transposee}
\enq
The explicit expression for the six-vertex $\op{R}$-matrix \eqref{ecriture matrice R six vertex} ensures that 
\beq
  \op{R}_{ab}\big( -\tfrac{\aleph}{N}\big) \, = \, \op{P}_{ab} + \op{N}_{ab} \;, 
\enq
where 
\beq
\op{N}_{a  b}\; = \;    \sul{i,k=1}{2} n_{ik} \cdot \op{E}_{a}^{ii}\, \op{E}_{b}^{kk} \qquad \e{with} \quad 
				\left\{ \ba{ccl}  n_{11}=n_{22} & = & \cosh\big( \tfrac{\aleph}{N}\big) -1 -\coth(\eta) \sinh\big( \tfrac{\aleph}{N}\big)  \vspace{2mm}  \\
						n_{12}=n_{21} & = &  -\sinh\big( \tfrac{\aleph}{N}\big) \big/\sinh(\eta)  	      \ea \right.  \;. 
\enq
Define the operators
\beq
\Pi_{ \ell } \, = \, \op{P}_{2\ell\,  0}^{ \mf{t}_{2\ell} } \op{P}_{0\, 2\ell-1}^{}
\enq
and
\beq
\op{W}_{\ell}\, = \,  \op{R}_{2\ell \, 0}^{\mf{t}_{2\ell}} \Big( -\tfrac{\aleph}{N}  \Big) \cdot  \op{R}_{0 \, 2\ell-1}\Big(  -\tfrac{\aleph}{N}\Big) \; - \;  \Pi_{ \ell } \;.  
\enq
Further, given $\ell \geq m$, define
\beq
\Om_{\ell;m} \; = \; \left\{ \ba{ccc}  
\Pi_{\ell} \cdots \Pi_{m+1}\cdot \ex{  \f{h}{2T} \sg^z_0 \, \de_{m 0} }  & \e{if}&   \ell \geq m+1 \vspace{2mm} \\
 \ex{  \f{h}{2T} \sg^z_0 \, \de_{m 0} }  & \e{if}&   \ell = m \ea \right.   \;, 
\label{definition operateur Om ell m}
\enq
where $\de_{ab}$ stands for the Kronecker symbol.

Upon  expressing each factor associated to  a pair of spaces $\mf{h}_{2\ell-1}\otimes \mf{h}_{2\ell}$ in \eqref{ecriture matrice de monodromie quantique} as
$\op{W}_{\ell}+\Pi_{\ell}$ one obtains that 
\beq
\op{T}_{\mf{q};0}(0) \; = \; \Om_{ N ; 0 } \; + \; \sul{ n=1 }{ N } \sul{ \bs{\ell} \in \mc{L}^{(n)}_N    }{} \bs{\mc{O}}_{\bs{\ell} } \qquad \e{with} \qquad 
\bs{\mc{O}}_{ \bs{\ell} } \; = \; \Om_{N;\ell_{n}} \cdot \op{W}_{\ell_n} \cdot \Om_{\ell_{n}-1;\ell_{n-1}} \, \cdots \, \op{W}_{ \ell_1 } \cdot \Om_{\ell_{1}-1;0} \; , 
\label{ecriture developpement QMM}
\enq
$\op{T}_{\mf{q};0}(\xi)$ as given in \eqref{ecriture matrice de monodromie quantique} and where the summation runs through
\beq
 \mc{L}^{(n)}_N \, = \, \Big\{  \bs{\ell} =(\ell_1,\dots, \ell_n) \; : \; 1 \leq\ell_1 <\dots<\ell_n \leq N \Big\} \;. 
\label{definition ensemble mathcal L n N}
\enq
The decomposition \eqref{ecriture developpement QMM} entails an analogous expansion for the quantum transfer matrix
\beq
\op{t}_{\mf{q}}\;=\; \bs{\om}_{N;0} + \de \op{t}_{\mf{q}} \;,
\label{ecriture decompotion qtm partie dominante et perturbation}
\enq
where 
\beq
\bs{\om}_{N;0} \; = \; \e{Tr}_{0}\Big[ \Om_{N;0} \Big]
\label{definition operateur dominant rang 1 pour QTM}
\enq
and
\beq
\de \op{t}_{\mf{q}} \, = \, \sul{ n=1 }{ N } \sul{ \bs{\ell} \in \mc{L}^{(n)}_N  }{} \op{O}_{ \bs{\ell} }  \qquad \e{with}  \qquad \op{O}_{ \bs{\ell} }\, = \, \e{tr}_0\big[ \bs{ \mc{O} }_{ \bs{\ell} }  \big]    \;. 
 \label{definition op perturbant QTM}
\enq

It appears  useful for further purposes to introduce a convention for writing vectors in $\mf{h}_{\mf{q}} \, = \,\bigotimes_{a=1}^{2N} \mf{h}_a $. Given vectors $\bs{v}_1,\dots, \bs{v}_{2N}$
 in $\mf{h}_1,\dots, \mf{h}_{2N}$, we shall write
\beq
\pl{a=1}{2N} \bs{v}_{a}^{(a)} \; = \; \bs{v}_1\otimes \cdots \otimes \bs{v}_{2N} \;, 
\enq
\textit{viz}. $ \bs{v}_{a}^{(a)}$ stands for the $a^{\e{th}}$ vector appearing in the full tensor product. Further, we shall denote by $\bs{e}_{a}$, $a=1,2$, the canonical basis of $\Cx^2$:
\beq
\bs{e}_1=\left( \ba{c} 1 \\ 0 \ea \right) \quad , \quad \bs{e}_2 = \left( \ba{c} 0 \\ 1 \ea \right)  \;. 
\enq

\begin{lemme}
 \label{Lemme pte proj dominant rang 1}
The operator $\bs{\om}_{N;0}$ introduced in \eqref{definition operateur dominant rang 1 pour QTM} has rank $1$ and takes the form 
\beq
\bs{\om}_{N;0} \; = \; \bs{v}\cdot \bs{w}^{\mf{t} } \quad  with  \quad
\left\{ \ba{ccl}  \bs{v} &=&  {\displaystyle \sul{ \bs{i}\in \{1,2\}^N }{} }  \ex{ \f{h}{2T} \veps_{i_{N}} } \pl{s=1}{N}\Big\{ \bs{e}_{i_s}^{(2s)} \bs{e}_{i_{s-1}}^{(2s-1)} \Big\}   \\ 
		 \bs{w} &=&   {\displaystyle \sul{ \bs{j}\in \{1,2\}^N }{} }    \pl{s=1}{N}\Big\{ \bs{e}_{j_s}^{(2s)} \bs{e}_{j_{s}}^{(2s-1)} \Big\}    \ea \right. \; , 
\enq
where $\veps_i = (-1)^{i-1}$ and where we made use of periodic boundary conditions for the indices of $\bs{i}=(i_1,\dots, i_{N})$, \textit{viz}. $i_0\equiv i_N$. Furthermore, it holds that
\beq
\big( \bs{w}, \bs{v} \big) \; = \; 2 \cosh\Big( \f{ h }{ 2 T } \Big) \quad, \qquad \norm{ \bs{v} }^2 =2^N \cosh\Big( \f{h}{T} \Big)  \qquad  and  \qquad  \norm{ \bs{w} }^2 =2^N \;. 
\enq

\end{lemme}

\Proof

Upon inserting the expression for the permutation operators in terms of elementary matrices \eqref{ecriture expression explicite matrice permutation et sa transposee}, one obtains that 
the operator $\Om_{\ell;m}$ introduced in \eqref{definition operateur Om ell m} can be expressed as
\beq
\Om_{ \ell ; m }  \; = \; \sul{ \substack{ i_s,  k_s \in \{1,2\} \\ s= 2m+1, \dots, 2 \ell}  }{} \pl{s=2m+1}{2\ell} \Big\{ \op{E}_{s}^{i_{s}k_{s}}   \Big\} 
\; \cdot \op{E}_{0}^{ i_{2\ell} k_{2\ell} } \, \op{E}_{0}^{k_{2\ell-1} i_{2\ell-1}}\cdots \op{E}_{0}^{k_{2m+1} i_{2m+1}}   \cdot \ex{\f{h}{2T} \veps_{i_{2m+1}}\de_{m 0} } \;. 
\enq
The algebra of elementary matrices $\op{E}^{ab}\op{E}^{cd}=\de_{bc}\op{E}^{ad}$ then allows one to simplify the product of elementary matrices over the auxiliary space $0$
as
\beq
 \op{E}_{0}^{ i_{2\ell} k_{2\ell} } \; \op{E}_{0}^{k_{2\ell-1} i_{2\ell-1}}\cdots \op{E}_{0}^{k_{2m+1} i_{2m+1}} \; = \;
	\op{E}_{0}^{i_{2\ell} i_{2m+1}} \cdot \pl{s=m+1}{\ell} \Big\{ \de_{k_{2s}k_{2s-1}} \Big\}  \cdot \pl{s=m+2}{\ell} \Big\{ \de_{i_{2s-1}i_{2s-2}} \Big\} \;. 
\enq
The Kronecker symbols  allow one to get rid of the summation over the odd labeled variables $i_{2s-1}$, $s=m+2,\dots, \ell$, and $k_{2s-1}$, $s=m+1,\dots, \ell$. Then, relabeling in the sum 
the even labeled variables as 
\beq
\{k_{2s}, i_{2s}\} \hookrightarrow \{ k_s^{\prime},i_s^{\prime}\} \; ,  \quad s=m+1,\dots, \ell \quad \e{and} \quad   i_{2m+1} \hookrightarrow  i_{m}^{\prime} \; ,
\enq
one obtains 
\beq
\Om_{ \ell ; m }  \; = \; \sul{ \{ i_s \}_{m}^{\ell}     }{} \sul{ \{ k_s \}_{m+1}^{\ell} }{}  \pl{s=m+1}{\ell} \Big\{ \op{E}_{2s}^{i_{s}k_{s}} \, \op{E}_{2s-1}^{i_{s-1}k_{s}} \Big\} 
\cdot \op{E}_{0}^{i_{\ell} i_{m}}  \cdot \ex{\f{h}{2T} \veps_{i_{m}}\de_{m 0} } \;. 
\label{ecriture representation Omega ell m}
\enq
Above, it is undercurrent that each $i_s$ or $k_s$ runs through the set $\{1,2\}$. At this stage, one can readily take the trace over the auxiliary space, hence leading to 
\beq
\bs{\om}_{N;0} \; = \;  \sul{ \{ i_s, k_s \}_{1}^{N}     }{} \ex{\f{h}{2T} \veps_{i_{N}}  }  \cdot  \pl{s=1}{N} \Big\{ \op{E}_{2s}^{i_{s}k_{s}} \, \op{E}_{2s-1}^{i_{s-1}k_{s}} \Big\}  \;. 
\enq
Here, we stress that the dependence on the $\e{id}$ operator on the auxiliary space $\mf{h}_0$ has been projected out in the elementary matrices appearing above.
Finally, observe that an elementary matrix on $\mc{M}_{2}(\Cx)$ can be recast as $\op{E}^{ab}=\bs{e}_a\cdot \big( \bs{e}_b \big)^{\mf{t}}$. This decomposition entails that 
\beq
 \pl{s=1}{N} \Big\{ \op{E}_{2s}^{i_{s}k_{s}} \, \op{E}_{2s-1}^{i_{s-1}k_{s}} \Big\}  \; = \;  \pl{s=1}{N} \Big\{  \bs{e}_{i_s}^{(2s)} \bs{e}_{i_{s-1} }^{(2s-1)}  \Big\}
\cdot  \Big\{ \pl{s=1}{N}   \bs{e}_{k_s}^{(2s)} \bs{e}_{k_{s} }^{(2s-1)}  \Big\}^{ \mf{t} } \;. 
\enq
The above factorisation allows one for a separation of the sums over the $i_a$'s and the $k_a$'s and leads to the claimed form of $\bs{\om}_{N;0}$. 
Finally, the value of the scalar product of $\bs{v}$, $\bs{w}$ along with their norms follows from direct calculations. \qed

Observe that one has the representation 

\beq
\op{W}_{\ell} \, = \, \sul{i,k=1}{2} \op{M}_{\ell}^{ik}\, \op{E}_{0}^{ik} \;, \qquad \e{where} \qquad
\op{M}_{\ell}^{ik} \; = \; \underbrace{ \e{id}\otimes \cdots \otimes \e{id} }_{ 2\ell-2 } \otimes \, \op{M}^{ik} \otimes \underbrace{ \e{id}\otimes \cdots \otimes \e{id} }_{ 2N-2\ell }  
\enq
and
\beq
\op{M}^{ik} \, = \, \sul{ \a = 1 }{ 4 } \bs{v}_{ik;\a} \cdot  \big( \bs{w}_{ik;\a} \big)^{ \mf{t} }
\enq
for some explicitly computable vectors $\bs{v}_{ik;\a}, \bs{w}_{ik;\a} $ which are normalised such that
\beq
\norm{ \bs{v}_{ik;\a} } \;= \; \frac{1}{2} \qquad \e{and} \qquad \norm{ \bs{w}_{ik;\a} } \; \leq \; C_{\bs{w}} \Big|\f{\aleph}{N} \Big| 
\label{ecriture estimation vecteurs v et w}
\enq
for a constant $C_{\bs{w}}>0$ and any $\a\in \intn{1}{4}$. 
%
%
%
%
%
%
%
%

\begin{lemme}
\label{Lemme propriete decomposition operateur o elln sur operateur rang 1}

Let $\bs{\ell}=(\ell_1,\dots, \ell_n)$. Then the operator $\op{O}_{ \bs{\ell}  }$ defined in \eqref{definition op perturbant QTM} can be decomposed as 
\beq
\op{O}_{ \bs{\ell}  }  \, = \, \sul{ \bs{i},\, \wh{\bs{j}}_{\bs{\ell} }  }{} \sul{ \bs{\a}_{\bs{\ell} } }{}  
\bs{x}^{ (\bs{\a}_{\bs{\ell} },\bs{\ell} ) }_{\bs{i}} \cdot \Big( \bs{y}^{ (\bs{\a}_{\bs{\ell} },\bs{\ell} ) }_{\bs{i}, \, \wh{\bs{j}}_{\bs{\ell} }  }\Big)^{\mf{t}} \;. 
\enq
Above, one sums over vectors 
\beq
\bs{\a}_{\bs{\ell} }\, = \, (\a_{\ell_1},\dots, \a_{\ell_n} \big) \in \intn{1}{4}^n \;, 
\label{definition alpha}
\enq
and
\beq
\bs{i}=(i_1,\dots, i_N)\in \{1,2\}^N \quad ,  \qquad \wh{\bs{j}}_{\bs{\ell}_n}  \, = \, (j_1,\dots,j_{\ell_1-1}, \wh{j_{\ell_1}},j_{\ell_1+1}, \dots, \wh{j_{\ell_n}},\dots, j_N)\in \{1,2\}^{N-n} \;  .
\enq
The  $\, \wh{} \, $ indicates the coordinates which are omitted.   Furthermore, one has
\beqa
 \bs{x}^{ (\bs{\a}_{\bs{\ell}},\bs{\ell} ) }_{\bs{i}} & = &  \ex{ \f{h}{2T} \veps_{i_N} } \pl{ \substack{ s=1 \\ \not= \ell_1,\dots, \ell_n}  }{ N } \Big\{ \bs{e}^{(2s)}_{i_s} \bs{e}^{(2s-1)}_{ i_{s-1} }  \Big\}
\cdot \pl{r=1}{n} \bs{v}_{ i_{\ell_r}   i_{\ell_r-1} ;\a_{\ell_r} }^{ (\ell_r)} \label{definition vecteur x} \\
\bs{y}^{ (\bs{\a}_{\bs{\ell} },\bs{\ell} ) }_{\bs{i}, \wh{\bs{j}}_{\bs{\ell}} } & = & 
\pl{ \substack{ s=1 \\ \not= \ell_1,\dots, \ell_n}  }{ N } \Big\{ \bs{e}^{(2s)}_{j_s} \bs{e}^{(2s-1)}_{ j_{s} }  \Big\}
\cdot \pl{r=1}{n} \bs{w}_{ i_{\ell_r}   i_{\ell_r-1} ;\a_{\ell_r} }^{(\ell_r)} \;. 
\label{definition vecteur y} 
\eeqa
Note that, in the above expression for $ \bs{x}^{ (\bs{\a}_{\bs{\ell} },\bs{\ell} ) }_{\bs{i}}$, we use periodic boundary conditions for the indices of $\bs{i}$, \textit{viz}. $i_N\equiv i_0$. 
\end{lemme}

In \eqref{definition vecteur x}, \eqref{definition vecteur y} $\bs{v}_{ ij ;\a}^{ (\ell)}$, resp. $\bs{w}_{ ij ;\a}^{ (\ell)}$, means that the vector $\bs{v}_{ij;\a}$, resp. $\bs{w}_{ij;\a}$, 
appears on the position reserved to the spaces $\mf{h}_{2\ell-1}\otimes\mf{h}_{2\ell}$ in the tensor product decomposition of the full vector.

\Proof 
Inserting the expressions for each operator and using the representation \eqref{ecriture representation Omega ell m} for $\Om_{\ell;m}$ yields that 
\bem
\bs{\mc{O}}_{ \bs{\ell}  } \; = \; \sul{  \big\{ \{ i_s^{\, (r)} \}_{s=\ell_r}^{\ell_{r+1}-1}\big\}_{r=0}^{n}  }{}  \sul{  \big\{ \{ j_s^{\, (r)} \}_{s=\ell_r+1}^{\ell_{r+1}-1}\big\}_{r=0}^{n}  }{} 
\sul{ \{q_t, p_t\}_1^{n} }{} \pl{r=0}{n} \pl{ s = \ell_r+1 }{ \ell_{r+1} - 1 } \Big\{  \op{E}_{2s}^{ i_{s}^{\,(r)} j_{s}^{\, (r)} } \, \op{E}_{2s-1}^{ i_{s-1}^{\,(r)} j_{s}^{\,(r)} } \Big\} 
\cdot \pl{t=1}{n} \op{M}_{ \ell_t }^{ q_t p_t } \\
\times \op{E}_{0}^{ i_{N}^{\, (n)} i_{\ell_n}^{\, (n)} }\, \op{E}_{0}^{ q_n p_n  } \, \op{E}_{0}^{ i_{\ell_n-1}^{\, (n-1)} i_{\ell_{n-1} }^{\, (n-1)}} \, \op{E}_{0}^{ q_{n-1} p_{n-1}  }
\cdots  \op{E}_{0}^{ q_{1} p_{1}  } \, \op{E}_{0}^{ i_{\ell_1-1}^{\, (0)} i_{0}^{\, (0)} }  \ex{\f{h}{2T} \veps_{ i_{0}^{\, (0)} } } \;. 
\end{multline}
Here, we agree upon $\ell_0=0$ and $\ell_{n+1}=N+1$. The last line can be simplified as
\beq
\op{E}_{0}^{ i_{N}^{\, (n)} i_{\ell_n}^{\, (n)} }\, \op{E}_{0}^{ q_n p_n  } \, \op{E}_{0}^{ i_{\ell_n-1}^{\, (n-1)} i_{\ell_{n-1} }^{\, (n-1)}} \, \op{E}_{0}^{ q_{n-1} p_{n-1}  }
\cdots  \op{E}_{0}^{ q_{1} p_{1}  } \, \op{E}_{0}^{ i_{\ell_1-1}^{\, (0)} i_{0}^{\, (0)} } 
\; = \; \op{E}_{0}^{ i_{N}^{\, (n)} i_{0}^{\, (0)} } \cdot \pl{s=1}{n} \Big\{ \de_{ i_{\ell_s}^{\, (s)} q_s} \Big\}   \cdot \pl{s=1}{n} \Big\{ \de_{ p_s i_{\ell_{s}-1}^{\, (s-1)} } \Big\} \;. 
\enq
These Kronecker symbols allow one to compute the sums over the $p_t$ and $q_t$ indices. In order to have a more compact expression it is 
useful to change the summation variables as 
\beq
\ba{ccccc} 		 i_a = i_a^{\, (r)}   &  \e{if}  & \ell_r \leq a \leq \ell_{r+1}-1   & \e{with} & a\in \intn{0}{ N }  \vspace{2mm} \\
		      j_a = j_a^{\, (r)}   &  \e{if}  & \ell_r \leq a \leq \ell_{r+1}-1   & \e{with} & a\in \intn{1}{ N } \setminus \{\ell_1,\dots, \ell_n\} \ea \;. 
\enq
 Then, one has
\beq
\bs{\mc{O}}_{ \bs{\ell}  } \; = \; \sul{   \{ i_a \}_{ a=0 }^{ N }  }{}  \sul{  \substack{  \{ j_a \}_{ a=1 }^{ N } \\  a \not= \ell_1,\dots,\ell_{n} }  }{ } \; 
\pl{ \substack{ s = 1 \\ \not= \ell_1,\dots, \ell_{n} } }{ N } \Big\{  \op{E}_{2s}^{ i_{s} j_{s} } \, \op{E}_{2s-1}^{ i_{s-1}  j_{s}  } \Big\} 
\cdot \pl{r=1}{n} \Big\{  \op{M}_{ \ell_r }^{ i_{\ell_r} i_{\ell_{r}-1} } \Big\} \cdot \op{E}^{ i_N i_0 }_{0} \, \ex{ \f{h}{2T} \veps_{i_0}  } \;. 
\enq
At this stage, one may readily take the trace over the auxiliary space $\mf{h}_0$. Upon projecting out the dependence on the $\e{id}$ operator on $\mf{h}_0$, it remains to observe that 
\bem
 \pl{ \substack{ s = 1 \\ \not= \ell_1,\dots, \ell_{n} } }{ N } \Big\{  \op{E}_{2s}^{ i_{s} j_{s} } \, \op{E}_{2s-1}^{ i_{s-1}  j_{s}  } \Big\} 
\cdot \pl{r=1}{n} \Big\{  \op{M}_{ \ell_r }^{ i_{\ell_r} i_{\ell_{r}-1} } \Big\}    \\ 
\; = \;  \sul{ \bs{\a}_{\bs{\ell}} }{} \pl{ \substack{ s=1 \\ \not= \ell_1,\dots, \ell_n}  }{ N } \Big\{ \bs{e}^{(2s)}_{i_s} \bs{e}^{(2s-1)}_{ i_{s-1} }  \Big\}
\cdot \pl{r=1}{n} \bs{v}_{ i_{\ell_r}   i_{\ell_r-1} ;\a_{\ell_r} }^{ (\ell_r)}  \cdot 
\Bigg\{   \pl{ \substack{ s=1 \\ \not= \ell_1,\dots, \ell_n}  }{ N } \Big\{ \bs{e}^{(2s)}_{j_s} \bs{e}^{(2s-1)}_{ j_{s} }  \Big\}
\cdot \pl{r=1}{n} \bs{w}_{ i_{\ell_r}   i_{\ell_r-1} ;\a_{\ell_r} }^{(\ell_r)}  \Bigg\}^{\mf{t}}  
\end{multline}
where $\bs{\a}_{\bs{\ell}}$ is as introduced in \eqref{definition alpha}. This entails the claim. \qed

\vspace{3mm}

Given two vectors $\bs{\ell} \in \mathbb{N}^n$, $\bs{r}\in \mathbb{N}^m$, introduce the sets 
\beq
\mc{S}_{ \bs{\ell}  } \, = \, \big\{\ell_1, \dots, \ell_n \big\} \quad, \qquad \mc{S}_{ \bs{r}  } \, = \, \big\{ r_1, \dots, r_m \big\} \;. 
\enq
Further define
\beq
\ba{ccc} 
\mc{S}_{ \bs{\ell}  \cap \bs{r}   } \; = \; \mc{S}_{ \bs{\ell} } \cap \mc{S}_{ \bs{r}  }  &  , \quad &
\mc{S}_{ \bs{\ell}  \cup \bs{r}   } \; = \; \mc{S}_{ \bs{\ell}  } \cup \mc{S}_{ \bs{r}  }  \vspace{2mm} \\
\mc{S}_{ \bs{\ell}    }^{ \e{c} } \; = \; \mc{S}_{ \bs{\ell} } \setminus  \mc{S}_{ \bs{\ell}  \cap \bs{r}    }
 &  , \quad &
\mc{S}_{  \bs{r}     }^{ \e{c} } \; = \; \mc{S}_{  \bs{r}   } \setminus  \mc{S}_{ \bs{\ell}  \cap \bs{r}   } 
\ea \;. 
\enq

\begin{lemme}
 
Let 
\beq
\mc{S}_{ \bs{\ell}  \cup \bs{r}   } \; = \; \big\{t_1,\dots, t_u \big\} \qquad with \qquad  1 \leq t_1<\cdots < t_u \leq N \;. 
\enq
Then
\bem
\op{O}_{ \bs{\ell} } \cdot \op{O}_{ \bs{r} } \, = \, \sul{ \bs{i} , \wh{ \bs{q} }_{ \bs{r}  }   }{ } \sul{ \bs{\a}_{ \bs{\ell}} , \bs{\be}_{ \bs{r}  }  }{ }
\Big( 2 \cosh\big[ \tfrac{h}{2T} \big] \Big)^{ \de_{n,0} \de_{m,0} }  \cdot 
\bs{x}^{ (\bs{\a}_{\bs{\ell} } ,\bs{\ell}  ) }_{\bs{i}}
\cdot 
\Bigg\{ \sul{ \{ p_{t_a} \}_{a=1}^{u} }{} \ex{\f{h}{2T} \veps_{ p_{t_u} } }  
\pl{ t_k \in \mc{S}_{ \bs{\ell} \cap \bs{r}  } }{} \bigg(  \bs{w}_{ i_{t_k}   i_{t_k-1} ;\a_{t_k} }  , \bs{v}_{ p_{t_k}   p_{t_{k-1}} ;\, \be_{t_k} }  \bigg)  \\ 
\times \pl{ t_k \in \mc{S}_{ \bs{\ell}  }^{\e{c}} }{} \bigg(  \bs{w}_{ i_{t_k}   i_{t_k-1} ;\a_{t_k} }  ,   \bs{e}_{ p_{t_{k-1}} }\otimes \bs{e}_{ p_{t_k} }    \bigg) 
\cdot \pl{ t_k \in \mc{S}_{ \bs{r}  }^{\e{c}} }{} \bigg(  \bs{u}  , \bs{v}_{ p_{t_k}   p_{t_{k-1}} ;\, \be_{t_k} }   \bigg)  \Bigg\}
\cdot \Big( \bs{y}^{ (\bs{\be}_{ \bs{r}  } ,\bs{r} ) }_{\bs{p}_{t},  \wh{\bs{q}}_{ \bs{r}  } }\Big)^{\mf{t}} \;. 
\label{resultat ecriture product deux operateurs}
\end{multline}
Above, we use periodic boundary conditions on the indices of $t_a$, \textit{viz}. $t_0=t_u$, we agree upon 
\beq
\bs{u}  \, = \, \bs{e}_1 \otimes \bs{e}_1 \, + \, \bs{e}_2 \otimes \bs{e}_2  \;,  
\enq
and  have set 
\beq
\bs{p}_t \, = \, \Big( \underbrace{ p_{t_u}, \dots, p_{t_u}}_{t_1-1}, \underbrace{ p_{t_1},\dots, p_{t_1} }_{t_2-t_1}, \dots, \underbrace{ p_{t_{u-1}},\dots, p_{t_{u-1}} }_{t_u-t_{u-1}}, 
\underbrace{ p_{t_u},\dots, p_{t_u} }_{N-t_u+1} \Big) \;. 
\enq

\end{lemme}

\Proof 
 
 Obviously, one has 
\beq
\op{O}_{ \bs{\ell}  } \cdot \op{O}_{ \bs{r}  } \, = \,  \sul{ \bs{i},\, \wh{\bs{j}}_{\bs{\ell} }  }{} \sul{ \bs{\a}_{\bs{\ell} } }{}  
 \sul{ \bs{p},\, \wh{\bs{q}}_{\bs{r} }  }{} \sul{ \bs{\be}_{\bs{r} } }{}  
\bs{x}^{ (\bs{\a}_{\bs{\ell} },\bs{\ell} ) }_{\bs{i}} \cdot
\bigg( \bs{y}^{ (\bs{\a}_{\bs{\ell} },\bs{\ell} ) }_{\bs{i}, \, \wh{\bs{j}}_{\bs{\ell} }  } \, , \bs{x}^{ (\bs{\be}_{\bs{r} },\bs{r} ) }_{\bs{p}} \bigg)   
\cdot \Big( \bs{y}^{ (\bs{\be}_{ \bs{r}  } ,\bs{r} ) }_{\bs{p} ,  \wh{\bs{q}}_{ \bs{r}  } }\Big)^{\mf{t}} \;.
\label{ecriture produit deux operateurs comme somme combinatoire}
\enq
It further holds that
\bem
 \bigg(  \bs{y}^{ (\bs{\a}_{\bs{\ell} },\bs{\ell} ) }_{\bs{i}, \, \wh{\bs{j}}_{\bs{\ell} }  } \, , \bs{x}^{ (\bs{\be}_{\bs{r} },\bs{r} ) }_{\bs{p}} \bigg) \; = \; 
 \ex{\f{h}{2T} \veps_{ p_{N} } }  \pl{  \substack{s=1 \\ \not\in \mc{S}_{ \bs{\ell} \cup \bs{r}  }   } }{ N } \Big\{ \de_{j_s p_{s-1}} \de_{p_s p_{s-1} }\Big\}
\pl{  k \in \mc{S}_{ \bs{\ell} \cap \bs{r}  } }{} \bigg(  \bs{w}_{ i_{k}   i_{k-1} ;\a_{k} }  , \bs{v}_{ p_{k}   p_{k-1} ;\, \be_{k} }  \bigg)  \\ 
\times \pl{  k \in \mc{S}_{ \bs{\ell} }^{\e{c}} }{} \bigg(  \bs{w}_{ i_{k}   i_{k-1} ;\a_{k} }  ,  \bs{e}_{ p_{k-1} }  \otimes \bs{e}_{ p_{k} }   \bigg) 
\cdot \pl{ s \in \mc{S}_{ \bs{r}  }^{\e{c}} }{} \bigg(   \bs{e}_{j_s}\otimes \bs{e}_{j_s}   , \bs{v}_{ p_{s}   p_{s-1} ;\, \be_{s} }  \bigg)   \;. 
\label{ecriture gros PS} 
\end{multline}
The product over Kronecker  symbols allows one to compute most of the sums over the $p_a$s and $j_a$s occurring in \eqref{ecriture produit deux operateurs comme somme combinatoire}. 
We first treat the case when $\mc{S}_{ \bs{\ell} \cup \bs{r}  }  \not= \emptyset$. 
To start with, consider the product 
\beq
\pl{  \substack{s=1 \\ \not\in \mc{S}_{ \bs{\ell} \cup \bs{r}  }   } }{ N } \de_{p_s p_{s-1} } \; = \;  \pl{ s =  1 }{ t_{1}-1 } \de_{p_s p_{s-1}}  \cdot \pl{v=2}{u} \pl{ s = t_{v-1}+1 }{ t_{v}-1 } \de_{p_s p_{s-1}} 
\cdot \pl{ s = t_{u}+1 }{ N } \de_{p_s p_{s-1}} \;. 
\enq
Thus, the above string of Kronecker deltas will set 
\beq
p_s=p_{t_v} \quad \e{for} \quad s\in \intn{t_v}{t_{v+1}-1} \quad v=1,\dots, u-1 \quad \e{and} \quad 
p_s=p_{t_u} \quad \e{for} \quad s\in \intn{1}{t_{1}-1}\cup\intn{t_u}{N} \;. 
\enq
Note that the disjoint interval for the $p_{t_u}$ variable comes from the boundary conditions $p_0=p_{N}$ on the indices of $\bs{p}$. 
Thus, the summation over $\bs{p}$ reduces to one over $p_{t_1}, \dots, p_{t_u}$.

Recall that the indices $\ell_1, \cdots , \ell_n$ are absent in $ \wh{ \bs{j} }_{ \bs{\ell}  }$. Thus, 
\beq
\pl{  \substack{s=1 \\ \not\in \mc{S}_{ \bs{\ell}  \cup \bs{r}  }   } }{ N } \de_{ j_s p_{s} } 
\enq
leaves free only the variables $j_a$ with $a \in  \mc{S}_{ \bs{r}  }^{\e{c}} $. These variables only appear in the last scalar product in \eqref{ecriture gros PS}. Hence,  by linearity one can pull the 
summation over each such variable into the corresponding scalar product. This yields the vectors $\bs{u} $  appearing in the last line of \eqref{resultat ecriture product deux operateurs}.

Finally, when $\mc{S}_{ \bs{\ell}\cup \bs{r} }  = \emptyset$, the summation over $\wh{\bs{j}}_{\bs{\ell} }$ can be explicitly performed, while the summation over then $p_a$s reduces
to the summation over $p_N$. Then, 
due to the presence of the weight factor $\ex{\f{h}{2T} \veps_{ p_{ N } } } $ in \eqref{ecriture gros PS},  one obtains $2 \cosh\big[ \tf{h}{2T} \big]$. 
  \qed

One can straightforwardly generalise the above result to the computation of  any product of $ \op{O}_{ \bs{\ell}  }$ operators for any choice of vectors $\bs{\ell}$.

\begin{cor}
\label{Corollaire produit ou trace M operateurs Ol}
 Let  $n_1,\dots,n_M \in \mathbb{N}$ be given and consider, for every $n_s>0$,  a vector  $\bs{\ell}^{(s)}=\big(\ell^{(s)}_1, \dots, \ell^{(s)}_{n_s} \big)$ with components $1\leq \ell^{(s)}_1< \dots < \ell^{(s)}_{n_s}  \leq N $. 
 Further, denote 
\beq
\mc{S}_{ \bs{\ell}^{(k)}\cup  \bs{\ell}^{(k+1)}  } \; = \; \Big\{ t_{1;k+1}, \dots, t_{u_{k+1};k+1 }  \Big\} \;. 
\enq
Then, upon denoting $ t_{0;k+1}= t_{u_{k+1};k+1 }$, it holds that
\bem
 \op{O}_{ \bs{\ell}^{(1)}  } \dots \op{O}_{ \bs{\ell}^{(M)}  }    \; = \; 
  \sul{ \big\{ \{ p^{(k)}_{ t_{a;k} } \}_{a=1}^{ u_{k } }  \big\}_{k=2}^{M} }{}
  \sul{  \substack{  \bs{\a}^{(k)}   \\ k : n_k>0 }  }{ }  \sul{ \bs{i} , \wh{ \bs{q} }_{ \bs{\ell}^{(M)} }   }{ } 
  \bs{x}^{ (\bs{\a}^{(1)}  ,\bs{\ell}^{(1)}    ) }_{\bs{i}}\cdot \Big( \bs{y}^{ (\bs{\a}^{(M)}  ,\bs{\ell}^{(M)}) }_{\bs{p}_{t},  \wh{ \bs{q} }_{ \bs{\ell}^{(M)} } }\Big)^{\mf{t}}
 \cdot \Big( 2 \cosh\big[ \tfrac{h}{2T} \big] \Big)^{ \de_{n_1,0} \de_{n_{M},0} }  \\
\times  \pl{s=1}{M-1}  \Bigg\{   \ex{\f{h}{2T} \veps_{ p^{(s+1)}_{ t_{u_{s+1};s+1} }   } } \cdot  
\Big( 2 \cosh\big[ \tfrac{h}{2T} \big] \Big)^{ \de_{n_s,0} \de_{n_{s+1},0} }  \cdot 
\pl{    t_{k;s+1}  \in \mc{S}_{ \bs{\ell}^{(s)} \cap \bs{\ell}^{(s+1)}  }    }{} 
\bigg(  \bs{w}_{ p_{  t_{k;s+1} }^{(s)}  p_{  t_{k-1;s+1} }^{(s)} ;\a_{   t_{k;s+1} }^{(s)} } , \bs{v}_{ p_{  t_{k;s+1} }^{(s+1)}  p_{  t_{k-1;s+1} }^{(s+1)} ;\a_{   t_{k;s+1} }^{(s+1)} }   \bigg)  \\ 
\times \pl{  t_{k;s }  \in \mc{S}_{ \bs{\ell}^{(s)} }\setminus \mc{S}_{ \bs{\ell}^{(s)} \cap \bs{\ell}^{(s+1)}  } }{}
\bigg(  \bs{w}_{ p_{  t_{k;s} }^{(s)}  p_{  t_{k-1;s} }^{(s)} ;\a_{   t_{k;s} }^{(s)} } ,    \bs{e}_{ p_{ t_{k-1;s} }^{(s+1)}  } \otimes \bs{e}_{ p_{ t_{k;s} }^{(s+1)} }   \bigg) 
\cdot \pl{  t_{k;s+1 }  \in \mc{S}_{ \bs{\ell}^{(s+1)} } \setminus \mc{S}_{ \bs{\ell}^{(s)} \cap \bs{\ell}^{(s+1)}  }  }{} 
\bigg(  \bs{u}   ,\bs{v}_{ p_{  t_{k;s+1} }^{(s+1)}  p_{  t_{k-1;s+1} }^{(s+1)} ;\a_{   t_{k;s+1} }^{(s+1)} }   \bigg) \Bigg\} \;. 
\label{expression produit multiple operateurs}
\end{multline}
Here, one parameterises
\beq
\bs{\a}^{(k)} \; = \; \Big(\a_{  t_{1;k} }^{(k)}, \dots, \a_{  t_{u_{k};k} }^{(k)}      \Big)
\enq
and agrees upon  the convention 
\beq
p^{(1)}_{t_{a;k}}=i_{ t_{a;k} } \; \quad and \quad 
\bs{p}_t \, = \, \Big( \underbrace{ p_{t_{u_M,M}}^{(M)}, \dots, p_{t_{u_M,M}}^{(M) }}_{t_{1,M}-1}, \underbrace{ p_{ t_{1,M}}^{(M)},\dots, p_{t_{1,M}}^{(M)} }_{t_{2,M}-t_{1,M} }, 
\dots, \underbrace{ p_{ t_{u_M-1,M} }^{(M)},\dots, p_{t_{u_M-1,M} }^{(M)} }_{t_{u_M,M} -t_{u_M-1,M} }, 
\underbrace{ p_{t_{u_M,M}}^{(M)},\dots, p_{t_{u_M,M}}^{(M)} }_{N-t_{u_M,M}+1} \Big) \;. 
\enq
Likewise, one has
\bem
\e{tr}_{ \mf{h}_{ \mf{q} } }\Big[ \op{O}_{ \bs{\ell}^{(1)}  } \dots \op{O}_{ \bs{\ell}^{(M)}  }  \Big] \; = \; 
  \sul{  \big\{ \{ p^{(k)}_{ t_{a;k} } \}_{a=1}^{ u_{k } }  \big\}_{k=1}^{M}  }{}  \sul{  \substack{  \bs{\a}^{(k)}   \\ k : n_k>0}  }{ }
\pl{s=1}{M} \Big( 2 \cosh\big[ \tfrac{h}{2T} \big] \Big)^{ \de_{n_s,0} \de_{n_{s+1},0} }  \\
\times\pl{s=1}{M}  \Bigg\{   \exp\bigg\{ \f{h}{2T} \veps_{ p^{(s+1)}_{ t_{u_{s+1};s+1} }   } \bigg\}  
\pl{    t_{k;s+1}  \in \mc{S}_{ \bs{\ell}^{(s)} \cap \bs{\ell}^{(s+1)}  }    }{} 
\bigg(  \bs{w}_{ p_{  t_{k;s+1} }^{(s)}  p_{  t_{k-1;s+1} }^{(s)} ;\a_{   t_{k;s+1} }^{(s)} } , \bs{v}_{ p_{  t_{k;s+1} }^{(s+1)}  p_{  t_{k-1;s+1} }^{(s+1)} ;\a_{   t_{k;s+1} }^{(s+1)} }   \bigg)  \\ 
\times \pl{  t_{k;s }  \in \mc{S}_{ \bs{\ell}^{(s)} }\setminus \mc{S}_{ \bs{\ell}^{(s)} \cap \bs{\ell}^{(s+1)}  } }{}
\bigg(  \bs{w}_{ p_{  t_{k;s} }^{(s)}  p_{  t_{k-1;s} }^{(s)} ;\a_{   t_{k;s} }^{(s)} } ,     \bs{e}_{ p_{ t_{k-1;s} }^{(s+1)}  } \otimes \bs{e}_{ p_{ t_{k;s} }^{(s+1)} }   \bigg) 
\cdot \pl{  t_{k;s+1 }  \in \mc{S}_{ \bs{\ell}^{(s+1)} }\setminus \mc{S}_{ \bs{\ell}^{(s)} \cap \bs{\ell}^{(s+1)}  }  }{} \bigg(  \bs{u}   ,\bs{v}_{ p_{  t_{k;s+1} }^{(s+1)}  p_{  t_{k-1;s+1} }^{(s+1)} ;\a_{   t_{k;s+1} }^{(s+1)} }   \bigg) \Bigg\} \;. 
\label{expression trace produit multiple operateurs}
\end{multline}
\end{cor}

\begin{prop}
\label{Proposition estimation rayon spectral delta tq}
 
There exists an $N$-independent constant $C>0$ such that the spectral radius of the operator $ \de \op{t}_{\mf{q}}$ appearing in the decomposition of the quantum transfer matrix \eqref{ecriture decompotion qtm partie dominante et perturbation}
is bounded as
\beq
r_{S}\Big(\de \op{t}_{\mf{q}} \Big) \; \leq \; C \cdot |\aleph| \:.  
\label{ecriture borne rayon spectral detla tq}
\enq
Furthermore, there exist constants $C, C^{\prime}>0$ such that,  for any $n\in \mathbb{N}$, 
\beq
\Big|\e{tr}_{ \mf{h}_{ \mf{q} } }\Big[ \bs{\om}_{N;0} \cdot \big( \de \op{t}_{\mf{q}} \big)^n \Big] \Big| \; \leq \;  C^{\prime} \cdot \big( C \cdot |\aleph| \big)^n \;,
\label{ecriture borne trace puissance n}
\enq
where $\bs{\om}_{N;0}$ has been introduced in \eqref{definition operateur dominant rang 1 pour QTM}. 

More generally,  there exist constants $C, C^{\prime}>0$ such that, for any $\ell_1, \dots, \ell_n \in \mathbb{N}$,
\beq
\Big|\e{tr}_{ \mf{h}_{ \mf{q} } }\Big[ \bs{\om}_{N;0} \cdot \big( \de \op{t}_{\mf{q}} \big)^{\ell_1}  \bs{\om}_{N;0} \cdots \bs{\om}_{N;0} \cdot \big( \de \op{t}_{\mf{q}} \big)^{\ell_n}\Big] \Big| \; \leq \; 
\big( C^{\prime}\big)^n \cdot \pl{a=1}{n}\big( C \cdot |\aleph| \big)^{\ell_a}  \;. 
\label{ecriture borne trace puissances mutliples delta tq et omega N}
\enq

\end{prop}

\Proof

It follows from the expansion \eqref{definition op perturbant QTM} that 
\beq
\Big( \de \op{t}_{\mf{q}} \Big)^M \; = \; \sul{  \substack{  \{ n_s \}_{s=1}^{M}  \\  n_s = 1 }  }{ N } \sul{   \substack{  \{ \bs{\ell}^{(s)} \}_{s=1}^{M} \\  \bs{\ell}^{(s)} \in \mc{L}^{(n_s)}_{N} }  }{}
\op{O}_{ \bs{\ell}^{(1)}} \dots \op{O}_{\bs{\ell}^{(M)}} \;. 
\enq
The expression obtained for the products of  operators $ \op{O}_{ \bs{\ell}^{(1)}} \dots \op{O}_{ \bs{\ell}^{(M)}} $ in Corollary \ref{Corollaire produit ou trace M operateurs Ol}, eqn. \eqref{expression produit multiple operateurs}, leads to the bound
\bem
\big| \big| \big|  \op{O}_{ \bs{\ell}^{(1)}} \dots \op{O}_{ \bs{\ell}^{(M)}} \big| \big| \big|   \; \leq \;
\sul{ \big\{ \{ p^{(k)}_{ t_{a;k} } \}_{a=1}^{ u_{k } }  \big\}_{k=2}^{M} }{}
  \sul{    \bs{\a}^{(k)}    }{ }  \sul{ \bs{i} , \wh{ \bs{q} }_{ \bs{\ell}^{(M)} }   }{ } 
 \big| \big|  \bs{x}^{ (\bs{\a}^{(1)}  ,\bs{\ell}^{(1)}    ) }_{\bs{i}} \big| \big|   \cdot  \big| \big|   \bs{y}^{ (\bs{\a}^{(M)}  ,\bs{\ell}^{(M)}) }_{\bs{p}_{t},  \wh{ \bs{q} }_{ \bs{\ell}^{(M)} } }  \big| \big|  \\
\times 
\pl{s=1}{M-1}  \Bigg\{   \exp\bigg\{ \f{h}{2T} \veps_{ p^{(s+1)}_{ t_{u_{s+1};s+1} }   }  \bigg\}   \cdot 
 \pl{  t_{k;s }  \in \mc{S}_{ \bs{\ell}^{(s)} }    }{}   \big| \big|  \bs{w}_{ p_{  t_{k;s} }^{(s)}  p_{  t_{k-1;s} }^{(s)} ;\a_{   t_{k;s} }^{(s)} }  \big| \big| 
\cdot \pl{  t_{k;s+1 }  \in \mc{S}_{ \bs{\ell}^{(s+1)} }  }{} \bigg\{ 2 \big| \big| \bs{v}_{ p_{  t_{k;s+1} }^{(s+1)}  p_{  t_{k-1;s+1} }^{(s+1)} ;\a_{   t_{k;s+1} }^{(s+1)} }   \big| \big|  \bigg\} \, \Bigg\} \;. 
\end{multline}
Here we took into account that $n_k>0$ for any $k$ and $\norm{\bs{u}}=2$. Thus, upon recalling the estimates \eqref{ecriture estimation vecteurs v et w} on $\bs{w}_{ij;\a}$, $\bs{v}_{ij;\a}$ and the expression for 
the vectors $ \bs{x}^{ (\bs{\a}   ,\bs{\ell}     ) }_{\bs{i}}$ \eqref{definition vecteur x} and  $\bs{y}^{ (\bs{\a}   ,\bs{\ell} ) }_{\bs{p} ,  \wh{\bs{q}}_{ \bs{r}_m } } $ \eqref{definition vecteur y}, one  obtains
\beq
  \sul{ \bs{i} }{ } \big| \big|  \bs{x}^{ (\bs{\a}^{(1)}  ,\bs{\ell}^{(1)}    ) }_{\bs{i}} \big| \big|  \; \leq \;  \f{2^{ N} }{ 2^{n_1} } \cdot \cosh\Big( \f{h}{2T} \Big) 
 \qquad \e{and} \qquad
  \sul{  \wh{ \bs{q} }_{ \bs{\ell}^{(M)} }   }{ }  \big| \big|   \bs{y}^{ (\bs{\a}^{(M)}  ,\bs{\ell}^{(M)}) }_{\bs{p}_{t},  \wh{ \bs{q} }_{ \bs{\ell}^{(M)} } }  \big| \big|  
  \; \leq \; \f{2^{ N} }{ 2^{n_M} } \cdot  \bigg( \f{ C_{\bs{w}}\,  |\aleph|}{N}  \bigg)^{n_M} \;   
\enq
and thus 
\beq
\big| \big| \big|  \op{O}_{ \bs{\ell}^{(1)}} \dots \op{O}_{ \bs{\ell}^{(M)}} \big| \big| \big|   \; \leq \; \pl{k=1}{M}\big\{  4^{n_k} \big\}  \cdot 2^{2N-n_M -n_1} 
 \cdot \cosh\Big( \f{h}{2T} \Big)  \cdot  \bigg( \f{ C_{\bs{w}} \, |\aleph|}{N}  \bigg)^{n_M}
\cdot \pl{s=1}{M-1} \Big\{  2^{u_s} \cosh\Big( \f{h}{2T} \Big) \cdot  \bigg( \f{ C_{\bs{w}} |\aleph|}{N}  \bigg)^{n_s}   \Big\}  \;. 
\enq
Since $u_k\leq n_k+n_{k+1}$, this eventually leads to 
\beq
\big| \big| \big| \op{O}_{ \bs{\ell}^{(1)}} \dots \op{O}_{ \bs{\ell}^{(M)}} \big| \big| \big|   \; \leq \; 2^{2N}  \cdot  \Big[\cosh\Big( \f{h}{2T} \Big) \Big]^{M}
\times \pl{s=1}{M} \Big\{   16  \cdot   \f{ C_{\bs{w}} |\aleph|}{N}     \Big\}^{n_s}  \;. 
\enq
Inserting the latter bound into the series and using that 
\beq
\sul{ 1\leq \ell_1<\dots<\ell_n \leq N }{} 1 \; \leq \;  \f{1}{n!} \sul{  \ell_1, \dots,\ell_n =1 }{N} 1  \;= \; \f{ N^n }{n!}
\enq
implies the estimate 
\beq
\big| \big| \big|  \, \Big( \de \op{t}_{\mf{q}} \Big)^M  \, \big| \big| \big|   \; \leq  \; 2^{2N} \cdot \Big[\cosh\Big( \f{h}{2T} \Big) \Big]^{M}
\sul{  \substack{  \{ n_s \}_{s=1}^{M}  \\  n_s = 1 }  }{ N }   \pl{s=1}{M}  \f{  \Big\{ 16  \cdot     C_{\bs{w}} |\aleph|     \Big\}^{n_s} }{ n_s!} 
\le 2^{2N} \Bigg\{  \cosh\Big( \f{h}{2T} \Big)\Big( \ex{|\aleph| \wt{C} } -1 \Big)\bigg\}^{M} \;, 
\enq
for some $N$-independent $ \wt{C} > 0$. Hence
\beq
r_{S}\Big(\de \op{t}_{\mf{q}} \Big) \; = \; \limsup_{M\tend +\infty}  \Big\{ \big| \big| \big|  \Big( \de \op{t}_{\mf{q}} \Big)^M  \big| \big| \big|^{\f{1}{M}} \Big\} \; \leq   \; 
\cosh\Big( \f{h}{2T} \Big)\Big( \ex{|\aleph| \wt{C} } -1 \Big)\;. 
\enq
This readily entails the bound \eqref{ecriture borne rayon spectral detla tq}.

The second bound \eqref{ecriture borne trace puissance n} is obtained in an analogous way upon using the representation \eqref{expression trace produit multiple operateurs}
and specialising the dimension $n_1$   associated with the vector $\bs{\ell}^{(1)}$ to   zero. The same strategy yields \eqref{ecriture borne trace puissances mutliples delta tq et omega N} as well.  \qed



%

\section{Existence of a dominant Eigenvalue and commutativity of limits}
\label{Section VP dominante et echange limites}

\subsection{Bound on the dominant and subdominant Eigenvalues}

Since the quantum transfer matrix $\op{t}_{\mf{q}}$ has real-valued entries, its characteristic polynomial has real coefficients which entails that 
the Eigenvalues of the quantum transfer matrix $\op{t}_{\mf{q}}$ are either real or appear in complex conjugate pairs.

\begin{prop}
\label{prop vp dominante et vp sous dominantes}
 The largest modulus  Eigenvalue $ \wh{\La}_{\e{max}}$ of the quantum transfer matrix $\op{t}_{\mf{q}} $  is non-degenerate,  real 
 and satisfies the estimate
\beq
\wh{\La}_{\e{max}}\, = \, 2+\e{O}\big( T^{-1} \big) \; , 
\label{ecriture estimee sur la VP maximale}
\enq
this uniformly in $N$.
 All the other Eigenvalues $ \wh{\La}_a$, $a=1,\dots,2^{2N}-1$ repeated according to their multiplicities, satisfy    
\beq
  \wh{\La}_{ a }   \,  =  \, \e{O}\big( T^{-1} \big)  \; \quad  uniformly  \;  in  \, N\;. 
\enq

\end{prop}

\Proof 

Proposition \ref{Proposition estimation rayon spectral delta tq} ensures that the operator $\de\op{t}_{\mf{q}}$ appearing in $\op{t}_{\mf{q}}\; = \; \om_{N;0}+\de\op{t}_{\mf{q}}$  has its spectral radius
such that 
\beq
r_S\big( \de\op{t}_{\mf{q}} \big)=\e{O}\big( T^{-1} \big) \; , \quad \e{uniformly}\, \e{in} \;  N \; .
\enq
Then, $\la-\de\op{t}_{\mf{q}}$ is invertible for any $\la \not\in \sg\big( \de\op{t}_{\mf{q}} \big)$, the spectrum of $\de\op{t}_{\mf{q}}$. Thus, for such  $\la$s, it holds that
\beq
  \e{det}\Big[\la -\op{t}_{\mf{q}} \Big] \; = \;  \e{det}\Big[\la - \de\op{t}_{\mf{q}} \Big] \cdot  \e{det}\Big[\e{id} - \Big(\la - \de\op{t}_{\mf{q}} \Big)^{-1}\bs{\om}_{N;0} \Big]  
\; = \; \e{det}\Big[\la - \de\op{t}_{\mf{q}} \Big] \cdot \Big\{ 1- \big(\bs{w}, \big[\la - \de\op{t}_{\mf{q}} \big]^{-1} \bs{v} \big) \Big\} \;, 
\enq
where we used the explicit expression for $\bs{\om}_{N;0}$ given in Lemma \ref{Lemme pte proj dominant rang 1}. The estimates obtained in Proposition \ref{Proposition estimation rayon spectral delta tq}, \eqref{ecriture borne trace puissance n}, can be recast as
\beq
\Big| \Big(\bs{w}, \big(\de\op{t}_{\mf{q}} \big)^n   \bs{v} \Big)  \Big|  \; \leq \;  \Big(  \f{C}{T} \Big)^n
\label{estimation trace tq n}
\enq
for some $N$-independent constant $C > 0$. Thus, the series 
\beq
\mc{S}(\la) \; = \; \f{1}{\la} \sul{n \geq 0 }{}  \Big(\bs{w}, \big(\la^{-1}  \de\op{t}_{\mf{q}}  \big)^{n} \bs{v} \Big)
\enq
converges uniformly on the set 
\beq
\Big\{ \la \in \Cx \; : \; |\la| > C (1+\eps) T^{-1} \Big\}\; , \quad \e{this} \, \e{for} \, \e{any} \; \eps>0 \; .
\enq
Furthermore, since the series
\beq
\f{1}{\la} \sul{n \geq 0 }{} \bigg( \f{ \de\op{t}_{\mf{q}} }{ \la } \bigg)^{n}
\enq
converges in the operator norm 
to $\big[\la - \de\op{t}_{\mf{q}} \big]^{-1}$ on the set 
\beq
\Big\{ \la \in \Cx \; : \; |\la| > ||| \de \op{t}_{\mf{q}} ||| \, \Big\} \;, 
\quad \e{one}\, \e{has} \, \e{that} \quad 
\mc{S}(\la) =  \big(\bs{w}, \big[\la - \de\op{t}_{\mf{q}} \big]^{-1} \bs{v} \big)
\enq
on this set. Thus, since $\la \mapsto  \big[\la - \de\op{t}_{\mf{q}} \big]^{-1}$ is analytic on $\Cx\setminus \sg\big( \de \op{t}_{\mf{q}} \big)$, by uniqueness of the analytic continuation,
it holds that
\beq
\mc{S}(\la) \; = \; \big(\bs{w}, \big[ \la - \de\op{t}_{\mf{q}} \big]^{-1} \bs{v} \big) \qquad \e{for} \; \e{any} \quad |\la| \, > \, C \cdot (1+\eps) \cdot T^{-1} \;. 
\label{ecriture definition S de Lambda}
\enq
The bound \eqref{estimation trace tq n} also entails that 
\beq
\mc{S}(\la) \, = \, \f{  1 }{ \la }  \cdot \overbrace{\big(\bs{w},  \bs{v} \big)}^{=2  \cosh(\frac{h}{2T}) }   \; + \; \e{O}\Big( \f{1}{T \la^2} \Big)
\label{ecriture estimee S}
\enq
with a differentiable remainder.

Let $\mc{S}_0(\la)=2/\la$. Then, using (\ref{ecriture estimee S}), it is easy to see that for some $C>0$ large enough
\beq
\Big|\mc{S}(\la)- \mc{S}_0(\la)  \Big| \, < \, \Big|1 \, - \,  \mc{S}_0(\la) \Big| \qquad \e{for} \; \; \la \in \Dp{}\mc{D}_{ 2 , \f{C}{T} } \;.
\enq
Since $1 - \mc{S}_0(\la)$ admits a unique zero in $\mc{D}_{ 2 , \f{C}{T} }$ at $\la=2$ and since $1 - \mc{S}_0(\la)$   does not vanish on $\Dp{}\mc{D}_{ 2 , \f{C}{T} }$, it follows by the Rouch\'{e}
theorem that $1-\mc{S}(\la)$ admits a unique zero in $\mc{D}_{ 2 , \f{C}{T} }$. 
An analogous reasoning based on (\ref{estimation trace tq n}) and applied to the domain $\Cx \setminus \Big\{ \mc{D}_{ 2 , \f{C}{T} } \cup \mc{D}_{ 0 , \f{C^{\prime}}{T} } \Big\}$ with $C^{\prime}>0$ large enough 
implies that $1-\mc{S}$ has no zeroes in this domain. Thus, since the zeroes of $  \la \mapsto \e{det}\Big[\la - \de\op{t}_{\mf{q}} \Big]$ belong to the spectrum   $\sg\big(\de \op{t}_{\mf{q}} \big)$ 
of the operator $\de \op{t}_{\mf{q}} $ and since $\sg\big(\de \op{t}_{\mf{q}} \big) \subset \mc{D}_{ 0 , \f{C^{\prime\prime}}{T} } $
for some $C^{\prime\prime}$, it follows that the characteristic polynomial $\e{det}\Big[\la -\op{t}_{\mf{q}} \Big]$ has a non-degenerate zero 
in the disk $\mc{D}_{ 2 , \f{C}{T} }$ and all its other zeroes are contained in the disk $\mc{D}_{ 0 , \f{ \wt{C}}{T} }$ with constants $C, \wt{C}$ being $N$-independent. 
This entails the claim. \qed

\subsection{Commutativity of the limits}

In this subsection, we establish  a theorem allowing one to exchange the Trotter and the thermodynamic limits when computing the free energy. 
One of the ingredients of the proof is the lemma below which was established by M. Suzuki in \cite{SuzukiArgumentsForInterchangeabilityTrotterAndVolumeLimitInPartFcton}. 

\begin{lemme} {\bf Suzuki} \cite{SuzukiArgumentsForInterchangeabilityTrotterAndVolumeLimitInPartFcton}
\label{Lemme Suzuki Exchange of limits}

 Let $a_{N,L}$ be a sequence in $\Cx$ such that 
\begin{itemize}
 
 \item for any  $L\in \mathbb{N}$,   $\lim_{N\tend + \infty}a_{N,L} \; = \; \a_L$;
 
 \item $\lim_{L \tend + \infty}\a_{L} \; = \; \a$;

 \item $\lim_{L\tend + \infty}a_{N,L} \; = \; z_N$, with a convergence holding uniformly in $N$. 

 \end{itemize}
 
 Then, $\lim_{ N \tend + \infty} z_N$ exists and equals $\a$.

\end{lemme}

\begin{theorem}
\label{Theorem echange limite thermo et Trotter}

 There exists $T_0 > 0$ such that, for any $T \geq T_0$, 
\beq
\lim_{L\tend + \infty} \lim_{ N \tend + \infty} \f{1}{L}\ln \e{tr}_{\mf{h}_{\mf{q}}} \Big[ \op{t}_{\mf{q}}^{L} \Big] \; = \; 
\lim_{ N \tend + \infty} \lim_{L\tend + \infty} \f{1}{L}\ln \e{tr}_{\mf{h}_{\mf{q}}} \Big[ \op{t}_{\mf{q}}^{L} \Big] \; .
\enq

\end{theorem}

\Proof 

The aim is to apply Lemma \eqref{Lemme Suzuki Exchange of limits} to the sequence 
\beq
\tau_{N,L} \; = \; \f{1}{L}\ln \e{tr}_{\mf{h}_{\mf{q}}} \Big[ \op{t}_{\mf{q}}^{L} \Big] \;. 
\enq
The equality 
\beq
\lim_{N\tend + \infty} \tau_{N,L} \; = \; \f{1}{L} \ln \e{tr}_{\mf{h}_{XXZ}} \Big[ \ex{-\frac{1}{T} \op{H} } \Big] 
\label{ecriture explicite limite Trotter}
\enq
holds from the very way the quantum transfer matrix is built. Indeed, one can show through elementary algebra based on the quantum inverse scattering method (see \cite{GohmannKlumperSeelFinieTemperatureCorrelationFunctionsXXZ}
for  details) that 
\beq
 \e{tr}_{\mf{h}_{\mf{q}}} \Big[ \op{t}_{\mf{q}}^{L} \Big] \; = \; \e{tr}_{\mf{h}_{XXZ}} \Big[ \ex{-\frac{1}{T} \op{H} + \op{A}_N  }   \Big] 
\enq
where $ ||| \op{A}_N ||| = \e{O}(N^{-1})$, but with an estimate that is \textit{non}-uniform in the volume $L$. 
 
 As discussed in the introduction, the existence  of the limit 
\beq
-T
 \lim_{L \tend + \infty} \f{1}{L} \ln \e{tr}_{\mf{h}_{XXZ}} \Big[ \ex{-\frac{1}{T} \op{H} } \Big] 
\enq
defining the \textit{per}-site free energy $f$ follows from standard considerations in rigorous statistical mechanics, see \cite{RuelleRigorousResultsForStatisticalMechanics}. 
Hence, it remains to establish the existence  of the limit $\lim_{L\tend + \infty}\tau_{N,L}$ and its uniformness in $N$.

Following the notations and conclusions of Proposition \ref{prop vp dominante et vp sous dominantes}, one has for $T$ large enough that
\beq
\e{tr}_{\mf{h}_{\mf{q}}} \Big[ \op{t}_{\mf{q}}^{L} \Big] \; = \; \wh{\La}_{\e{max}}^L \; + \; \sul{a=1}{2^{2N}-1} \wh{\La}_{a}^L \;. 
\enq
Thus, it is clear that 
\beq
\lim_{L\tend + \infty} \tau_{N,L} \; = \; \ln \Big[  \wh{\La}_{\e{max}} \Big] \;. 
\label{ecriture resultat limite volume infini sur approx Trotter}
\enq
In order to establish the uniformness in $N$ of this convergence, one should provide sharp bounds  on the sum over the sub-dominant Eigenvalues.

Let $\mf{P}$ denote the projector on the subspace of $\mf{h}_{\mf{q}}$ given as the direct sum of the Eigenspaces of $\op{t}_{\mf{q}}$
associated with the subdominant Eigenvalues. Then, it holds that
\beq
\e{tr}_{\mf{h}_{\mf{q}}} \Big[ \Big( \mf{P} \op{t}_{\mf{q}} \mf{P} \Big)^{L} \Big] \; = \;\sul{a=1}{2^{2N}-1} \wh{\La}_{a}^L \;. 
\enq
By using that $\mf{P}^2=\mf{P}$ and the cyclicity of the trace, one obtains
\beq
\e{tr}_{\mf{h}_{\mf{q}}} \Big[ \Big( \mf{P} \op{t}_{\mf{q}} \mf{P} \Big)^{L} \Big] \; = \;  \e{tr}_{\mf{h}_{\mf{q}}} \Big[ \Big( \mf{P} \op{t}_{\mf{q}}  \Big)^{L} \Big]   \;. 
\enq

The projector $\mf{P}$ can be computed in an explicit form. Namely, standard considerations of functional calculus on the spectrum of an operator and 
 Proposition \ref{prop vp dominante et vp sous dominantes} ensure  that, for $T$ large enough, it holds that
\beq
 \mf{P} \, = \, \Oint{ \Dp{}\mc{D}_{0,1} }{} \f{\dd \la }{ 2\i \pi } \f{1}{ \la  - \op{t}_{\mf{q}} } \; .
\enq
Recall that $\op{t}_{\mf{q}}= \bs{\om}_{N;0} \, + \,  \de \op{t}_{\mf{q}} $ with $r_S(\de\op{t}_{\mf{q}} ) = \e{O}(T^{-1})$ as established in Proposition \ref{Proposition estimation rayon spectral delta tq}. 
The latter ensures that the operator $ \la  - \de \op{t}_{\mf{q}}$ is invertible for any $\la \in \Dp{}\mc{D}_{0,1}$, hence leading to the representation 
\beq
 \f{1}{ \la - \op{t}_{\mf{q}} }  \; = \;  \f{1}{ \la  - \de \op{t}_{\mf{q}} }  \cdot  \f{1}{ \e{id}  - \big[ \la  - \de \op{t}_{\mf{q}} \big]^{-1}\cdot \bs{\om}_{N;0} }  \;. 
\enq
Furthermore,
\beq
\big[ \la  - \de \op{t}_{\mf{q}} \big]^{-1}\cdot \bs{\om}_{N;0} = \bs{u} \cdot \bs{w}^{\mf{t}} \; , \quad  \e{with} \quad  \bs{u} \, = \, \big[ \la  - \de \op{t}_{\mf{q}} \big]^{-1}\cdot  \bs{v} \;, 
\enq
and by virtue of \eqref{ecriture definition S de Lambda}-\eqref{ecriture estimee S}, one has $  (\bs{w}, \bs{u} ) = 2\cdot \la^{-1} + \e{O}(T^{-1})$  uniformly in $\la \in \Dp{}\mc{D}_{0,1}$ and in $N$. 
Then, it is easy to check that 
\beq
\f{1}{ \e{id}  - \big[ \la  - \de \op{t}_{\mf{q}} \big]^{-1}\cdot \bs{\om}_{N;0} }  \; = \; \e{id} +  \f{1}{1-\mc{S}( \la ) }\big[ \la   - \de \op{t}_{\mf{q}} \big]^{-1}\cdot \bs{\om}_{N;0}
\enq
with $\mc{S}$ as defined in \eqref{ecriture definition S de Lambda}. This yields
\beq
 \mf{P} \, = \, \Oint{ \Dp{}\mc{D}_{0,1} }{} \f{\dd \la }{ 2\i \pi }  \f{1}{ \la  - \de \op{t}_{\mf{q}} } \cdot \Bigg\{  \e{id} +  \f{ 1 }{ 1 - \mc{S}(\la) }\big[ \la  - \de \op{t}_{\mf{q}} \big]^{-1}\cdot \bs{\om}_{N;0}  \Bigg\} \;. 
\label{ecriture rep int pour projecteur}
\enq
It is established in the proof of Proposition  \ref{prop vp dominante et vp sous dominantes} that $1-\mc{S}$
admits a unique zero in $\Cx \setminus \mc{D}_{0,1}$ which corresponds to $\wh{\La}_{\e{max}}$. Thence, by taking the integral \eqref{ecriture rep int pour projecteur} by the residues lying outside of $\mc{D}_{0,1}$, one obtains
\beq
 \mf{P} \, = \, \e{id}  \; +\; \f{ \big[ \wh{\La}_{\e{max}} \,  - \, \de \op{t}_{\mf{q}} \big]^{-2}\cdot \bs{\om}_{N;0}  }{\mc{S}^{\prime}( \wh{\La}_{\e{max}} ) }    \;. 
\enq
Finally, one has the decomposition
\beq
\mf{P} \op{t}_{\mf{q}}   \, = \, \mf{T}_0 \, + \, \de \mf{T} \qquad \e{with} \quad
		    \left\{ \ba{ccc}   \mf{T}_0 & = &  \mf{P}  \cdot  \bs{\om}_{N;0}   \\ 
			      \de \mf{T} & = &   \mf{P}  \cdot  \de \op{t}_{\mf{q}}     \ea \right.  \;, 
\enq
which entails the expansion
\beq
\e{tr}_{\mf{h}_{\mf{q}}} \Big[ \Big( \mf{P} \op{t}_{\mf{q}} \mf{P} \Big)^{L} \Big] \; = \; \sul{n=0}{ L }  \sul{ \bs{\ell}\in \mc{L}^{(n)}_L }{ }  
\e{tr}_{\mf{h}_{\mf{q}}} \Big[ \mf{T}_0^{\ell_1-1} \cdot \de\mf{T} \cdot  \mf{T}_0^{\ell_2-\ell_1-1}   \cdot \de\mf{T} \, \cdots \,    \mf{T}_0^{\ell_{n}-\ell_{n-1}-1}   \cdot \de\mf{T} \cdot   \mf{T}_0^{L-\ell_n}  \Big]\;. 
\label{ecriture dvpmt trace projection QTM sur traces elementaires}
\enq
Here $\bs{\ell}=(\ell_1,\dots,\ell_n)$  and $\mc{L}^{(n)}_L$ is as introduced in \eqref{definition ensemble mathcal L n N}.

In order to provide estimates for the summand, it is convenient to first
recast $\mf{T}_0$. By using that $\mc{S}( \wh{\La}_{\e{max}} ) \, = \, 1$, one  obtains that 
\beq
- \mc{S}^{\prime}( \wh{\La}_{\e{max}} ) \, = \, \f{ 1 }{ \wh{\La}_{\e{max}}  } \cdot (1+s_T) \quad \e{with} \quad s_T \, = \, \f{ 1 }{ \wh{\La}_{\e{max}}  }  \sul{n\geq 1}{} n  \Big( \bs{w}, \big[\,  \wt{\de \op{t}}_{\mf{q}} \big]^{n} \bs{v} \Big)
\quad \e{and} \quad 
   \wt{\de \op{t}}_{\mf{q}}   \, = \, \f{ \de \op{t}_{\mf{q}} }{ \wh{\La}_{\e{max}} } \;. 
\enq
The estimates \eqref{ecriture borne trace puissance n} and \eqref{ecriture estimee sur la VP maximale} entail that $s_{T}=\e{O}(T^{-1})$ uniformly in $N$. 
Furthermore, one has the rewriting
\beq
2 \cosh\Big( \tfrac{h}{2T}\Big) \cdot \big(  1+ s_T \big)^{-1}  \, = \,  \wh{\La}_{\e{max}} \, ( 1 + w_T )  
\enq
where $w_T= \e{O}\big( T^{-1} \big)$ owing to \eqref{ecriture estimee sur la VP maximale}. This yields
\beq
 \mf{T}_0 \, = \,  \Bigg\{ \e{id} -   \big[ \e{id}  -  \wt{\de \op{t}}_{\mf{q}}  \big]^{-2} (1+w_T) \Bigg\} \cdot  \bs{\om}_{N;0}   
 \, = \, - \Bigg\{ w_T +   \wt{\de \op{t}}_{\mf{q}} \cdot  \f{ 2    -   \wt{ \de \op{t}}_{\mf{q}}  }{ \big[ \e{id}  -  \wt{\de \op{t}}_{\mf{q}}  \big]^{2} } \cdot (1+w_T) \Bigg\} \cdot  \bs{\om}_{N;0} \;. 
\enq

 This representation for $\mf{T}_0$ is already enough so as to bound the summands in \eqref{ecriture dvpmt trace projection QTM sur traces elementaires}. 
Since the operator products appear under the trace, as in the proof of Proposition \ref{prop vp dominante et vp sous dominantes}, one 
justifies that it is licit to expand all expressions of the form $\big[ \e{id}  -  \wt{\de \op{t}}_{\mf{q}}  \big]^{-k}$ for some $k\in \mathbb{N}$
into power series in $\wt{\de \op{t}}_{\mf{q}}$ and to commute the trace with the summation symbols. Upon expanding each such factor, writing up the issuing sums, and then, in the very end,
applying the bounds based on \eqref{ecriture borne trace puissances mutliples delta tq et omega N} followed by various resummations of the resulting sums, 
one easily concludes that bounding the trace 
\beq
\e{tr}_{\mf{h}_{\mf{q}}} \Big[ \mf{T}_0^{\ell_1-1} \cdot \de\mf{T} \cdot  \mf{T}_0^{\ell_2-\ell_1-1}   \cdot \de\mf{T} \, \cdots \,    \mf{T}_0^{\ell_{n}-\ell_{n-1}-1}   \cdot \de\mf{T} \cdot   \mf{T}_0^{L-\ell_n}  \Big] 
\enq
amounts to dropping the trace symbol and replacing each appearance of the operator $\bs{\om}_{N;0}$ in the product contained under the trace or by a constant $C_1>0$, each appearance of $w_T$ or $s_T$ by $C_2/T$ with a constant $C_2>0$, and each appearance of  
$ \wt{\de \op{t}}_{\mf{q}} $ or $ \de \op{t}_{\mf{q}} $ by $C_3/T$ with a constant $C_3>0$, \textit{viz}. 
\beq
\bigg| \e{tr}_{\mf{h}_{\mf{q}}} \Big[ \mf{T}_0^{\ell_1-1} \cdot \de\mf{T} \cdot  \mf{T}_0^{\ell_2-\ell_1-1}   \cdot \de\mf{T} \, \cdots \,    \mf{T}_0^{\ell_{n}-\ell_{n-1}-1}   \cdot \de\mf{T} \cdot   \mf{T}_0^{L-\ell_n}  \Big]  \bigg| \; \leq \; 
\mc{Z}_0^{\ell_1-1} \cdot \de\mc{Z} \cdot  \mc{Z}_0^{\ell_2-\ell_1-1}   \cdot \de\mc{Z} \, \cdots \,    \mc{Z}_0^{\ell_{n}-\ell_{n-1}-1}   \cdot \de\mc{Z} \cdot   \mc{Z}_0^{L-\ell_n} 
\enq
where 
\beq
\mc{Z}_0 \; = \; C_1 \Bigg\{  \f{C_2}{T} \, + \, \f{C_3}{T} \f{ 2 + \tfrac{C_3}{T} }{ \big[ 1 - \tfrac{C_3}{T} \big]^2 } \cdot \Big( 1+\tfrac{C_2}{T} \Big)    \Bigg\}
\quad \e{and} \quad 
 \de\mc{Z} \; = \;  \Bigg\{ 1 \, + \,  C_1 \f{   |\wh{\La}_{\e{max}} |^{-1}  }{ ( 1 - \tfrac{C_2}{T} ) \cdot \big[ 1 - \tfrac{C_3}{T} \big]^2 } \Bigg\} \cdot \f{C_3}{T} \;. 
\enq
This immediately yields, for some $C, C^{\prime}>0$, the estimate 
\beq
\bigg| \e{tr}_{\mf{h}_{\mf{q}}} \Big[ \mf{T}_0^{\ell_1-1} \cdot \de\mf{T} \cdot  \mf{T}_0^{\ell_2-\ell_1-1}   \cdot \de\mf{T} \, \cdots \,    \mf{T}_0^{\ell_{n}-\ell_{n-1}-1}   \cdot \de\mf{T} \cdot   \mf{T}_0^{L-\ell_n}  \Big]  \bigg| 
\; \leq \;  \Big( C^{\prime} \Big)^{L-n} \Big( \f{C}{T} \Big)^L \;. 
\enq
Thus, one obtains
\beq
\Big| \e{tr}_{\mf{h}_{\mf{q}}} \Big[ \Big( \mf{P} \op{t}_{\mf{q}} \mf{P} \Big)^{L} \Big]  \Big| \; \leq  \; \sul{n=0}{ +\infty }  \f{ L^{n} }{n!} \cdot   \big[ C^{\prime} \big]^{L-n} \cdot \bigg( \f{C}{T} \bigg)^L
= \bigg( \f{C C^{\prime} \ex{ \frac{1}{C^{\prime}}} }{ T } \bigg)^L \;. 
\enq
This bound allows one to obtain the estimate
\beq
\Big| \tau_{N,L}-\wh{\La}_{\e{max}} \Big|\; =  \; \f{1}{L} \Big| \ln \Big[ 1+ \big( \, \wh{\La}_{\e{max}} \big)^{-L} \cdot \sul{a=1}{2^{2N}-1} \wh{\La}_{a}^L  \Big]  \Big|  \; \leq \; 
\f{1}{L} \bigg( \f{  C^{\prime\prime}   }{ T } \bigg)^L
\enq
for some $N$-independent constant $C^{\prime\prime}>0$, hence ensuring a uniform in $N$ convergence in \eqref{ecriture resultat limite volume infini sur approx Trotter}. 
One is then able to conclude by virtue of Lemma \ref{Lemme Suzuki Exchange of limits}. 
\qed

\section{The non-linear integral equation based description of the spectrum}
\label{Section connection spectre qtm et NLIE}

\subsection{Preliminary discussion}

The main advantage of rewriting the partition function as the Trotter limit involving the quantum transfer matrix \eqref{definition QTM} 
is that one may construct the Eigenvectors of $\op{t}_{\mf{q}}$ through the Bethe Ansatz \cite{BetheSolutionToXXX,FaddeevSklyaninTakhtajanSineGordonFieldModel}. We stress that it has not been established whether the Bethe Ansatz 
provides one with a complete set of Eigenstates or not, although one may explicitly check that it is so when $\De=0$. 
The question of the completeness of the Bethe Ansatz is however not so relevant in that the computation of $f$ only demands to be able to construct the largest Eigenvalue by means of the Bethe Ansatz.
Further information, such as the finite temperature correlation lengths, can also be accessed at least for the sub-dominant Eigenvalues which can be constructed by the Bethe Ansatz.

Consider $M$ complex numbers $\la_1,\dots, \la_M \in \Cx$, distinct or not. If some of these numbers coincide, the total amount of these that is equal to a given complex number $z$ is called their multiplicity and denoted $k_{z}$. 
 One then defines the set 
\beq
\{ \la_a \}_{1}^{M} \; = \; \Big\{ (\la, k_{\la} ) \; : \; \la \in \{\la_1,\dots, \la_M \}  \Big\} \;. 
\enq

In the Bethe Ansatz approach, one looks for Eigenstates of $\op{t}_{\mf{q}}$ in the form of combinatorial sums parametrised in terms of $M$ complex numbers  $\la_1,\dots, \la_M$, 
\beq
\bs{\Psi}\Big( \{ \la_a\}_{a=1}^{M} \Big) \;. 
\label{ecriture vecteur de Bethe} 
\enq
In order for $\bs{\Psi}\Big( \{ \la_a\}_{a=1}^{M} \Big)$ to give rise to an Eigenstate of the quantum transfer matrix $\op{t}_q$, the roots $ \la_1,\dots, \la_M$ 
should be admissible, namely satisfy  
\begin{itemize}
\item  $\la_{a} \ne\la_{b} \pm \i \zeta \; \e{mod} \; \i \pi \, \mathbb{Z}$ for any $a,b$, 
\item $\la_a\not\in \big\{ \pm \tf{\aleph}{N}, \pm \tf{\aleph}{N} \pm \i\zeta \big\}$ for any $a$, 
\end{itemize}
and, for any $a=1,\dots, M$, solve the set of Bethe Ansatz equations:
\beq
\ex{-\tfrac{h}{T}} (-1)^{s} \f{ \Dp{}^p }{ \Dp{} \xi^p} \Bigg\{ \pl{k=1}{M} \bigg\{ \f{\sinh( \i\zeta - \xi + \la_k) }{  \sinh( \i \zeta  +  \xi - \la_k) }  \bigg\}
\cdot  \bigg\{ \f{ \sinh(   \xi - \tf{\aleph}{N} ) \sinh( \i\zeta +   \xi + \tf{\aleph}{N} ) }{  \sinh(   \xi + \tf{\aleph}{N} ) \sinh( \i\zeta -  \xi + \tf{\aleph}{N} )  }    \bigg\}^{N} \Bigg\}_{\mid \xi=\la_a}
\; = \; -\de_{p,0} \; ,  \quad p=0, \dots, k_{\la_a}-1
\label{ecriture eqns Bethe Trotter fini}
\enq
where $k_{\la_a}$ is the multiplicity of $\la_a$  and one adds the subsidiary condition that the derivative does not vanish for $p=k_{\la_a}$.
Further, $\aleph$ corresponds to a reparametrisation of the temperature:
\beq
 \aleph=-\i J \f{ \sin (\zeta) }{T }  
\enq
and $s = N - M$ is called the spin. Note that, if $\la_1,\dots, \la_M$ are all pairwise distinct and admissible, the system of Bethe Ansatz equations reduces to the usually encountered form 
\beq
\ex{-\tfrac{h}{T}} (-1)^{s} \pl{k=1}{M} \bigg\{ \f{\sinh( \i\zeta - \la_a + \la_k) }{  \sinh( \i \zeta  + \la_a - \la_k) }  \bigg\}
\cdot  \bigg\{ \f{ \sinh(  \la_a - \tf{\aleph}{N} ) \sinh( \i\zeta +  \la_a +\tf{\aleph}{N} ) }{  \sinh(  \la_a + \tf{\aleph}{N} ) \sinh( \i\zeta -\la_a + \tf{\aleph}{N} )  }    \bigg\}^{N} \; = \; -1 \; , \quad a=1, \dots, M \;. 
\enq

When all of the above conditions are fulfilled, the vector \eqref{ecriture vecteur de Bethe} is associated with the Eigenvalue  
\bem
 \tau\big( \xi \mid \{\la_k\}_1^M \big) \, = \,  (-1)^N \ex{\f{h}{2T} }\pl{k=1}{M} \Bigg\{ \f{ \sinh(\xi-\la_k + \i \zeta) }{ \sinh(\xi-\la_k )  }  \Bigg\} \cdot 
\Bigg( \f{ \sinh(\xi + \tf{\aleph}{N} ) \sinh(\xi - \tf{\aleph}{N} -\i \zeta) }{ \sinh^2(-\i\zeta) } \Bigg)^N   \\ 
\, + \,  (-1)^N  \ex{-\f{h}{2T} } \pl{k=1}{M}  \Bigg\{ \f{ \sinh(\xi-\la_k - \i \zeta) }{ \sinh(\xi-\la_k ) }  \Bigg\} \cdot 
 \Bigg(  \f{ \sinh(\xi + \tf{\aleph}{N} +\i\zeta) \sinh(\xi - \tf{\aleph}{N})  }{ \sinh^2(-\i\zeta) }  \Bigg)^N  
\label{ecriture valeur propre qtm}
\end{multline}
of the trace of the quantum monodromy matrix $ \e{tr}_{\mf{h}_0} \big[ \op{T}_{\mf{q};0}(\xi) \big]$.  See, \textit{e.g.} \cite{KlumperNLIEfromQTMDescrThermoXYZOneUnknownFcton},
for more details.

The system of Bethe Ansatz equation is highly non-trivial to solve, with the exception of the case $\zeta=\tf{\pi}{2}$, \textit{viz}. $\De=0$, 
where the equations decouple and can thus be solved explicitly. 
Furthermore, it appears hopeless to take directly the infinite Trotter number limit on the level of the Bethe equations \eqref{ecriture eqns Bethe Trotter fini}. 
An alternative has been proposed in the literature  \cite{DestriDeVegaAsymptoticAnalysisCountingFunctionAndFiniteSizeCorrectionsinTBAFirstpaper,KlumperNLIEfromQTMDescrThermoXYZOneUnknownFcton}.
The idea consists in putting the problem of finding solutions to \eqref{ecriture eqns Bethe Trotter fini} in correspondence 
with the one of solving certain non-linear integral equations. The non-linear integral equations appear easier to deal with, be it with respect to 
a numerical calculation of the spectrum or relatively to various formal manipulations thereof, for instance the calculation of the infinite Trotter number limit.

In the following, we will make use of the notations 
\beq
\mc{S}_{\a} \, = \, \Big\{ z \in \Cx \, : \, \big| \Im(z) \big| < \a \Big\} \quad \e{and} \quad \zeta_{\e{m}}\, = \, \e{min} \big\{ \zeta, \pi-\zeta \big\} \;. 
\enq
Also, in order to state the non-linear integral equation based characterisation, it is convenient to introduce 
\beq
\th(\la) \; = \; \left\{      \ba{ccc}         \i \ln \bigg( \f{ \sinh(\i\zeta+\la) }{ \sinh(\i\zeta-\la) }\bigg)   & \e{for} & |\Im(\la)| \, < \, \zeta_{\e{m}}    \vspace{2mm} \\
				  -\pi \e{sgn}\big(\pi-2\zeta \big)+\i \ln \bigg( \f{ \sinh(\i\zeta+\la) }{ \sinh(\la - \i\zeta) }\bigg)   & \e{for} &  \zeta_{\e{m}} \, < \, |\Im(\la)| \, < \, \tf{\pi}{2}
\ea \right. 				  
\enq
where "$\ln$" corresponds to the principal branch of the logarithm. 
The above definition makes $\th$ an $\i\pi$-periodic holomorphic function on $\Cx \setminus \Big\{ \R^+ \, \pm \, \i \zeta_{\e{m}} \, +\,  \i\pi \mathbb{Z} \Big\}$
with cuts on $ \R^+ \, \pm \, \i \zeta_{\e{m}} \, +\,  \i\pi \mathbb{Z}$. In the following, $\th_+(z)$ will stand for the $+$ boundary value of $\th$, \textit{viz}. the limit
$\th_+(z) \, = \, \lim_{\eps\tend 0^+} \th(z+\i\eps)$. This regularisation is only needed if $z \in \Big\{ \R^+ \, \pm \, \i \zeta_{\e{m}} \, +\,  \i\pi \mathbb{Z} \Big\}$. 

\noindent Further, we introduce
\beq
K(\xi) \, = \, \f{1}{2\pi} \th^{\prime}(\xi) \, = \, \f{ \e{sgn}(\pi-2\zeta) }{2\i\pi} \Big\{  \coth(\xi-\i\zeta_{\e{m}}) \, - \, \coth(\xi + \i\zeta_{\e{m}}) \Big\} \;. 
\enq
Finally, a set $\mc{M}$ will be called admissible if, for any $x, x^{\prime}\in \mc{M}$ it holds $x\not= x^{\prime}\pm \i\zeta$ mod $\i\pi \mathbb{Z}$ and if $\pm \tf{\aleph}{N}$
and $\pm \tf{\aleph}{N} \pm \i \zeta$ do not belong to $\mc{M}$, mod $\i\pi \mathbb{Z}$.

\begin{prop}

Let the set $\{\la_a\}_{a=1}^{M}$ be built up form an admissible solution to the Bethe Ansatz equations \eqref{ecriture eqns Bethe Trotter fini}. Then, there exists a bounded domain $\mc{D} \subset \mc{S}_{\tf{ \zeta_{\e{m}} }{ 2} }$
containing $- \tf{\aleph}{N}$, such that $\{\la_a\}_{a=1}^{M}$ allows one to construct a solution $\big(\wh{\mf{A}}, \wh{\mf{X}}, \wh{\mc{Y}} \big)$ to the below non-linear problem. 

\noindent Find

\begin{itemize}

 \item $\wh{\mf{A}}$ piecewise continuous on $\Dp{}\mc{D}$;
 
 \item a set $\wh{\mf{X}}=\{\wh{x}_a\}_{1}^{|\wh{\mf{X}}|}$ with $\wh{x}_a \in  \mc{D} \setminus  \Big\{  \tfrac{\aleph}{N}, -\tfrac{\aleph}{N} \Big\}$, for $a=1,\dots, | \wh{\mf{X}}|$;
 
 \item a set $\wh{\mc{Y}}=\{\wh{y}_a\}_{1}^{|\wh{\mc{Y}}|}$  with  $\wh{y}_a \in  \mc{S}_{\tf{\pi}{2}}\setminus \ov{\mc{D}}$   for $a=1,\dots, | \wh{\mc{Y}}|$, and $ \Big\{\wh{y}_1,\dots, \wh{y}_{ |\wh{\mc{Y}}| } \Big\}$ being an admissible set;

\end{itemize}

\noindent such that 
\begin{itemize}

 \item $\ex{ \wh{\mf{A}} }$ extends to a meromorphic, $\i\pi$-periodic function on $\Cx$ whose only poles in $\mc{D}$ build up exactly the set 
\beq
\wh{\mc{Y}}_{\e{sg}} \; = \; \Big\{ (y- \i \, \e{sgn}\big( \pi-2\zeta) \cdot \zeta_{\e{m}} , k_{y})\; : \;  (y , k_{y})\in \wh{\mc{Y}} \;\;  and  \; \;  y - \i \, \e{sgn}\big( \pi-2\zeta) \cdot \zeta_{\e{m}} \in \mc{D}\Big\} \; ,
\label{definition ensemble Y sg}
\enq
the order of the pole at $y$ being given by the multiplicity $k_y$ of  $(y,k_y) \in \wh{\mc{Y}}$;

 \item $1+\ex{ \wh{\mf{A}} }$ does not vanish on $\Dp{}\mc{D}$;

\item for any $(x,k_x) \in \wh{\mf{X}}$, resp. $(y,p_y) \in \wh{\mc{Y}}$: 
\beq
\ba{cccc} 
\Dp{\xi}^{r} \Big\{ \ex{ \wh{\mf{A}}( \xi ) } \Big\}_{\mid \xi=x} = - \de_{r,0} & \quad for \; any & \quad x \in \wh{\mf{X}},  &  r=0,\dots, k_x-1 \\ 
\Dp{\xi}^{r} \Big\{ \ex{ \wh{\mf{A}}( \xi ) } \Big\}_{\mid \xi=y} = -\de_{r,0} & \quad for \; any & \quad x \in \wh{\mc{Y}},  &  r=0,\dots, p_y-1 \ea \; 
\label{equation quantification particles et tous cas Trotter fini}
\enq
and the derivatives do not vanish for $r=k_x$, resp. $r=p_y$; 

\item $\wh{\mf{A}}$ is subject to the monodromy constraint 
\beq
\Oint{ \Dp{}\mc{D} }{ }  \f{  \wh{\mf{A}}^{\, \prime}\!(u) }{ 1+  \ex{- \wh{\mf{A}}(u)} }  \cdot \f{ \dd u }{2\i\pi} \; = \; -s - |\wh{\mc{Y}}|-|\wh{\mc{Y}}_{\e{sg}}| + |\wh{\mf{X}}| \;. 
\enq
for some $s \in \mathbb{Z}$; 

\item $\wh{\mf{A}}$ solves the non-linear integral equation
\beq
\wh{\mf{A}}(\xi) \, = \,  -\f{h}{T} \, +\, \mf{w}_N(\xi)  \; - \; \i \pi s  \; + \; \i \sul{ y \in \wh{\mathbb{Y}}_{\kappa} }{} \th_+(\xi-y)
\; + \; \Oint{ \Dp{} \mc{D}  }{} K(\xi-u) \cdot  \mc{L}\mathrm{n}\Big[ 1+  \ex{ \wh{\mf{A}} } \, \Big](u)  \cdot \dd u   
\label{ecriture eqn NLI forme primordiale}
\enq
with $\xi \in \mc{S}_{\zeta_{\e{m}}/2}$ and where, for $v\in \Dp{}\mc{D}$, one has
\beq
\mc{L}\mathrm{n} \Big[ 1+  \ex{ \wh{\mf{A}} } \, \Big](v) \, = \, \Int{\kappa}{v} \f{  \wh{\mf{A}}^{\, \prime}\!(u) }{ 1+  \ex{- \wh{\mf{A}}(u)} }  \cdot \dd u \; + \; \ln \Big[ 1+  \ex{ \wh{\mf{A}}(\kappa) } \, \Big] \;.
\enq
Here $\kappa$ is some point on $\Dp{}\mc{D}$ and the integral is taken, in the positive direction along $\Dp{}\mc{D}$, from $\kappa$ to $v$. The function $"\ln"$ appearing above corresponds to the principal branch of the logarithm 
extended to $\R^{-}$ with the convention $\e{arg}(z) \in \intfo{-\pi}{\pi}$. 

Finally, we have set  
\beq
\mf{w}_N(\xi) \; = \;  N \ln \bigg(  \f{ \sinh(  \xi - \tf{ \aleph }{N} ) \sinh(  \xi + \tf{ \aleph }{N} -\i\zeta) }{  \sinh(  \xi + \tf{ \aleph }{N} ) \sinh( \xi - \tf{ \aleph }{N} -\i\zeta )  }   \bigg)    \;.
\enq

\end{itemize}

The sets appearing in the non-linear integral equation \eqref{ecriture eqn NLI forme primordiale} are defined as  
\beq
\wh{\mathbb{Y}}_{\kappa} \, = \, \wh{\mathbb{Y}} \ominus \big\{     \kappa\big\} ^{ \oplus (s+| \wh{\mathbb{Y}} |) } \qquad with \qquad  
 \wh{\mathbb{Y}} \, = \, \wh{\mc{Y}}\oplus  \wh{\mc{Y}}_{\e{sg}}  \ominus   \wh{\mf{X}}  \;, 
\label{definition ensemble hat Y de kappa}
\enq
where we employed the conventions introduced in \eqref{definition difference algebrique ensemble} and above \eqref{definition ensemble repetee}. Finally, \eqref{ecriture eqn NLI forme primordiale}
also builds on the summation convention introduced in \eqref{defintion convention somme produit et cardinalite ensembles}-\eqref{definition ensemble repetee}.

\vspace{3mm}

Conversely, any solution to the above non-linear problem gives rise to an admissible solution of the Bethe Ansatz equations \eqref{ecriture eqns Bethe Trotter fini}. 

\end{prop}

The above proposition was first formulated and argued in \cite{DestriDeVegaAsymptoticAnalysisCountingFunctionAndFiniteSizeCorrectionsinTBAFirstpaper} and subsequently developed
in many works. Since the construction plays an important role in our analysis, we reproduce the proof for the reader's convenience. 

Note also that  $\mc{L}\mathrm{n}\big[ 1+  \ex{ \wh{\mf{A}} } \, \big]$ introduced above is well-defined since $1+  \ex{ \wh{\mf{A}} }$ does not 
vanish on $\Dp{}\mc{D}$. Furthermore, the definition of $\mc{L}\mathrm{n}$ implies that 
\beq
\ex{ \mc{L}\mathrm{n} \big[ 1+  \ex{ \wh{\mf{A}} } \, \big](v) } \; = \;  1+  \ex{ \wh{\mf{A}}(v) }
\enq
meaning that it provides one with a determination for the logarithm of $1+  \ex{ \wh{\mf{A}} }$.

\Proof

Given a distribution of roots $\{\la_a\}_1^{M}$ solving the Bethe equations \eqref{ecriture eqns Bethe Trotter fini}, one introduces an auxiliary function 
 $\wh{\mf{a}}$ by the formula
\beq
\wh{\mf{a}}\big( \xi \big) \, = \, \ex{-\tfrac{h}{T}} (-1)^{s} \pl{k=1}{M} \bigg\{ \f{\sinh( \i\zeta - \xi + \la_k) }{  \sinh( \i \zeta  + \xi - \la_k) }  \bigg\}
\cdot  \bigg\{ \f{ \sinh(  \xi - \tf{ \aleph }{N} ) \sinh( \i\zeta +  \xi +\tf{ \aleph }{N} ) }{  \sinh(  \xi + \tf{ \aleph }{N} ) \sinh( \i\zeta - \xi + \tf{ \aleph }{N} )  }    \bigg\}^{N} \;. 
\label{definition fct auxiliaire a}
\enq
Obviously, by definition of the Bethe roots, it holds that $ \big( \Dp{\xi}^p\, \wh{\mf{a}} \,\big)\big( \la_a \big)=- \de_{p,0} $ for $a=1,\dots, M$ and $p=0,\dots, k_{\la_a}-1$ where $k_{\la_a}$
is the multiplicity of the root $\la_a$, while  $ \big( \Dp{\xi}^{k_{\la_a}}\, \wh{\mf{a}} \, \big)\big( \la_a \big) \not= 0$. Since
 \begin{itemize}
  \item the total order of $\wh{\mf{a}}$\,'s poles in a fundamental strip of width $\pi$ is $M+2N$,
  \item $\lim_{\Re(\la) \tend \pm \infty } \mf{a}(\la) \, = \, \ex{- \f{h}{T} + 2\i \zeta (N-M) }$,
 \end{itemize}
the function $1+\wh{\mf{a}}$ admits $2N$  more zeroes, counted with multiplicities, in a strip of width $\pi$.

\vspace{2mm}

\noindent One then picks a domain $\mc{D} \subset \mc{S}_{ \tf{ \zeta_{\e{m}} }{2 } }$ such that 
\begin{itemize}
\item the $N$-fold pole at $-\tf{\aleph}{N}$ of $\wh{\mf{a}}$ is contained in $\mc{D}$;
\item $1+\wh{\mf{a}}$ does not vanish on $\Dp{}\mc{D}$ and also, has no poles on this boundary;
\item $\ov{\mc{D}}$ is bounded.
\end{itemize}

A domain $\mc{D}$ being chosen, the auxiliary function $\wh{\mf{a}}$ associated with the  Bethe roots $\{\la_a\}_1^{M}$ will be such that $1+\wh{\mf{a}}$ might have zeroes in $\mc{D}$. 
Some of these zeroes correspond to the subset of $\{ \la_a\}_1^{M}$ contained in $\mc{D}$, but there may be other zeroes not corresponding to Bethe roots. These will form the set $\wh{\mf{X}}$. Namely, 
one denotes $\wh{x}_1,\dots, \wh{x}_{|\wh{\mf{X}}|}$ the zeroes of $1+\wh{\mf{a}}$ in $\mc{D}$ which do not belong to the set  $\{\la_a\}_1^{M}$ and which are repeated according to their multiplicities. One defines
\beq
\wh{\mf{X}} \, = \, \{ \wh{x}_a \}^{|\wh{\mf{X}}|}_1  \, = \, \Big\{ (z,k_z) \; : \;  z \in \mc{D} \; \e{with} \;\; z\not= \la_1,\dots, \la_M  \; \e{and} \; z \, \e{is}\, \e{a} \, \e{zero} \, \e{of} \, 1+\wh{\mf{a}} \, \e{of} \, \e{order} \; k_z \Big\} \;. 
\enq

The roots $\la_a$ which are not contained in $\ov{\mc{D}}$ are denoted as $\wh{y}_1,\dots, \wh{y}_{ |\wh{\mc{Y}}| }  $ and repeated according to their multiplicity. They are gathered in the set
\beq
\wh{\mc{Y}}=\{ \, \wh{y}_a \}^{|\wh{\mc{Y}}|}_1   \, = \, \Big\{ (z, k_z)  \in \{\la_a\}_1^{M} \; : \; z \not \in \ov{\mc{D}} \; \e{mod} \; \i \pi \, \mathbb{Z}  \Big\} \;. 
\enq

Since the Bethe roots are admissible, so is the set $\big\{ \wh{y}_1,\dots, \wh{y}_{ |\wh{\mc{Y}}| }   \big\}$. One further defines the set  $\wh{\mc{Y}}_{\e{sg}}$ as in \eqref{definition ensemble Y sg}. Then the poles
of $\wh{\mf{a}}$ in $\mc{D}$ are located at $y$, with $(y,k_y) \in \wh{\mc{Y}}_{\e{sg}}$. The order of the pole if then given by the 
multiplicity $k_{y}$ of $y$.

The crucial observation of \cite{DestriDeVegaAsymptoticAnalysisCountingFunctionAndFiniteSizeCorrectionsinTBAFirstpaper} is that one may express 
a logarithm $\wh{\mf{A}}$ of $\wh{\mf{a}}$, \textit{viz}. a function $\wh{\mf{A}}$ satisfying to the relation 
\beq
\wh{\mf{a}}=\exp\Big\{ \, \wh{\mf{A}} \, \Big\} \;, 
\enq
by means of a direct calculation of residues. Indeed, observe that for any  $|\Im(\xi)|\leq \tf{ \zeta_{\e{m}} }{2}$, the map $u\mapsto \th(\xi-u)$  is holomorphic on $\mc{D}$ due to $\mc{D} \subset \mc{S}_{ \tf{ \zeta_{\e{m}} }{2}}$ 
what ensures that the function's cuts are located away from that domain. By definition  of the sets $\wh{\mf{X}}$, $\wh{\mc{Y}}$, and $\wh{\mc{Y}}_{\e{sg}}$, this then yields 

\beq
 \wh{\mf{A}}(\xi) \; = \; -\f{h}{T} \, +\, \mf{w}_N(\xi) \, - \, \i \pi s  \; + \; \i \sul{ y \in \wh{\mathbb{Y}} }{} \th_+(\xi-y)
\; + \; \i  \Oint{ \Dp{}  \mc{D}  }{} \th(\xi-u)  \f{  \wh{\mf{a}}^{\, \prime}\!(u) }{ 1+ \wh{\mf{a}}\,(u) } \cdot \f{ \dd u }{ 2\i \pi }  \;. 
\label{ecriture rep int hat A via hat a}
\enq
Here,  the summation set $\wh{ \mathbb{Y} }$ is as defined in \eqref{definition ensemble hat Y de kappa}. 
Finally, if for some $(y,k_y) \in \wh{\mathbb{Y}}$, $\xi-y$ lies on a cut of $\th$, then the formula should be understood in the sense of +boundary values\symbolfootnote[2]{ 
While we make the choice of such a prescription, it is not so relevant in that the choice of any boundary value would still lead to the same definition of $\wh{\mf{a}}$.}.

Upon using that $\wh{\mf{A}}$ is a logarithm for $\wh{\mf{a}}$, one may recast \eqref{ecriture rep int hat A via hat a} into a non-linear integral equation for $\wh{\mf{A}}$
\beq
 \wh{\mf{A}}(\xi) \; = \; -\f{h}{T} \, +\, \mf{w}_N(\xi) \, - \, \i \pi s  \; + \; \i \sul{ y \in \wh{\mathbb{Y}} }{} \th(\xi-y)
\; + \; \i  \Oint{ \Dp{} \mc{D} }{} \th(\xi-u)  \f{  \wh{\mf{A}}^{\, \prime}\!(u) }{ 1+  \ex{- \wh{\mf{A}}(u)} } \cdot \f{ \dd u }{ 2\i \pi }\;.   
\label{ecriture forme directe NLIE hat A}
\enq
It remains to recast \eqref{ecriture forme directe NLIE hat A} into its equivalent form given by \eqref{ecriture eqn NLI forme primordiale}. To achieve this, 
one should first establish the monodromy condition 
\beq
\Oint{ \Dp{}\mc{D} }{ }  \f{  \wh{\mf{A}}^{\, \prime}\!(u) }{ 1+  \ex{- \wh{\mf{A}}(u)} }  \cdot \f{ \dd u }{2\i\pi} \; = \; M-N - |\wh{\mc{Y}}|-|\wh{\mc{Y}}_{\e{sg}}| + |\wh{\mf{X}}|
\enq
which is a simple consequence of a residue calculation. Then, an integration by parts yields 
\beq
 \Oint{  \Dp{} \mc{D} }{} \th(\xi-u)  \cdot \f{  \wh{\mf{A}}^{\, \prime}\!(u) }{ 1+  \ex{- \wh{\mf{A}}(u)} } \cdot \f{ \dd u }{ 2  \pi }   \; = \;
  \Oint{  \Dp{} \mc{D} }{} K(\xi-u)  \mc{L}\mathrm{n} \Big[ 1+  \ex{ \wh{\mf{A}} } \, \Big](u)  \cdot   \dd u  \, -\i \Big( s + |\wh{\mathbb{Y}}| \Big) \th(\xi-\kappa)\;. 
\enq
This concludes the proof of the first part of the statement.

\vspace{4mm}

It remains to prove the second part of the statement, namely that a solution to the non-linear problem gives rise to an admissible solution 
to the Bethe Ansatz equations. Thus, assume that one is given a solution $\big( \wh{\mf{A}}$, $\wh{\mf{X}}$, $\wh{\mc{Y}}$) to the non-linear problem. 

The very form of the non-linear integral equation satisfied by $\wh{\mf{A}}$, ensures that $\wh{\mf{a}}(\xi) \; = \; \ex{ \wh{\mf{A}}(\xi)}$ is a meromorphic function 
on $\mc{D}$ that extends to a meromorphic, $\i\pi$-periodic function on $\Cx$ by the formula
\bem
\wh{\mf{a}}(\xi) \; = \; (-1)^s \cdot \pl{\eps=\pm}{}  \Big[  1+  \ex{ \wh{\mf{A}} ( \xi-\eps \i \zeta_{\e{m}}) } \Big]^{\eps   \e{sgn}\big(\pi -  2\zeta \big)  \bs{1}_{ \mc{D} } \big( \xi-\eps \i \zeta_{\e{m}} \big) } \cdot 
 \pl{ y \in \wh{\mathbb{Y}}_{\kappa} }{} \bigg\{ \f{ \sinh(\i\zeta + y- \xi ) }{ \sinh(\i\zeta + \xi - y ) } \bigg\}   \\
\bigg(  \f{ \sinh(  \xi - \tf{ \aleph }{N} ) \sinh(  \xi + \tf{ \aleph }{N} -\i\zeta) }{  \sinh(  \xi + \tf{ \aleph }{N} ) \sinh( \xi - \tf{ \aleph }{N} -\i\zeta )  }   \bigg)^N     \cdot 
 \exp\bigg\{ -\f{h}{T} \; + \; \Oint{ \Dp{} \mc{D}  }{} K(\xi-u) \cdot  \mc{L}\mathrm{n}\Big[ 1+  \ex{ \wh{\mf{A}} } \, \Big](u)  \cdot \dd u    \bigg\} \;. 
\end{multline}
Here, $\bs{1}_{A}$ stands for the indicator function of the set $A$.
In particular, $1+\wh{\mf{a}}$ has no poles or zeroes on $\Dp{}\mc{D}$ and the only singularities of $\wh{\mf{a}}$ in $\mc{D}$ consist of an $N^{\e{th}}$ order pole 
at $-\tf{ \aleph }{ N }$ and poles at $y$ modulo $\i\pi \mathbb{Z}$, with $(y,k_y)\in \wh{\mc{Y}}_{\e{sg}}$, whose order corresponds to the multiplicity $k_y$ of $y$. 
Being meromorphic on $\ov{\mc{D}}$, $1+\wh{\mf{a}}$ admits a finite number $N_{\mc{Z}}$ of zeroes on $\mc{D}$, which are repeated according to their multiplicities $\mu_1,\dots, \mu_{N_{\mc{Z}} }$. 
Then, by construction, one has $\wh{\mf{X}} \subset \{\mu_a \}_{1}^{N_{\mc{Z}}}$. 

Integrating by parts the contour integral in \eqref{ecriture eqn NLI forme primordiale} with the help of the monodromy condition and, then taking the resulting integral 
by residues, ensures that, for $\xi \in \mc{D}$, it holds 
\beq
\wh{\mf{A}}(\xi) \, = \,  -\f{h}{T} \, +\, \mf{w}_N(\xi)  \; - \; \i \pi s  \; + \; \i \sul{ y \in \wh{\mathbb{Y}} }{} \th_+(\xi-y)
-\i N  \th\Big( \xi + \tfrac{\aleph}{N} \Big) \, - \, \i \sul{ y \in \wh{\mc{Y}}_{\e{sg}} }{}\th(\xi-y) \, + \, 
 \i \sul{ a=1}{ N_{\mc{Z}} } \th(\xi-\mu_a) \;. 
\enq
The set  $\{ \mu_a\}_{1}^{N_{\mc{Z}} } \oplus \wh{\mc{Y}} \ominus \wh{\mf{X}}$ unambiguously defines the set $\{ \la_a \}_{1}^{M} $ built up from $M$ complex numbers $\la_1,\dots, \la_M$. 
The cardinality $M$ of this set can be related to the parameter $s$
by taking explicitly the monodromy condition, leading to the constraint $s=N-M$. Then, the above ensures that $\wh{\mf{a}}$ is expressed in terms of $\la_1,\dots, \la_M$
exactly as given in \eqref{definition fct auxiliaire a}. Furthermore, it holds that $(\Dp{\xi}^p\wh{\mf{a}}\,)(\la_a)= - \de_{p,0}$ for $a=1,\dots, M$ and $p=0,\dots, k_{\la_a}-1$, where $k_{\la_a}$ 
is the multiplicity of the root $\la_a$ and that $(\Dp{\xi}^{ k_{\la_a} } \wh{\mf{a}}\,)(\la_a) \not=0$. Thus, the $\la_a$'s satisfy the system of Bethe Ansatz equations.

It remains to establish that the roots  $\la_1,\dots, \la_M$ so constructed are admissible. Since $\{\mu_a\}_{1}^{N_{\mc{Z}}} \subset \mc{D} \subset \mc{S}_{\tf{ \zeta_{\e{m}} }{2} }$, the roots $\mu_a$
are mutually admissible. Also, by the properties of solutions to the non-linear problem, so are the roots $\wh{\mc{Y}}$. Thus, the only possibility for $\{\la_a\}_{1}^{M}$
to be a non-admissible solution set is that there exists a root $\mu_a$ such that $\mu_a=y\pm \i\zeta_{\e{m}}$. Since the set $\wh{\mc{Y}}_{\e{sg}}$ gathers 
the poles of $\wh{\mf{a}}$ inside of $\mc{D}$, which obviously cannot give rise to zeroes of $1+\wh{\mf{a}}$, the only possible solution is that $\mu_a=\la_b+\i \e{sgn}(\pi-2\zeta) \zeta_{\e{m}}$,
for some $(\la_b,k_{\la_b}) \in \wh{\mc{Y}}$. However, one has that 
\bem
\wh{\mf{a}}\,\Big( \la_b+\i \e{sgn}(\pi-2\zeta) \zeta_{\e{m}} + \eps \Big) \; = \; 
\f{(-1)^s \ex{-\f{h}{T}} \sinh(-\eps)  }{ \sinh\big[2 \i  \zeta + \eps \big] } \cdot 
\pl{ \substack{ k=1 \\ \not= b } }{ M } \bigg\{ \f{ \sinh(\la_k-\la_b - \eps) }{ \sinh(2\i\zeta + \la_b-\la_k+ \eps)  }  \bigg\}  \\
\times \bigg(  \f{ \sinh(  \la_b - \tf{ \aleph }{N} + \i\zeta + \eps  ) \sinh(  2\i\zeta + \la_b + \tf{ \aleph }{N} +\eps) }
{  \sinh(  \la_b + \tf{ \aleph }{N} + \i\zeta + \eps  )  \sinh( -\la_b + \tf{ \aleph }{N} -\eps )  }   \bigg)^N   \;. 
\end{multline}
In order for the $\eps \tend 0$ limit not to give zero, one needs the $\sinh(\eps)$ prefactor to be compensated by some $\eps^{-1}$ singularity 
which would stem from the remaining factors. By definition of $\wh{\mc{Y}}$ being admissible, the last term cannot blow up in the $\eps\tend 0$
limit. Hence, the only possibility is that $2\i\zeta + \la_b-\la_k=0$ for some $k$. This however means that $\mu_a= \la_k-\i \e{sgn}(\pi-2\zeta) \zeta_{\e{m}}$,
what is a contradiction to the previous argument. \qed

\vspace{5mm}

It turns out that the knowledge of the $\wh{\mf{A}}$ function allows one to compute all the quantities of interest in the model. 
For instance, one may recast the Eigenvalue of the transfer matrix associated with a distribution of roots $\{\la_a\}_{a=1}^{M}$ 
in terms of the associated function $\wh{\mf{A}}$ as  
\bem
\tau\big( 0 \mid \{\la_a\}_{a=1}^{M} \big) \, = \, \pl{ y \in \wh{\mathbb{Y}}_{\kappa} }{} \bigg\{ \f{ \sinh(y -\i\zeta)  }{ \sinh(y)  } \bigg\} \cdot 
\bigg( \f{ \sinh(\tf{\aleph}{N}+\i\zeta) }{ \sinh(\i\zeta) }  \bigg)^{2N} \\
\times \exp \Bigg\{  \f{h}{2T}      \ - \,  \Oint{ \Dp{}\mc{D}  }{ }  \f{ \sin(\zeta) \,   \mc{L}\mathrm{n}\big[ 1+  \ex{ \wh{\mf{A}} } \, \big](u) }{ \sinh(u-\i \zeta)\,  \sinh(u) } \cdot \f{ \dd u }{ 2\pi }   \Bigg\} \;.
\label{ecriture forme vp QTM}
\end{multline}
 See \textit{e.g.} \cite{DestriDeVegaAsymptoticAnalysisCountingFunctionAndFiniteSizeCorrectionsinTBAFirstpaper} for the details of the algebraic manipulations. 
Also, in \eqref{ecriture forme vp QTM}, we employed the convention for  products which was introduced in \eqref{defintion convention somme produit et cardinalite ensembles}.
Finally, by changing the integration point from $\kappa$ to $\kappa^{\prime}$ in the definition of $\ln \big[ 1+  \ex{ \wh{\mf{A}} } \, \big]$ one may readily check that 
\eqref{ecriture forme vp QTM} does not depend on the choice of $\kappa$, as it should be.

It is clear from the proof that for a given set of admissible Bethe roots $\{\la_a\}_{1}^{M}$, there is a very large choice of domains $\mc{D}$ and  that different choices of domains 
$\mc{D}$ may lead to very different sets $\wh{\mf{X}}$ and $\wh{\mc{Y}}$. Thus, apparently very different non-linear integral equations may lead to exactly the same solution to 
the Bethe Ansatz equations. Thus, in order to hope finding all distinct solutions to the original problem of solving Bethe equations, a reasonable strategy seems to fix, once and 
for all, the domain $\mc{D}$ and to look for different solutions giving rise to distinct sets $\wh{\mf{X}}$ and $\wh{\mc{Y}}$. Such a construction is also dictated by the physical content
of the non-linear integral equation.

As shown earlier on in Proposition \ref{prop vp dominante et vp sous dominantes}, $\op{t}_{q}$  admits a non-degenerate dominant Eigenvalues $\wh{\La}_{\e{max}}$: all the other
Eigenvalues are smaller in modulus. It was argued in the literature \cite{DestriDeVegaAsymptoticAnalysisCountingFunctionAndFiniteSizeCorrectionsinTBAFirstpaper,DestriDeVegaAsymptoticAnalysisCountingFunctionAndFiniteSizeCorrectionsinTBAFiniteMagField,
GohmannKlumperSeelFinieTemperatureCorrelationFunctionsXXZ,KlumperNLIEfromQTMDescrThermoXYZOneUnknownFcton} and this property will be established rigorously in the following, that this dominant Eigenvalue is described by a solution set 
$\{\la_a^{(\e{max})}\}_1^{N}$ to the Bethe equations containing exactly $N$ distinct admissible Bethe roots, $\la_1^{(\e{max})}, \dots, \la_N^{(\e{max})}$.

Let $\wh{\mf{a}}_{\e{max}}$ stand for the auxiliary function associated with this distribution of roots. 
One then considers a domain $\mc{D}$ along with  its canonically oriented boundary $\Dp{}\mc{D}$ such that 
\begin{itemize}
\item  $\{\la_a^{(\e{max})}\}_1^{M} \subset \mc{D}$ ;
\item  no other zero of $1+\wh{\mf{a}}_{\e{max}}$, \textit{viz}. one that would not be a Bethe root, is contained in $\ov{\mc{D}}$;
\item the $N$-fold pole at $-\tf{\aleph}{N}$ is contained in $\mc{D}$;
\item no other pole of $\wh{\mf{a}}_{\e{max}}$, is contained in $\ov{\mc{D}}$.
\end{itemize}
Note that there is still a rather large freedom of choice of the domain $\mc{D}$, although for practical purposes, some may turn out to be
more suited than others. Traditionally, owing to its important physical interpretation, this domain $\mc{D}$ is then used to study the non-linear problem 
involving non-empty sets $\wh{\mf{X}}$ and $\wh{\mc{Y}}$ for which the associated
   solution, if existing, will give rise to some Eigenvalue of the quantum transfer matrix 
differing from the dominant one.

The domain $\mc{D}$ so chosen for the dominant state's Bethe roots could, in principle, be located outside of $\mc{S}_{ \tf{ \zeta_{\e{m}} }{2 } }$.
However, as will be shown in a later stage of this work, for temperature high enough, it is not so and, in fact, $\mc{D}$ can be taken as a small disc centred at the 
origin. 

The domain $\mc{D}$ associated with the dominant Eigenvalue's Bethe roots being fixed, the sets $\wh{\mf{X}}$ and $\wh{\mc{Y}}$ subordinate to it are usually called
the hole set -for $\wh{\mf{X}}$- and the particle set -for $\wh{\mc{Y}}$-.

\subsection{The formal infinite Trotter number limit}

The main advantage of the non-linear integral equation based description, relies on the possibility to easily take, at least on formal grounds, 
the Trotter limit on the level of the integral representation for the Eigenvalues \eqref{ecriture forme vp QTM}
and for the solution to the non-linear integral equations \eqref{ecriture eqn NLI forme primordiale} driving these. 
Indeed, the parameter $N$ only appears in the driving term  $\mf{w}_N$ of \eqref{ecriture eqn NLI forme primordiale}. 
Thus, \textit{assuming} that $\wh{\mf{A}} \limit{N}{+\infty} \mf{A}$ pointwise on $\Dp{}\mc{D}$, and that all properties of the non-linear problem are preserved under this limit,  one may readily characterise the limit of 
$\tau\big( 0 \mid \{\la_a\}_{a=1}^{M} \big)$ as well as derive a non-linear integral equation satisfied by 
$\mf{A}$. Indeed, in the infinite Trotter number limit, one has 
\beq
 \mf{w}_N(\xi) \;  \limit{N}{+\infty}\;  - 2 \aleph \Big( \coth(\xi)-\coth(\xi-\i\zeta) \Big)= \f{2J \sin^2(\zeta) }{ T \sinh(\xi)\sinh(\xi-\i\zeta) } \;. 
\enq
In addition one \textit{assumes} the existence of the limit of the particle and hole roots
\beq
\wh{x}_a\tend x_a \; , \;\; a=1,\dots,|\mf{X}| \qquad \e{and} \qquad  \wh{y}_a \tend y_a \; , \;\; a=1,\dots,|\mc{Y}| \;. 
\enq
All these assumptions result in the non-linear integral equation satisfied by the limit function on $\mc{S}_{\zeta_{\e{m}}/2}$:
\beq
\mf{A}(\xi) \, = \, -\f{1}{T} e_{0}(\xi) \; - \; \i \pi s  \; + \; \i \sul{ y \in \mathbb{Y}_{\kappa} }{} \th(\xi-y)
\; + \; \Oint{ \Dp{} \mc{D} }{} K(\xi-u) \cdot   \mc{L}\mathrm{n}\Big[ 1+ \ex{\mf{A}} \Big](u) \cdot \dd u   \;.  
\label{ecriture NLIE Trotter infini}
\enq
Here, 
\beq
e_0(\xi) \, = \, h \, - \, \f{2J \sin^2(\zeta) }{ \sinh(\xi) \sinh(\xi-\i\zeta) } \;. 
\enq
Furthermore,   
\beq
\mathbb{Y}_{\kappa} \, = \, \mathbb{Y} \ominus \big\{     \kappa\big\} ^{\oplus  (s+| \mathbb{Y} |)} \quad \e{with} \quad 
\mathbb{Y} \, = \, \mc{Y}\oplus   \mc{Y}_{\e{sg}}  \ominus   \mf{X}   \; , 
\label{premiere definition de Y kappa}
\enq
and where 
\beq
\mf{X} \; = \; \big\{ x_a \big\}_{a=1}^{ | \mf{X} | } \qquad \e{and} \qquad \mc{Y} \; = \; \big\{ y_a \big\}_{a=1}^{ | \mc{Y} | } \;. 
\enq
The non-linear integral equation at infinite Trotter number is to be supplemented with the constraints
\beq
\Oint{ \Dp{}\mc{D} }{} \f{ \mf{A}^{\prime}(u) } { 1+ \ex{-\mf{A}(u)} }  \cdot \f{\dd u}{2\i\pi}  \; = \;|\mf{X}|  - |\mc{Y}| -|\mc{Y}_{\e{sg}}| - s\quad,  
\label{ecriture condition subsidiaires solution NLIE spectre}
\enq
and
\beq
\left\{ \ba{cccc}  \Big\{ \Dp{\xi}^r \ex{\mf{A}(\xi)} \Big\}_{\mid \xi=x} \, = \, -1  & \e{for}  &  (x,k_x) \,  \in \mf{X} \;,   &  \, r=0,\dots, k_x-1 \; ,\vspace{2mm} \\
   \Big\{ \Dp{\xi}^r \ex{\mf{A}(\xi)} \Big\}_{\mid \xi=y} \, = \, -1  & \e{for}  &  (y,p_y) \,  \in \mc{Y}  \; ,  & r=0,\dots, p_y-1  \ea \right. \;  
\enq
where $k_x$, resp. $p_y$, is the multiplicity of $x$ in respect to the roots $x_1,\dots, x_{|\mf{X}|}$, resp. of $y$ in respect to the roots $y_1,\dots, y_{|\mc{Y}|}$.

Similar handlings lead to 
\bem
\lim_{N\tend + \infty} \Big\{ \tau\big( 0 \mid \{\la_a\}_{a=1}^{M} \big) \Big\}  \, = \, \pl{ y \in \mathbb{Y}_{\kappa} }{} \bigg\{ \f{ \sinh(y -\i\zeta)  }{ \sinh(y)  } \bigg\}  \\
\times \exp \Bigg\{  \f{h}{2T}  -  \f{2 J}{T} \cos(\zeta)   \, - \,  \Oint{ \Dp{}\mc{D}  }{ }  \f{ \sin(\zeta) \,  \mc{L}\mathrm{n}\big[ 1+  \ex{ \mf{A} } \, \big](u) }{ \sinh(u-\i \zeta)\,  \sinh(u) } \cdot \f{ \dd u }{ 2\pi }  \Bigg\} \;.
\label{ecriture forme vp QTM limite}
\end{multline}

\section{Existence and uniqueness of solutions to the NLIE for large T }
 \label{Section existance et unicite sols NLIE}

Recall the notations $\mc{S}_{\a}$ for the strip of width $\a$ centred around $\R$
and $\mc{D}_{z_0,\a}$ for the disc of radius $\a$ centred at $z_0 \in \Cx$,  both introduced in \eqref{definition bande largeur alpha et disque}. 
 In this section, we develop a rigorous framework allowing one to lay a firm ground to the heuristics outlined in the previous section.

\subsection{\boldmath An auxiliary problem grasping the leading high-temperature behaviour}

The first step of the proof consists in providing a characterisation of the properties of an auxiliary function which will subsequently be shown to grasp the leading large-$T$
asymptotic behaviour of $\mf{A}$.

For further convenience it appears useful to single out a specific class of "particle"-"hole" sets. 

\begin{defin}
\label{Definition C eps alpha rho class}
 The sets $\mf{X}$, $\mc{Y}$ are said to belong to the class $\mc{C}^{\eps}_{\a, \varrho}$ with cardinalities $n_x, n_y$ if

 \begin{itemize}
 
 \item   $\mf{X}= \{x_1,\dots, x_{|\mf{X}|} \}\subset \mc{D}_{0,\eps}$  with  $|\mf{X}|=n_x$ and all $x_a$'s are pairwise distinct; 
 
 \item $\mc{Y} \subset \mc{S}_{ \frac{\pi}{2} } \setminus  \Big\{ \mc{D}_{0,\eps} \cup_{\ups=\pm } \mc{D}_{ \ups \i \zeta_{\e{m}}, \a} \Big\}$  and $\mc{Y}= \{y_1,\dots, y_{|\mc{Y}|} \} $ with $|\mc{Y}|=n_y$
 and the $y_a$'s being pairwise distinct; 
 
 \item the elements of $\mc{Y}$ are subject to the constraint
\beq
\Big|(-1)^s  \pl{y \in \mc{Y} }{}  \f{ \sinh(\i\zeta + y) }{ \sinh(\i\zeta - y) } + 1 \Big| \, > \, \varrho \;.
\label{condition eloignement racines de 0}
\enq

 \end{itemize}

\end{defin}

Note that  for $\eps <\a$, all sets $\mf{X}$ and $\mc{Y}$ in the class   $\mc{C}^{\eps}_{\a, \varrho}$  are such that there are no singular roots $\mc{Y}_{\e{sg}}$ relative to $\mc{D}_{0,\eps}$, 
\textit{viz}. $y-\i \zeta_{\e{m}} \not\in \mc{D}_{0,\eps}$ modulo $\i\pi \, \mathbb{Z}$, for any $y \in \mc{Y}$.

\begin{lemme}
 \label{Ecriture Lemme no zeroes a infty}

Fix integers $n_x, n_y\geq 0$ and $n_{\kappa}\in \mathbb{Z}$.  Let $\a, \varrho>0$ be given. Then, there exists $\eps>0$, $\tf{\a}{2}> \eps > 0$ such that 
\begin{itemize}
 
 \item for any $\mf{X}$, $\mc{Y}$ in the class $\mc{C}^{\eps}_{\a, \varrho}$ with cardinalities $n_x, n_y$,
\item for any $\kappa \in  \ov{\mc{D}}_{0,\eps}$,
 
\end{itemize}
the function $1+f$ with 
\beq
f(\la) \; = \; (-1)^s\pl{  \substack{ y \in \mc{Y}   \\ \ominus \big\{ \mf{X}\oplus \{\kappa\}^{ \oplus  n_{\kappa} } \big\}  } }{} \f{ \sinh(\i\zeta + y - \la) }{ \sinh(\i\zeta + \la- y) } \;, 
\enq
has no zeroes inside of $ \ov{\mc{D}}_{0,\eps}$ and, uniformly in $\ov{\mc{D}}_{0,\eps}$, is subject to the lower bound
\beq
\big| 1 \, + \, f(\la) \big| \, \geq \, \f{ \varrho }{2} \;. 
\label{ecriture borne inf 1+f}
\enq

\end{lemme}

\proof 

On may recast 
\beq
1+f(\la) \; = \; \bigg( (-1)^s  \pl{y \in \mc{Y} }{}  \f{ \sinh(\i\zeta + y) }{ \sinh(\i\zeta - y) } + 1 \bigg) \, \ex{u(\la)} \; + \, 1-\ex{ u(\la)}
\enq
with 
\beq
u(\la) \; = \; \i \sul{ y \in \mc{Y} }{} \big[ \wt{\th}(\la-y) - \wt{\th}(-y) \big] \, - \, \i \sul{ y \in   \mf{X}\oplus \{\kappa\}^{ n_{\kappa} } }{}   \wt{\th}(\la-y)  \;. 
\enq
Here, $\wt{\th}$ corresponds to  a choice of the branch of the logarithm defining $\th$ in such a way that  $\la \mapsto  \wt{\th}(\la-y)$ is holomorphic on $\mc{D}_{0,\eps}$. 
Note that the branch may differ for different choices of $y \in \mc{Y}$. We point out that such a choice is possible since singular roots do not exist due to 
$\mc{Y} \subset  \msc{B}_{\a,\eps}$ with 
\beq
\msc{B}_{\a,\eps} \, = \, \mc{S}_{ \frac{\pi}{2} } \setminus \Big\{ \mc{D}_{0,\eps} \cup_{\ups=\pm } \mc{D}_{ \ups \i \zeta_{\e{m}}, \a} \Big\}
\enq
 for $\eps>0$ small enough. Observe that if $0<\eps< \tf{\a}{2} $ and $\a$ is small enough, then for any $\la \in \mc{D}_{0,\eps}$, 
\beq
y \, \in   \msc{B}_{\a,\eps}  \quad \Rightarrow \quad 
\la-y \, \in  \, \mc{S}_{ \frac{\pi}{2} +\eps } \setminus \underset{\ups=\pm }{\cup}  \mc{D}_{ \ups \i \zeta_{\e{m}}, \a-\eps }  \;. 
\enq
Thus, by the mean value theorem and $\i \pi $- periodicity of $K$,
\beqa
\big| u(\la) \big| & \leq &  \eps \,  |\mc{Y} | \,   \underset{s \in \mc{D}_{0,\eps} }{\e{sup}}\,  \underset{ w \in \msc{B}_{\a,\eps} }{ \e{sup} } |2\pi K(s-w)| 
\, + \,    2\eps \, \Big(   |\mf{X}| + |n_{\kappa}| \Big) \,  \underset{s \in \mc{D}_{0,2 \eps} }{\e{sup}}| 2\pi K(s )|  \\
& \leq & 4 \pi  \eps  \Big( |\mc{Y} |   + |\mf{X}| + |n_{\kappa}| \Big) \,  \underset{ w \in  \msc{B}_{ \frac{\a}{2},0 }  }{\e{sup}} |K( w)|  \; . 
\eeqa
The above bounds ensure the existence of an $\eps$-independent constant $C>0$  such that 
\beq
\big| 1+f(\la) \big| \, \geq \, 
\big|  \rho {\rm e}^{-|u(\la)|} - |u(\la)|  {\rm e}^{|u(\la)|}  \big|  
\, \geq \, \rho {\rm e}^{- \eps C} - \eps C {\rm e}^{\eps C}   \geq  \, \f{\varrho}{2}
\enq
provided that $\eps$ is small enough. \qed

\vspace{3mm}

The above lemma has an immediate corollary.

\begin{cor}

For given $n_x, n_y \in \mathbb{N}$ there exists $\eps>0$
such that for any $\mf{X}$ and $\mc{Y}$ in the class $\mc{C}^{\eps}_{\a, \varrho}$ with cardinalities $n_x, n_y$ and, for any $\kappa \in \Dp{}\mc{D}_{0,\eps}$,   upon denoting
\beq
\mathbb{Y}_{\kappa} \; = \; 
\mc{Y}  \ominus \Big\{  \mf{X} \oplus \{\kappa\}^{\oplus s+|\mc{Y}|+|\mc{Y}_{\e{sg}}|-|\mf{X}|}\Big\}   \; , 
\label{defintion Y kappa}
\enq
the function 
\beq
\mf{A}_{\infty}(\xi) \, = \,-\i \pi s \, + \, \i \sul{ y \in \mathbb{Y}_{\kappa} }{} \th ( \xi-y) 
\label{definition fct A infty}
\enq
 has no zeroes in $\ov{\mc{D}}_{0,\eps}$, satisfies the lower bound 
\beq
\big| 1+\ex{\mf{A}_{\infty}(\la)}\big| \geq \f{\varrho}{2} \quad on \quad  \ov{\mc{D}}_{0,\eps} 
\label{ecriture lower bound exp A infty}
\enq
and solves the non-linear integral equation
\beq
\mf{A}_{\infty}(\xi) \, = \,-\i \pi s \, + \, \i \sul{ y \in \mathbb{Y}_{\kappa} }{} \th ( \xi-y) 
 \; + \;  \Oint{ \Dp{}\mc{D}_{0,\eps} }{} K(\xi-u) \mc{L}\mathrm{n}\Big[ 1+ \ex{\mf{A}_{\infty}} \Big](u) \cdot \dd u  \;,  
\label{ecriture equation pour a infty}
\enq
for $\xi \in \mc{S}_{\zeta_{\e{m}}/2}$,
 where the logarithm is defined as 
\beq
\mc{L}\mathrm{n}\Big[ 1+ \ex{\mf{A}_{\infty}} \Big](u) \; = \; \Int{ \kappa }{ u}  \f{ \mf{A}_{\infty}^{\prime} (s) }{  1+ \ex{ - \mf{A}_{\infty}(s)}  } \cdot \dd s  \; + \; \ln \Big[  1+ \ex{\mf{A}_{\infty}(\kappa)}  \Big] \, 
\enq
and the integral runs along $\Dp{}\mc{D}_{0,\eps}$, from $\kappa$ to $u$.

Finally, $\mf{A}_{\infty}$ satisfies the  zero-monodromy condition
\beq
 \Oint{ \Dp{}\mc{D}_{0,\eps} }{  }  \f{ \mf{A}_{\infty}^{\prime} (s) }{  1+ \ex{ - \mf{A}_{\infty}(s)} } \cdot \f{ \dd s }{ 2\i \pi }  \; = \; 0 \;. 
\label{condition no monodromy for exp A infty}
\enq

\end{cor}

Note that, \eqref{ecriture equation pour a infty} is well-defined since the condition \eqref{ecriture lower bound exp A infty} ensures that $\mc{L}\mathrm{n}\Big[ 1+ \ex{\mf{A}_{\infty}} \Big]$
can be extended to an analytic function on $\ov{\mc{D}}_{0,\eps}$; in particular, the result does not depend on the path used for the continuation. 
Also, we recall that due to $\mc{Y} \subset \mc{S}_{ \frac{\pi}{2} } \setminus \Big\{ \mc{D}_{0,\eps} \cup_{\ups=\pm } \mc{D}_{ \ups \i \zeta_{\e{m}}, \a} \Big\}$, one has that $\mc{Y}_{\e{sg}}=\emptyset$ and hence 
the definitions of $\mathbb{Y}_{\kappa}$ given in \eqref{defintion Y kappa} and \eqref{premiere definition de Y kappa} are  consistent.

\Proof 

The zero-monodromy condition follows from the fact that $\mf{A}_{\infty}$ is holomorphic on $\mc{D}_{0,\eps}$ and that one has the lower bound \eqref{ecriture lower bound exp A infty}.

Using that $K(\xi - u)$ is meromorphic in $\mc{S}_{ \frac{\pi}{2} }$
with simple poles at $u = \xi \pm {\rm i} \zeta_{\e{m}} $ mod $\i\pi$n we conclude by means of the
residue theorem that, for $\xi \in \mc{S}_{ \tf{\zeta_{\e{m}}}{2} }$
\beq
\Oint{ \Dp{}\mc{D}_{0,\eps} }{} K(\xi-u) \mc{L}\mathrm{n}\Big[ 1+ \ex{\mf{A}_{\infty}}\Big](u) \cdot \dd u \; = \; 0 \; . 
\enq

\qed

\subsection{An auxiliary fixed point problem}

Recall the definition \eqref{definition fct A infty} of the function $\mf{A}_{\infty}(\xi)$, and introduce
\beq
\msc{L}(\nu,x) \; = \; \f{1}{ 1 + \ex{-\mf{A}_{\infty}(\nu)-x} } \;. 
\enq
Further set 
\beq
\varpi_{N}(\la) \, = \, h-T\mf{w}_N(\la)  
\enq
and, given $\eps>0$, define
\beq
 \chi_{N;\eps}(\la) \, = \, -\Oint{ \Dp{}\mc{D}_{0,\eps}}{} \hspace{-1mm}  \f{ K(\la-u)  }{ 1 + \ex{-\mf{A}_{\infty}(u)} } \cdot \varpi_{N}(u)  \cdot \dd u \;. 
\label{fonction chi definissant partie cste op ONT}
\enq
Next,  given $\xi \in\mc{O}(\mc{S}_{\tf{\zeta_{\e{m}} }{2}})$, denote
\beq
 G[ \ga ](\nu,t) \, = \,    \ga(\nu)\ga^{\prime}(\nu) \,  \Dp{2}\msc{L}\bigg( \nu, \f{t \ga (\nu) }{ T } \bigg)  \; + \; (1-t) \, \ga^2(\nu) \, \mf{A}_{\infty}^{\prime}(\nu) \, 
\Dp{2}^2 \msc{L}\bigg(\nu, \f{t \ga(\nu) }{T} \bigg)   \;. 
\label{ecriture fonction G gamma}
\enq
 Finally, given $\ga \in\mc{O}(\mc{S}_{\tf{\zeta_{\e{m}} }{2}}) $, define the map $\op{O}_{T,N}$ 
\beq
\op{O}_{T,N}[\xi](\la) \; = \; \chi_{N;\eps}(\la) \, + \,     \f{1}{ T}  \Oint{  \Dp{} \mc{D}_{0,\eps} }{}\hspace{-1mm} \dd u   K(\la-u) 
  \Int{\kappa}{u}  \dd v  \Int{0}{1} \dd t \;   G\big[ \xi  - \varpi_N \big](\nu,t)  \;. 
\label{definition operateur contractant}
\enq
Above, the $v$ integral runs along $ \Dp{} \mc{D}_{0,\eps}$. 

We are going to establish that the operator $\op{O}_{T,N}$
\begin{itemize}
 
 \item  is well-defined on an appropriate functional space, provided $T$ and $N$ are large enough;
 \item is closely related to the problem of interest, in that it provides one with a rewriting of the non linear integral equation \eqref{ecriture eqn NLI forme primordiale}. 
 
\end{itemize}

We first introduce the appropriate functional space. Given $r>0$, set 
\beq
\mc{B}_{ r } \; = \; \Big\{ \xi \in \mc{O}(\mc{S}_{\tf{\zeta_{\e{m}} }{2}}) \, : \, \norm{ \xi }_{ L^{\infty}(\mc{S}_{\tf{ \zeta_{\e{m}} }{2}})} \, \leq \, r   \Big\}
\quad \e{with} \quad \norm{g}_{L^{\infty}(U)} \; = \; \e{supess}_{u \in U} |g(u)| \;. 
\enq
By virtue of Montel's theorem, for any $r>0$, $\mc{B}_{ r } $ is a complete metric space  with respect to the distance 
\beq
d(f,g) \; = \; \norm{ f-g }_{ L^{\infty}(\mc{S}_{\tf{ \zeta_{\e{m}} }{2}})}\;. 
\enq

 In the following set
\beq
\mf{c} \, = \, 2 \norm{ \chi_{ \infty ;\eps }}_{L^{\infty}(\mc{S}_{\tf { \zeta_{\e{m}} }{2}})}  \;. 
\label{definition constante c}
\enq

\begin{prop}
\renewcommand{\labelenumi}{(\roman{enumi})}
\label{Proposition correspondance solutions NLIE originelle et NLIE type point fixe}
For given $n_x, n_y \in \mathbb{N}$ there exist $T_0>0$, $N_0\in \mathbb{N}$ and $\eps>0$ 
such that for any $\mf{X}$ and $\mc{Y}$ in the class $\mc{C}^{\eps}_{\a, \varrho}$ with cardinalities $n_x, n_y$ and for any $\kappa \in \Dp{}\mc{D}_{0,\eps}$
\begin{enumerate}
\item  $\op{O}_{T,N}$ is well-defined on $\mc{B}_{ \mf{c} }$, with $\mf{c}$ as given in \eqref{definition constante c};
\item $\op{O}_{T,N}$  stabilises  $\mc{B}_{ \mf{c} }$, \textit{viz}. $\op{O}_{T,N}\big[\mc{B}_{ \mf{c} } \big] \subset \mc{B}_{ \mf{c} }$;
\item
the solutions to the non-linear integral equation  \eqref{ecriture eqn NLI forme primordiale} associated with the sets $\mc{Y}$ and $\mf{X}$ are in one-to-one correspondence with the fixed points of the operator $\op{O}_{T,N}$. 
\end{enumerate}
\end{prop}

\Proof

We start by establishing (iii). Upon making the change of unknown function 
\beq
\wh{\mf{A}}(\la) \, = \, \f{1}{T} \Big( \xi(\la)\, - \, \varpi_N(\la) \Big)\, + \, \mf{A}_{\infty}(\la) \; ,
\label{definition lien hat A et pt fixe}
\enq
where $\mf{A}_{\infty}(\la)$ was introduced in \eqref{definition fct A infty},  one recasts the NLIE \eqref{ecriture eqn NLI forme primordiale} in the form
\beq
\xi(\la) \; = \; T \Oint{  \Dp{} \mc{D}_{0,\eps} }{} K(\la-u)\cdot \bigg\{ \mc{L}\mathrm{n}\Big[ 1+  \ex{\mf{A}_{\infty} +\frac{\xi-\varpi_N }{ T} } \Big](u)  -   \mc{L}\mathrm{n}\Big[ 1+  \ex{ \mf{A}_{\infty} } \Big](u) \bigg\} \;,
\enq
defining $\xi(\la)$ not only on $\Dp{} \mc{D}_{0,\eps}$ but for all $\la \in \mc{S}_{\tf{\zeta_{\e{m}} }{2}}$.
Then, the Taylor-integral expansion
\beq
\msc{L}(\nu,x) \; = \; \msc{L}(\nu,0) \, + \, x  \Dp{2}\msc{L}(\nu,0)  \, + \, x^2 \Int{0}{1} (1-t) \Dp{2}^2\msc{L}(\nu,tx) \cdot \dd  t
\enq
leads to the expression
\bem
\mc{L}\mathrm{n}\Big[ 1+  \ex{\mf{A}_{\infty} +\frac{\ga }{ T} } \Big](u) \; = \; \mc{L}\mathrm{n}\Big[ 1+  \ex{\mf{A}_{\infty}} \Big](u)\; + \frac{1}{T}\biggl[ \; \f{ \ga(u) }{  1+  \ex{-\mf{A}_{\infty}(u) }   } 
\, - \,  \f{ \ga(\kappa) }{  1+  \ex{-\mf{A}_{\infty}(\kappa) }   } \biggr] \\
\;+\; \f{1}{T^2} \Int{\kappa}{u}\dd \nu \Int{0}{1} \dd t \Bigg\{ \ga^{\prime}(\nu)\, \ga(\nu)  \, \Dp{2} \msc{L}\bigg(\nu, \f{t \ga(\nu) }{T} \bigg) \, + \,  (1-t) \, \ga^2(\nu) \, \mf{A}_{\infty}^{\prime}(\nu) \, 
\Dp{2}^2 \msc{L}\bigg(\nu, \f{t \ga(\nu) }{T} \bigg)  \Bigg\} \;. 
\end{multline}
Altogether, this entails that $\xi$ is a fixed point  of $\op{O}_{T,N}$:
\beq
\xi(\la) \, = \,  \op{O}_{T,N}[\xi](\la) \;. 
\enq
By tracing the above steps backwards, one  concludes that any fixed point $\xi$ of  $\op{O}_{T,N}$ gives rise to a solution $\wh{\mf{A}}$ of \eqref{ecriture eqn NLI forme primordiale}. 

 The first claim of the proposition follows once  a few preparatory bounds are set. 
 As shown in Lemma \ref{Ecriture Lemme no zeroes a infty}, uniformly in the parameters forming the sets $\mc{Y}, \mf{X}$, it holds that
\beq
\Big|  1+\ex{\mf{A}_{\infty}(\nu)} \Big| \, \geq \, \frac{\varrho}{2} \quad \e{and} \quad   \Big| \ex{\mf{A}_{\infty}(\nu)} \Big| \, \leq \, C
\enq
for any $\nu \in \op{\mc{D}}_{0,\eps}$, provided that $\eps>0$ is small enough. 
Then, there exists $x_0$ such that for any $|x| \, \leq \, 2x_0$, one has $|1-\ex{-x}|\leq \tf{\varrho}{4}$. 
Thence, for any $|x| \, \leq \, 2x_0$ and $\nu \in \op{\mc{D}}_{0,\eps}$ one has the lower bound
\beq
\Big|  1+\ex{-x-\mf{A}_{\infty}(\nu) } \Big| \, \geq \, \f{\varrho}{4}  \Big| \ex{ - \mf{A}_{\infty}(\nu)} \Big| \qquad \e{leading} \; \e{to} \qquad 
\Big| \msc{L}(\nu,x) \Big| \, \leq \, \f{ 4 C }{ \varrho } \;. 
\enq
Since $x \mapsto \msc{L}(\nu,x) $ is holomorphic on $\mc{D}_{0,2x_0}$ it follows that
\beq
\norm{  \Dp{2}^k \msc{L}(\nu, \cdot )   }_{ L^{\infty} (\mc{D}_{0,x_0} ) } \; \leq \;  C_k \norm{   \msc{L}(\nu, \cdot )   }_{ L^{\infty} (\Dp{}\mc{D}_{0, 2 x_0} ) } \; \leq \; \wt{C}_k
\label{estell}
\enq
for some constants $C_k, \wt{C}_k$\ uniformly in $\nu \in \op{\mc{D}}_{0,\eps}$.

Let $\xi \in \mc{B}_{ \mf{c} }$ and define
\beq
\ga_{\xi} \, = \, \xi-\varpi_{N} \; .
\enq
Then there exists $N_0 \ge 0$ such that for all $N \ge N_0$ the function $\varpi_N$
is holomorphic in an annulus containing $\Dp{}\mc{D}_{0,\eps}$ and
$\Dp{}\mc{D}_{0,2\eps}$. Thus, for $N \ge N_0$, there are constants
$C_k'$ such that
\beq
\norm{  \ga_{\xi}^{(k)}}_{ L^{\infty} (\Dp{}\mc{D}_{0,\eps})} \; \leq \;
C_k' \norm{  \ga_{\xi}}_{ L^{\infty} (\Dp{}\mc{D}_{0,2\eps})} \; \leq \;
C_k' \Bigl( \mf{c} + \norm{\varpi_{N}}_{ L^{\infty} (\Dp{}\mc{D}_{0,2\eps} ) } \Bigr)
\label{estgammaxi}
\enq
where $\ga_{\xi}^{(0)} = \ga_{\xi}$ and $\ga_{\xi}^{(k)}$ for $k > 0$ stands
for the $\rm k^{th}$ derivative of $\ga_{\xi}$. Due to \eqref{estgammaxi} there
exists $T_0$ such that $T^{-1}\cdot \norm{  \ga_{\xi}}_{ L^{\infty}
(\Dp{}\mc{D}_{0,\eps} ) } \; \leq \; x_0$ for all $T > T_0$.
Using \eqref{ecriture fonction G gamma}, \eqref{definition operateur contractant},
\eqref{estell} and \eqref{estgammaxi}, one establishes the bound
\bem
\norm{ \op{O}_{T,N}[\xi]  }_{ L^{\infty}(\mc{S}_{ \tf{\zeta_{\e{m}}  }{2} }) } \, \leq  \, \norm{ \chi_{N;\eps }  }_{ L^{\infty}(\mc{S}_{ \tf{ \zeta_{\e{m}} }{2} }) } \\
+ \, \f{4 \pi^2 \eps^2}{T} \cdot \norm{ K }_{ L^{\infty}(\mc{S}_{ \tf{ \zeta_{\e{m}} }{2} +\eps }) } \cdot   C_0'  
\Bigl( \mf{c} + \norm{\varpi_{N}}_{ L^{\infty} (\Dp{}\mc{D}_{0,2\eps} ) } \Bigr)^2
\cdot \bigg\{ C_1^{\prime} \wt{C}_1 \, + \, \frac{  C_0'  \wt{C}_2 }{2}    \cdot
\norm{  \mf{A}_{\infty}^{\prime}}_{ L^{\infty} (\Dp{}\mc{D}_{0,\eps ) }} \bigg\}   \;. 
\end{multline}

Then, since $\chi_{N,\mc{D}}=\chi_{\infty;\eps} + \e{O}\Big( \tfrac{ 1 }{ N T }\Big)$, it is enough to take $T$ and $N$ large enough so as to get 
\beq
\norm{ \op{O}_{T,N}[\xi]  }_{ L^{\infty}(\mc{S}_{ \tf{\zeta_{\e{m}}}{2} }) } \, \leq  \,  2 \norm{ \chi_{ \infty ;\eps}  }_{ L^{\infty}(\mc{S}_{ \tf{ \zeta_{\e{m}} }{ 2 } }) }   \;. 
\enq
This entails the claim. \qed

\vspace{2mm}

Let $\op{O}_{T}$ be defined as 
\beq
\op{O}_{T }[\xi](\la) \; = \; \chi_{\infty;\eps}(\la) \, + \,     \f{1}{ T}  \Oint{  \Dp{} \mc{D}_{0,\eps} }{}\hspace{-1mm} \dd u   K(\la-u) 
  \Int{\kappa}{u}  \dd v  \Int{0}{1} \dd t \;   G\big[ \xi  - e_0 \big](\nu,t) 
\label{ecriture rep int operateur limite OT}
\enq
with
\beq
  \chi_{\infty;\eps}(\la) \, = \, -\Oint{ \Dp{}\mc{D}_{0,\eps}}{} \hspace{-1mm}  \f{ K(\la-u)  }{ 1 + \ex{-\mf{A}_{\infty}(u)} } \cdot e_0(u)  \cdot \dd u
\enq
and the $v$ integral in \eqref{ecriture rep int operateur limite OT} running along $ \Dp{} \mc{D}_{0,\eps}$.

\begin{theorem}
 \label{Theorem existence et unicite point fixe de OTN}
 
 There exist $N_0, T_0$ such that, for any $N>N_0$ and $T>T_0$:
\begin{itemize}
 
\item  the operator $\op{O}_{T,N}$ admits a unique fixed point in $\mc{B}_{\mf{c}}$;
\item the fixed point is continuous in $N\geq N_0$ and 
converges, when $N\tend +\infty$, to the unique fixed point of $\op{O}_{T}$ in  $\mc{B}_{\mf{c}}$.

\end{itemize}

\noindent The existence of a unique fixed point for $\op{O}_{T}$ is part of the conclusions. 

\end{theorem}

\Proof

 One  starts by showing that $\op{O}_{T,N}$ is strictly contractive on $\mc{B}_{\mf{c}}$, \textit{viz}. for any $\xi_1,\xi_2 \in \mc{B}_{\mf{c}}$,
\beq
\norm{ \op{O}_{T,N}[\xi_1] - \op{O}_{T,N}[\xi_2]   }_{ L^{\infty}(\mc{S}_{ \tf{ \zeta_{\e{m}} }{2} }) } \, \leq  \, \f{C}{T} \cdot \norm{ \xi_1-\xi_2  }_{ L^{\infty}(\mc{S}_{ \tf{ \zeta_{\e{m}} }{2} }) }  
\enq
for some constant $C>0$.  Notice that 
\beq
\ga_{\xi_{1}}^2-\ga_{\xi_{2}}^2 \, = \, \big( \xi_1-\xi_2\big) \big( \xi_1+\xi_2-2\varpi_N \big) \;. 
\enq
Likewise, one has 
\bem
\ga_{\xi_{1}}(\nu) \ga_{\xi_{1}}'(\nu) \cdot \Dp{2} \msc{L} \bigg( \nu, \f{t \ga_{\xi_1} ( \nu ) }{ T } \bigg) \, - \,\ga_{\xi_{2}}(\nu) \ga_{\xi_{2}}'(\nu) \cdot \Dp{2} \msc{L} \bigg( \nu, \f{t \ga_{\xi_2} ( \nu ) }{ T } \bigg)  \\
\; = \;  \f{1}{2}\Dp{\nu}\Big[ \ga_{\xi_{1}}^2(\nu)-\ga_{\xi_{2}}^2(\nu)   \Big]\cdot \Dp{2}\msc{L}\bigg(\nu, \f{t \ga_{\xi_1} (\nu) }{ T } \bigg)
\, + \, \f{1}{2} \Dp{\nu}\Big[ \ga_{\xi_{2}}^2(\nu) \Big] \cdot \Bigg\{  \Dp{2}\msc{L}\bigg(\nu,  \f{t \ga_{\xi_1} (\nu) }{ T } \bigg)\, - \,  \Dp{2}\msc{L}\bigg( \nu, \f{t \ga_{\xi_2} (\nu) }{ T } \bigg)  \Bigg\} \, ,
\end{multline}
which allows one to express the difference involving the $\msc{L}$'s by a first order Taylor integral formula 
\beq
\msc{L}(\nu,x) \; = \; \msc{L}(\nu,0) \, + \,  x  \Int{0}{1}   \Dp{2} \msc{L}(\nu,tx) \dd t \;.
\enq
By proceeding analogously relatively to the second term present in \eqref{ecriture fonction G gamma} one obtains
\bem
 G[ \ga_{\xi_{1}} ](\nu,t) \, - \,  G[ \ga_{\xi_{2}} ](\nu,t) \; = \; 
 \f{1}{2} \Dp{ \nu}\Big\{\big( \xi_1-\xi_2\big)(\nu) \big( \xi_1+\xi_2-2\varpi_N\big)(\nu) \Big\}\cdot \Dp{2} \msc{L}\bigg( \nu, \f{t \ga_{\xi_1} (\nu) }{ T } \bigg)  \\
\, + \, \f{t}{ 2 T} \big(\xi_1-\xi_2\big)(\nu) \Dp{\nu}\Big[ \ga_{\xi_{2}}^2(\nu)\Big] \cdot \Int{0}{1}  \Dp{2}^2\msc{L}\bigg( \nu, \f{t \ga_{\xi_2} ( \nu ) }{ T } +  \f{t x  }{ T } \big[ \ga_{\xi_1} (\nu)-\ga_{\xi_2} (\nu) \big] \bigg) \dd x \\
\, + \, (1-t) \mf{A}_{\infty}^{\prime}(\nu) \Bigg\{  \big( \xi_1-\xi_2\big)(\nu) \big( \xi_1+\xi_2-2\varpi_N\big)(\nu) \cdot \Dp{2}^2 \msc{L}\bigg( \nu, \f{t \ga_{\xi_1} (\nu) }{ T } \bigg)  \\
\, + \, \f{t}{   T} \big(\xi_1-\xi_2\big)(\nu)   \,  \ga_{\xi_{2}}^2(\nu)  \cdot \Int{0}{1}  \Dp{2}^3\msc{L}\bigg( \nu, \f{t \ga_{\xi_2} ( \nu ) }{ T } +  \f{t x  }{ T } \big[ \ga_{\xi_1} (\nu)-\ga_{\xi_2} (\nu) \big] \bigg) \dd x \Bigg\}  \;. 
\end{multline}
This representation readily allows one to infer that there exists a constant $C>0$ such that 
\beq
\norm{ \op{O}_{T,N}[\xi_1] \, - \, \op{O}_{T,N}[\xi_2]   }_{ L^{\infty}(\mc{S}_{ \tf{\zeta_{\e{m}}  }{2} }) } \, \leq  \, 
\f{C}{T} \norm{ \xi_1-\xi_2  }_{  L^{\infty} (\Dp{}\mc{D}_{0,  \eps} ) }
%
%
%
%
%
%
%
\enq
from where the strict contractivity of $\op{O}_{T,N}$ follows provided that $T$ is large enough. Thus,  by the Banach fixed point theorem, $\op{O}_{T,N}$ admits a unique fixed point in $\mc{B}_{\mf{c}}$. 

Furthermore, by dominated convergence and the previous bounds, one has that, for any $\xi \in \mc{B}_{\mf{c}}$, the map 
$N\mapsto \op{O}_{T,N}[\xi]$ is continuous with a uniform in $N$ strict contractivity constant. This ensures that the 
solution is continuous in $N$ as well. The claims relative to the operator $\op{O}_T$ follow from this continuity. \qed

\section{The free energy and correlation lengths  at high temperatures}
\label{Section caracterisation VP a haute tempe} 
 
 \subsection{Identification of the dominant Eigenvalue}
 
 \begin{theorem}
  \label{Theorem VP QTM a T grand et N generique}
  There exist $T_0>0$, $N_0>0$  such that, uniformly in $N\geq N_0$ and $T \geq T_0$, the quantum transfer matrix admits a non-degenerate dominant Eigenvalue $\wh{\La}_{\e{max}}$. 
 There exist $\eps>0$  such that $\wh{\La}_{\e{max}}$ admits the integral representation 
\beq
\wh{\La}_{\e{max}} \, = \,   \bigg( \f{ \sinh(\tf{\aleph}{N}+\i\zeta) }{ \sinh(\i\zeta) }  \bigg)^{2N}
\exp \Bigg\{  \f{h}{2T}      \, - \,  \Oint{ \Dp{}\mc{D}_{0,\eps}  }{ }  \f{ \sin(\zeta) \cdot \mc{L}\mathrm{n} \big[ 1+  \ex{ \wh{\mf{A}}_{\e{max}} } \, \big](u) }{ \sinh(u-\i \zeta)\,  \sinh(u) } \cdot \f{ \dd u }{ 2\pi }  \Bigg\} 
\enq
which involves the unique solution to the non-linear integral equation on $\mc{B}_{c}$:
\beq
\wh{\mf{A}}_{\e{max}}(\xi) \, = \,  -\f{h}{T} \, +\, \mf{w}_N(\xi)   
\; + \; \Oint{ \Dp{} \mc{D}_{0,\eps}  }{} K(\xi-u) \cdot \mc{L}\mathrm{n} \Big[ 1+  \ex{ \wh{\mf{A}}_{\e{max}} } \, \Big](u)  \cdot \dd u   \;. 
\enq

 \end{theorem}

 \Proof 
 Let $\wh{\xi}_{\e{max}}$ be the unique fixed point of the operator $\op{O}_{T,N}$ in \eqref{definition operateur contractant}  associated with the choices $s=0$, $\mc{Y}=\emptyset$  and $\mf{X}=\emptyset$. 
 Since $\mf{A}_{\infty}=0$, and since $\varpi_N$ is bounded for $T$ large enough, uniformly in $N$ on $\Dp{}\mc{D}_{0,\eps}$, it follows that the 
 associated function $\wh{\mf{A}}_{\e{max}}$ \eqref{definition lien hat A et pt fixe} solving the non-linear integral equation \eqref{ecriture eqn NLI forme primordiale} is such that 
\beq
 \wh{\mf{A}}_{\e{max}} \; = \; \e{O}\Big( T^{-1} \Big) \qquad \e{uniformly}\; \e{on} \quad  \Dp{}\mc{D}_{0,\eps} \;. 
\enq
Furthermore, it holds that 
\beq
 \exp\Big\{ \wh{\mf{A}}_{\e{max}}(\la) \Big\} \, = \,  \bigg(  \f{  \la - \tf{ \aleph }{N}   }{   \la +   \tf{ \aleph }{N}   } \bigg)^N \cdot   \ex{  \f{1}{T} \wh{\mf{A}}_{\e{eff}}(\la) }  
\enq
where, upon defining $\e{sinhc}(\la) = \tf{ \sinh(\la) }{ \la }$, we have set 
\beq
 \wh{\mf{A}}_{\e{eff}}(\la) \, = \,  \wh{\xi}_{\e{max}}(\la)  \; + \; N  T  \ln \bigg( \f{  \sinh(\la   +  \tf{ \aleph }{N}-\i \zeta) }{ \sinh(\la - \tf{ \aleph }{N}-\i \zeta) }  \bigg)
\; + \; N  T  \ln \bigg( \f{  \e{sinhc}(\la - \tf{ \aleph }{N}) }{ \e{sinhc}(\la + \tf{ \aleph }{N} ) }  \bigg)  -h \;. 
\label{reecriture hat A eff}
\enq
It is easy to see that $\norm{ \wh{\mf{A}}_{\e{eff}} }_{ L^{\infty}(\mc{D}_{0,\eps})}<C$ uniformly in $N$ and $T$ large enough.  
This representation shows explicitly that $\ex{  \wh{\mf{A}}_{\e{max}}  }$ admits an $N^{ \rm th }$-order pole at $ - \tf{ \aleph }{N}$ and no other singularities on $\mc{D}_{0,\eps}$. Hence, the zero monodromy condition 
\beq
\Oint{ \Dp{} \mc{D}_{0,\eps} }{} \f{ \dd \la }{2\i \pi }  \f{  \wh{\mf{A}}_{\e{max}}^{\prime}(\la)  }{1 \, + \, \ex{ -\wh{\mf{A}}_{\e{max}}(\la) } } \; = \; 0
\enq
entails that $1+\ex{ \wh{\mf{A}}_{\e{max}} }$ admits  $N$ zeroes, counted with multiplicities, inside of $\mc{D}_{0,\eps}$. 
Let $z$ be any such zero. The explicit form for $\wh{\mf{A}}_{\e{eff}}$ given in \eqref{reecriture hat A eff}, leads, upon taking the $N^{\e{th}}$ root, to
\beq
z-\f{ \aleph }{N} \; = \; \Big( z+\f{ \aleph }{N} \Big) \cdot \ex{\i \psi_{N} } \quad \e{with} \quad    \psi_{N} \, = \,  \f{ 1 }{ N } \Big\{  \big( 2p-1  \big) \pi + \f{\i }{  T } \wh{\mf{A}}_{\e{eff}}(z) \Big\} 
\enq
for some $ p \in  Z_N$, where $Z_N = \intn{-N/2 + 1}{N/2}$ if $N$ is even, $Z_N = \intn{- (N-1)/2}{(N-1)/2}$ if $N$ is odd. One then readily concludes that 
\beq
z-\f{ \aleph }{N} \; = \; \f{2 \aleph \cdot \ex{\i \psi_{N} } }{ N \cdot \big[ 1 - \ex{\i \psi_{N} } \big] } \qquad \e{and} \qquad z+\f{\aleph }{N} \; = \; \f{2  \aleph }{ N \cdot \big[ 1 - \ex{\i \psi_{N} } \big] }  \;. 
\enq
The root $z$ has multiplicity higher or equal to 2 if and only if 
\beq
\Big( 1 + \ex{ \wh{\mf{A}}_{\e{max}}  } \Big)^{\prime}(z) \; = \; -\wh{\mf{A}}_{\e{max}}^{\prime}(z) \; = \; 0\;.  
\enq
The above allows one to compute $\wh{\mf{A}}_{\e{max}}^{\prime}(z)$ explicitly in the form 
\beqa
\wh{\mf{A}}_{\e{max}}^{\prime}(z) & = &  2 \f{N^2 }{ \aleph }  \cdot \sin^2\Big( \tfrac{\psi_N}{2}\Big)  + \f{1}{T} \wh{\mf{A}}_{\e{eff}}^{\prime}(z) \\
&=&   \f{\pi^2 ( 2p - 1)^2 }{ 2 \aleph  } \bigg\{ 1 + \i \f{ \wh{\mf{A}}_{\e{eff}}(z)  }{  \pi T (2p - 1)  }  \bigg\}^2 \cdot \bigg( \f{ \sin\big[ \tf{\psi_N}{2} \big] }{ \tf{\psi_N}{2} } \bigg)^2  
\; + \; \f{1}{T} \wh{\mf{A}}_{\e{eff}}^{\prime}(z) \;. 
\eeqa
Pick $\tf{1}{2}>c>0$ small enough, then uniformly in $N,T$ large enough and for any $p\in Z_N $
\beq
\Re( \psi_N )\in   \intff{-\pi(1+c) }{\pi(1+c)} \quad \e{and} \quad  \Im( \psi_N ) \in \intff{ -\tfrac{c}{T} }{ \tfrac{c}{T} } 
\quad viz. \quad 
 \bigg| \f{ \sin\big[ \tf{\psi_N}{2} \big] }{ \tf{\psi_N}{2} } \bigg| >C^{\prime} 
\enq
for some $N$ and $T$ independent constant $C^{\prime}>0$. This is enough  to conclude that $\wh{\mf{A}}_{\e{max}}^{\prime}(z) \ne 0$ and 
that all the zeroes of $1 + \ex{ \wh{\mf{A}}_{\e{max}}  }$ on $\mc{D}_{0,\eps}$ are necessarily simple. Let $ \la_1^{(\e{max})}, \dots, \la_N^{(\e{max})} $ denote these distinct zeroes. 

By tracing backwards the steps that led to the construction of the non-linear integral equation, \textit{viz}. taking the non-linear terms by means of residues, one obtains
\beq
\ex{ \wh{\mf{A}}_{\e{max}}( \xi)  } \; = \;  \ex{-\tfrac{h}{T}}   \pl{k=1}{N} \bigg\{ \f{\sinh\big( \i\zeta - \xi + \la_k^{(\e{max})} \big) }{  \sinh\big( \i \zeta  + \xi - \la_k^{(\e{max})} \big) }  \bigg\}
\cdot  \bigg\{ \f{ \sinh(  \xi - \tf{ \aleph }{N} ) \sinh( \i\zeta +  \xi +\tf{ \aleph }{N} ) }{  \sinh(  \xi + \tf{ \aleph }{N} ) \sinh( \i\zeta - \xi + \tf{ \aleph }{N} )  }    \bigg\}^{N} \;. 
\enq
Thus, it exactly coincides with the functional form of   the  auxiliary function   introduced in \eqref{definition fct auxiliaire a}. 
Since the zeroes $ \la_1^{(\e{max})}, \dots, \la_N^{(\e{max})}$ are pairwise distinct, the vector $\Psi\Big( \big\{ \la_a^{(\e{max})} \big\}_{a=1}^{N} \Big)$
introduced in \eqref{ecriture vecteur de Bethe} does indeed produce an Eigenvector of $\op{t}_{\mf{q}}$. The associated Eigenvalue \eqref{ecriture valeur propre qtm} 
then has the integral representation 
\beq
\tau\big( 0 \mid \{\la_a\}_{a=1}^{M} \big) \, = \, 
\bigg( \f{ \sinh(\tf{ \aleph }{N}+\i\zeta) }{ \sinh(\i\zeta) }  \bigg)^{2N}
\exp \Bigg\{  \f{h}{2T}      \, - \,  \Oint{ \Dp{}\mc{D}_{0,\eps}  }{ }  \f{ \sin(\zeta) \,  \mc{L}\mathrm{n} \big[ 1+  \ex{ \wh{\mf{A}}_{\e{max}} } \, \big](u) }{ \sinh(u-\i \zeta)\,  \sinh(u) } \cdot \f{ \dd u }{ 2\pi }  \Bigg\} \;. 
\label{ecriture forme vp QTM b}
\enq
Since the result does not depend on the choice of $\kappa$, one may as well set $\kappa=\eps$, which allows one to get rid of the last term. 
It is then readily seen that the integral term appearing in the Eigenvalue of $\op{t}_{\mf{q}}$ associated with this solution exhibits the large-$T$ behaviour
\beq
 \Oint{ \Dp{}\mc{D}_{0,\eps}  }{ }  \f{ \sin(\zeta) \,  \mc{L}\mathrm{n} \big[ 1+  \ex{ \wh{\mf{A}}_{\e{max}} } \, \big](u) }{ \sinh(u-\i \zeta)\,  \sinh(u) } \cdot \f{ \dd u }{ 2\pi }   \; = \; -\ln 2 \; + \; \e{O}\big( T^{-1} \big) \;. 
\enq
Thence, \eqref{ecriture forme vp QTM b} entails that the associated Eigenvalue of the quantum transfer matrix takes the form 
\beq
\tau\big( 0 \mid \{\la_a^{(\e{max})}\}_{a=1}^{N} \big) \, = \,  2 \; + \; \e{O}\big( T^{-1} \big)\;. 
\enq
%
%
 Thus, by virtue of Proposition \ref{prop vp dominante et vp sous dominantes},   $\tau\big( 0 \mid \{\la_a^{(\e{max})}\}_{a=1}^{N} \big)$  does coincide with the dominant Eigenvalue $\wh{\La}_{\e{max}}$. \qed

 \vspace{3mm}
 
 All is now set in place so as to allow for the proof of Theorem \ref{Theorem principal aticle f a T fini} stated in the introduction

\vspace{2mm}

 \Proof 
 
 It is a corollary of Proposition \ref{prop vp dominante et vp sous dominantes}, Theorem \ref{Theorem echange limite thermo et Trotter}, the classical result for the existence of the limit \eqref{definition energie libre}
 and the one for the limit \eqref{ecriture explicite limite Trotter} that 
\beq
-\f{f}{T} \; = \; \lim_{N\tend + \infty} \ln \wh{\La}_{\e{max}} \;. 
\enq
The existence of the Trotter limit of the solution $\wh{\mf{A}}_{\e{max}}$ to the non-linear integral equation, and the convergence, uniformly on $\Dp{}\mc{D}_{0,\eps}$ 
of $\wh{\mf{A}}_{\e{max}}$ to the solution $\mf{A}_{\e{max}}$ adjoined to the integral representation for $\wh{\La}_{\e{max}}$ provided in Theorem \ref{Theorem VP QTM a T grand et N generique}
then allows one to conclude. \qed

\subsection{Sub-dominant Eigenvalues at high temperatures}

In this section we study the solvability of the non-linear integral equation problem associated with the infinite Trotter number limit, namely the 
existence of solutions $\mf{A}$ to the non-linear integral equation \eqref{ecriture NLIE Trotter infini} subject to additional constraints \eqref{ecriture condition subsidiaires solution NLIE spectre} on the monodromy of the solution and 
on the parameters building up the particle-hole sets. We focus only on the characterisation of a subset of solutions, namely those whose sets $\mf{X}$ and $\mc{Y}$
belong to the class $\mc{C}^{\eps}_{\a,\varrho}$ with fixed cardinalities $n_x,n_y$ and parameters $\eps, \a, \varrho$ small enough but finite. 
While it is not a problem to carry out the same analysis at finite Trotter number, such an analysis would be of limited use in that, from the point of view of physical applications,  
one is interested, in the end, in results at infinite Trotter number.

Recall that, by virtue of Proposition \ref{Proposition correspondance solutions NLIE originelle et NLIE type point fixe} and Theorem \ref{Theorem existence et unicite point fixe de OTN},
given $\a, \varrho>0$, $n_x, n_y \in \mathbb{N}$ fixed, there exist $T_0>0$ and $\eps>0$ such that 
any solution to the non-linear integral equation \eqref{ecriture NLIE Trotter infini} with $\mf{X}$, $\mc{Y}$ belonging to the class $\mc{C}^{\eps}_{\a,\varrho}$ with cardinalities $n_x,n_y$ 
takes the form 
\beq
\mf{A}=\mf{A}_{\infty} \, + \, \tfrac{1}{ T} \mf{b} \qquad \e{with} \qquad \mf{b}=\xi - e_0 \;. 
\label{ecriture forme explicite fct A}
\enq
Here $\xi$ is the unique fixed point of the operator $\op{O}_{T}$ \eqref{ecriture rep int operateur limite OT} associated with the roots $\mf{X}$, $\mc{Y}$ and
the function $\mf{b}$ is bounded on $\Dp{}\mc{D}_{0,\eps}$. Below we study the solvability, in the large-$T$ regime, 
of the additional constraints \eqref{ecriture condition subsidiaires solution NLIE spectre} assuming that $\mf{X}$, $\mc{Y}$ belong to the class $\mc{C}^{\eps}_{\a,\varrho}$ with cardinalities $n_x,n_y$.   

\vspace{3mm}

Let $\mc{Y}= \big\{ y_1,\dots,  y_{n_y} \big\}$ and $\mc{Y}^{\prime}= \big\{  y_1^{\prime},\dots,  y_{n_y}^{\prime}  \big\}$ be two sets of equal cardinality. 
One defines the distance between the sets $\mc{Y}$ and $\mc{Y}^{\prime}$ as
\beq
\e{d}\big(\mc{Y},\mc{Y}^{\prime} \big) \, = \, \underset{ \sg \in \mf{S}_{n_y} }{\e{min}} \Big\{ \sul{a=1}{n_y} |y_a-y^{\prime}_{\sg(a)} | \Big\}\;. 
\enq
Further, define the collection of solution sets to the Bethe Ansatz equations connected with the ordinary transfer matrix of the spin-$1$ XXZ chain:
\beq
\sg_{\infty} \; = \; \Bigg\{ \mc{Y}_{\infty}=\{ y_{\infty;1}, \dots, y_{\infty;n_y} \} \, : \, \forall a \in \intn{1}{n_y}  \; , 
\quad (-1)^{s+1} \lim_{u\tend y_{\infty;a} } \pl{ b=1}{n_y} \bigg\{ \f{  \sinh(\i\zeta + y_{\infty;b} - u) }{ \sinh(\i\zeta +u - y_{\infty;b} ) } \bigg\} 
\cdot  \bigg( \f{  \sinh(\i\zeta + u ) }{ \sinh(\i\zeta - u ) } \bigg)^{n_x}  =1 \Bigg\} \;. 
\enq
Note that this way of defining the solution set allows, in principle, for roots located at $\infty$. Indeed, the limit of some roots going to $\infty$ is well-defined within the 
prescription used for defining the full solution set $\sg_{\infty}$.

One defines the distance of a set $\mc{Y}$ of cardinality $n_y$ to $\sg_{\infty}$ as
\beq
\e{d}\big(\sg_{\infty}, \mc{Y} \big) \, = \,  \underset{ \mc{Y}_{\infty} \in \sg_{\infty} }{\e{min}} \Big\{ \e{d}\big(\mc{Y}_{\infty}, \mc{Y} \big)  \Big\} \; .
\enq

\begin{prop}
 \label{Proposition forme solution NLIE avec part trous a gde T}
Let  $\a, \varrho>0$, $n_x, n_y \in \mathbb{N}$ be fixed. Let $\mf{X}$ and $\mc{Y}$ belong to the class $\mc{C}^{\eps}_{\a,\varrho}$ with cardinalities $n_x,n_y$ 
and let $T_0>0$ and $\eps>0$ be such that the non-linear integral equation \eqref{ecriture NLIE Trotter infini} admits a unique solution. 

If $\mf{X}=\{x_1,\dots, x_{n_x}\}$ and $\mc{Y}=\{y_1,\dots, y_{n_y} \}$ are $T$-dependent sets satisfying the subsidiary conditions \eqref{ecriture condition subsidiaires solution NLIE spectre} uniformly in $T$
large enough, then necessarily 
\begin{itemize}
 
 \item $|\mf{X}|=|\mc{Y}|+s$;
 
 \item there exist integers  $k_a \in \mathbb{Z}$ such that the "hole roots" take the form 
\beq
x_a\; = \; \f{-2 J \sin(\zeta) }{  T \Big[  (2k_a + 1 +s)\pi \, - \,   \sul{y \in \mc{Y} }{} \th(-y) \Big]  }  \; + \; \e{O}\big( T^{-2} \big)  \quad a=1,\dots, n_x \; ;
\enq
\item $\e{d}\big(\sg_{\infty}, \mc{Y} \big)=\e{o}(1)$, where the control on the remainder only depends on $\eps, \varrho, \a, n_x$ and $n_y$. 
\end{itemize}

\end{prop}

Note that the integers $k_a$ appearing in the statement of the proposition may, in principle, dependent on $T$.

\Proof 

The expansion \eqref{ecriture forme explicite fct A} ensures that 
\beq
\f{ \mf{A}^{\prime} (\la) }{ 1 + \ex{-\mf{A}(\la) }  } \; = \; \f{ \mf{A}^{\prime}_{\infty} (\la) }{   1 + \ex{-\mf{A}_{\infty}(\la) }  }
\; + \; \f{1}{T} \Dp{\la} \bigg\{  \f{ \mf{b}(\la)  }{  1 + \ex{-\mf{A}_{\infty}(\la)}  } \bigg\}
\; + \; \e{O}\Big( \f{1}{T^2}\Big) \;, 
\enq
with the $\e{O}\big(T^{-2} \big)$ remainder  being uniform on $\Dp{}\mc{D}_{0,\eps}$. 

Upon inserting the above expansion into the monodromy constraint \eqref{ecriture condition subsidiaires solution NLIE spectre}, one obtains that 
\beq
\Oint{ \Dp{}\mc{D}_{0,\eps} }{} \f{ \mf{A}^{\prime}_{\infty} (\xi) }{   1 + \ex{-\mf{A}_{\infty}(\xi) }  }\cdot \f{\dd u}{2\i \pi}  \; + \;\e{O}\Big( \f{1}{T}\Big)  \; = \;|\mf{X}|  - |\mc{Y}|  - s  \;, 
\label{ecriture conditions auxiliaires monodromie et emplacement racines}
\enq
since the sets $\mf{X}$, $\mc{Y}$, belonging to  the class $\mc{C}^{\eps}_{\a,\varrho}$ with $\eps>0$ small enough, do not give rise to singular roots.
However, due to \eqref{ecriture lower bound exp A infty}, the first integral vanishes. Since the \textit{lhs} is integer valued, and  since the sole dependence on $\mf{X}$ , $\mc{Y}$
of the remainder $\e{O}\big( T^{-1} \big)$ is bounded by $|\mf{X}|  + |\mc{Y}|$, one concludes that, for any $T> T_0^{\prime}$ for some $T_0^{\prime}$ large enough, 
in case of sets $\mf{X}$, $\mc{Y}$ belonging to the class  $\mc{C}^{\eps}_{\a,\varrho}$, one may produce solutions to the joint problem \eqref{ecriture condition subsidiaires solution NLIE spectre} 
if and only if the sets' cardinalities are constrained as 
\beq
 |\mf{X}|  = |\mc{Y}|  + s \;. 
\enq

We now analyse more precisely the structure of the sets $\mf{X}$ and $\mc{Y}$ satisfying to the subsidiary constraints in \eqref{ecriture condition subsidiaires solution NLIE spectre}. 

\subsubsection*{\boldmath $\bullet$ The hole set $\mf{X}$}

One  starts by determining elements forming the hole set $\mf{X}$. By virtue of \eqref{ecriture forme explicite fct A}, $\mf{b}$ is meromorphic on $\mc{D}_{0,\eps}$ with a single pole at $\xi=0$ and such that 
\beq
\e{Res}\big(\mf{b}(\xi)\dd \xi, \xi=0 \big) \, = \,   -  2\i J   \sin(\zeta) \;. 
\enq
As a consequence, there exits a holomorphic function $\mf{b}^{(r)}$ on $\mc{D}_{0,\eps}$ that is bounded uniformly in $T$ large enough and such that 
\beq
\mf{b}(\xi) \, = \,  - \f{ 2\i J }{   \xi  } \sin(\zeta) \, + \,     \mf{b}^{(r)}(\xi) \;. 
\label{ecriture locale voisinage trous de b}
\enq
Any hole root $x$ will satisfy  the equation 
\beq
\mf{A}_{\infty}(x)\,+\,\f{ \mf{b}(x) }{ T } \, = \, (2k_x + 1 )\i\pi \qquad \e{for} \, \e{some} \; k_x \in \mathbb{Z}\;. 
\label{ecriture eqn definissant trous a haute temperature}
\enq
By virtue of \eqref{ecriture lower bound exp A infty}, it holds that $\e{d}\Big( \mf{A}_{\infty}(x), \i \pi + 2\i\pi \mathbb{Z}\Big)>C_{\varrho}$
for some $C_{\varrho}>0$ depending on the parameter $\varrho>0$. Hence, any solution to \eqref{ecriture eqn definissant trous a haute temperature}
has to be such that $| \tf{ \mf{b}(x) }{ T } | >C_{\varrho}$. The form of the behaviour around the origin \eqref{ecriture locale voisinage trous de b}
then entails that $T x$ has to be uniformly bounded. Thus, one reparametrises $x=\tf{u_x}{T}$, with $u_{x}$ bounded in $T$, which leads to the equation
\beq
 (2k_x + 1 )\i\pi \, = \,  \mf{A}_{\infty}(0) - \f{2\i J}{u_x} \sin(\zeta) \; + \; \e{O}\big( T^{-1} \big) \;. 
\enq
Thus
\beq
u_x\,=\, \f{-2\i J \sin(\zeta) }{   (2k_x + 1 )\i\pi \, - \,    \mf{A}_{\infty}(0)  }  \; + \; \e{O}\big( T^{-1} \big) \;. 
\enq
This entails that all hole roots converge to $0$ with speed at least $\e{O}\big(T^{-1}\big)$. Note that this convergence may be faster than $T^{-1}$
in case of integers $k_x$ also going to infinity with $T$.

\subsubsection*{\boldmath $\bullet$ The particle set $\mc{Y}$}

The very definition of the class $\mc{C}^{\eps}_{\a,\varrho}$ entails that particle roots $y \in \mc{Y}$ are such that $y\pm \i\zeta \not\in \mc{D}_{0,\eps}$. Then, one has
\beq
\ex{\mf{A}(\la)}\,=\, \ex{\mf{A}_{\infty}(\la) } \cdot \ex{ \tfrac{\mf{b}(\la)}{T} } \qquad \e{for} \quad 
\la \in \mc{S}_{\frac{\pi}{2}} \setminus \bigg\{ \mc{D}_{0,\eps} \cup_{\ups = \pm }^{} \mc{D}_{\ups \i \zeta_{\e{m}}, \a } \bigg\}
\enq
and thus any root $y\in \mc{Y}$ satisfies 
\beq
(-1)^{s+1} \cdot \lim_{u\tend y}  \Bigg\{ \pl{ y^{\prime} \in \mc{Y}\ominus \mf{X} }{} \f{  \sinh(\i\zeta + y^{\prime} - u) }{ \sinh(\i\zeta +u - y^{\prime} ) }  \Bigg\} \cdot \ex{ \e{O}(T^{-1}) } \; = \; 1 \;. 
\enq
Note that the limit procedure provides one with a prescription for treating roots at $\infty$ and also for regularising potential zeroes and poles which cancel eventually out between the numerator and the denominator. 
 It follows from the above that $x=\e{O}(T^{-1})$ for any $x \in \mf{X}$. 
Thus, the properties of the class $\mc{C}^{\eps}_{\a,\varrho}$ allow one to recast the above equation in the form 
\beq
(-1)^{s+1} \cdot \lim_{u\tend y}  \Bigg\{ \pl{ y^{\prime} \in \mc{Y}  }{} \f{  \sinh(\i\zeta + y^{\prime} - u) }{ \sinh(\i\zeta +u - y^{\prime} ) }  \Bigg\} \cdot
 \bigg\{   \f{ \sinh(\i\zeta + y  ) }{  \sinh(\i\zeta  - y) }  \bigg\}^{n_x}    \cdot \ex{ \e{O}(T^{-1}) } \; = \; 1 \;. 
\label{ecriture equation effective racines Y}
\enq
Now assume that $\mc{Y}$ is a one parameter $T$ family of particle sets solving the constraints \eqref{ecriture condition subsidiaires solution NLIE spectre} and such that $\e{d}(\mc{Y},\sg_{\infty})\not\tend 0$ as $T\tend \infty$. Thus, one may extract a sequence of sets $\mc{Y}_n$ associated with a sequence of temperatures $T_n\tend +\infty$ such that 
$\e{d}(\mc{Y}_n,\sg_{\infty})>\ga $ for some $\ga>0$. By taking the $n\tend +\infty$ on the level of the associated equation \eqref{ecriture equation effective racines Y}, one gets that the limiting set 
$\mc{Y}_{\infty}\in \sg_{\infty}$, which is a contradiction to $\e{d}(\mc{Y}_n,\sg_{\infty})>\ga $. This entails the claim. \qed

\vspace{2mm}

Basically, Proposition \ref{Proposition forme solution NLIE avec part trous a gde T} states that, for temperatures large enough, for the solutions to the non-linear integral equations describing the Eigenvalues of the quantum transfer matrix
with the hole and particle sets $\mf{X}$, $\mc{Y}$ belonging to a class $\mc{C}^{\eps}_{\a,\varrho}$ with cardinalities $n_x,n_y$, any element of the hole set collapses to $0$ with speed $T^{-1}$, and the leading behaviour of this collapse 
is parametrised by a collection of integers. In their turn, the elements of the particle set $\mc{Y}$ essentially collapse onto the zero set, 
on $\mc{S}_{\frac{\pi}{2}} \setminus \Big\{ \mc{D}_{0,\eps} \cup_{\ups = \pm} \mc{D}_{\i \ups \zeta_{\e{m}}, \a  } \Big\}$, of the map  
\beq
\bs{f} \; : \; \left\{ \ba{ccc}  \Cx^{ |\mc{Y}| } & \tend & \Cx^{ |\mc{Y}| }   \\
		  \bs{y}   & \mapsto & \bs{f}(\bs{y}) \ea  \right.  \qquad 
\bs{f}(\bs{y}) \, = \, \big(\bs{f}_1(\bs{y}), \cdots, \bs{f}_{ |\mc{Y}| }(\bs{y})  \big) \;, 
\enq
with 
\beq
\bs{f}_b(\bs{y}) \; = \; 1+ (-1)^{s} \cdot  \pl{ a=1  }{ |\mc{Y}|} \f{  \sinh(\i\zeta + y_a - y_{b}) }{ \sinh(\i\zeta +y_{b} - y_{a} ) }  
\cdot \Bigg( \f{ \sinh(\i\zeta +y_{b}   ) }{  \sinh(\i\zeta  - y_{b}) } \Bigg)^{ |\mc{Y} | + s } \;. 
\enq
So as to be more precise relatively to the speed of the convergence of the set $\mc{Y}$, one would need to have more information on the behaviour of $\bs{f}$ in the vicinity of its zeroes. 
In particular, if $\bs{y}_{\infty} = \big( y_{\infty;1},\cdots, y_{\infty;|\mc{Y}|} \big)$ is a zero of $\bs{f}$ and the differential of $\bs{f}$ at $\bs{y}_{\infty}$ is invertible, 
then if $\bs{y} = \big( y_1, \dots, y_{|\mc{Y}|} \big)$ is the vector built up from the elements of the particle set $\mc{Y}$, and if it holds that $\bs{y}=\bs{y}_{\infty} + \e{o}(1)$, where the remainder is 
to be understood coordinate-wise
when $T\tend +\infty$, then it is easy to show using the implicit function theorem, that, in fact, 
\beq
y_a \, = \, y_{\infty;|\mc{Y}|}  \, + \, \e{O}\Big( T^{-1} \Big) \;. 
\enq

 Thus, our analysis shows that, in the high-temperature regime, at least part of the subdominant Eigenvalues of the quantum transfer matrix, and hence the associated correlation lengths, are parameterised  by solutions to the Bethe Ansatz equations
  of the spin-$1$ XXZ spin chain.

\section{Numerical illustration}
\label{Section analyse numerique}

In this section, we illustrate the conclusions of the analysis carried out in the previous Sections \ref{Section existance et unicite sols NLIE}, \ref{Section caracterisation VP a haute tempe}
by providing a numerical analysis of the solutions to the Bethe Ansatz equations for finite Trotter numbers. The fact that the results are obtained at finite and relatively small Trotter numbers 
is not a problem when $T$ is high enough, in that the convergence of the infinite Trotter number limit results is controlled as $\e{O}([TN]^{-1})$. Our analysis deals with the Trotter numbers $N= 4, 5, 6$, and
we employed an algorithm proposed in \cite{AlbertiniDasmahapatraMcCoySpectrumAndCompleteness3StatePottsComputerStudy}.
The latter combines a numerical diagonalisation of the quantum transfer matrix and the use of an expansion of Baxter's TQ relation.
The algorithm enables us, for generic $\zeta$, to locate the Bethe roots associated with {\it any} Eigenstate and then 
to find the zeros of the Eigenvalues of the quantum transfer matrix associated with the given Eigenstate easily. This information provides the data for  $\wh{\mf{X}}$ and $\wh{\mc{Y}}$.

\subsection{Examples of of Bethe and hole roots}
We first present a few examples of the  Bethe roots and hole roots $\wh{\mf{X}}$.

For the first example,  we have determined the solutions to \eqref{ecriture eqns Bethe Trotter fini} with the choice of parameters $N=5, M=5, T=100, \zeta=\frac{\pi}{7}$. 
Here, for the sake of brevity, we only list three solutions. 
The Eigenvalues $\wh{\La}_{a} $, associated Bethe roots and set $\wh{\mf{X}}$ for the 2nd,  the 12-th  and 83-th subdominant Eigenvalues of $\op{t}_{\mf{q}}(0)$
are gathered in Table \ref{tab:N10M5T100}.  Note that the Eigenvalues are labelled in respect to the non-increasing order\symbolfootnote[2]{In principle, degeneracies of $|\wh{\Lambda}_a|$ may occur and then 
the choice of the ordering for the various Eigenvalues with fixed modulus is taken in the direction of increasing arguments, with $\e{arg} \in \intfo{-\pi}{\pi}$.} of  $|\wh{\Lambda}_a|$.

\begin{table}[!h]
\begin{center}
\begin{tabular}{|l|c|c|c|c|c|}
\hline     &&&\\
    &                  $\wh{\Lambda}_a $&      
      $ \Big\{ \frac{\lambda_a}{\zeta} \Big\}$ &      
       $ \Big\{ \frac{\wh{x}_a}{\zeta} \Big\} $     \\
       &&&\\
\hline
$\text{ 2nd}$&       $ -9.19523\times 10^{-3}$ &       
\begin{tabular}{c}
 $ \{\underline{\frac{\pi}{2\zeta} \i},   \pm 1.329782\times 10^{-3},$\\
 $ \pm 3.14044 \times 10^{-4} $    \}       
 \end{tabular}&
  $\{\pm6.4790 \times 10^{-2}\}$                     \\
\hline
$\text{ 12-th}$&       
 \begin{tabular}{c}
$ -4.07262\times 10^{-4}+$\\
 $ \phantom{abc} 9.0811\times 10^{-5} \i$ 
 \end{tabular}&      
\begin{tabular}{c}
  $\{ 2.18463 \times 10^{-7},     7.02268 \times 10^{-4}, $\\
  $2.98366 \times 10^{-3}$,\\
  $\underline{-1.14080 \times 10^{-3} \pm 0.575835 \i} \}$
  \end{tabular}&        
   \begin{tabular}{c}  
    $\{-2.97906  \times 10^{-3},$        \\
    $ 7.02935 \times 10^{-4} \}$ 
    \end{tabular} 
                 \\
\hline
$\text{ 83-th}$&       
$ 2.01835\times 10^{-6}$&      
  $\{ 0, \pm 3.140243 \times 10^{-4}, \underline{\pm \i} \} $&        

   \begin{tabular}{c}  
    $\{\pm 1.32951 \times 10^{-3},$        \\
    $ \pm 5.65988 \times 10^{-2} \}$ 
    \end{tabular} 
                 \\
\hline
\end{tabular}
\caption{Eigenvalues $\wh{\La}_{a} $, associated Bethe roots and sets $\wh{\mf{X}}$ and $\wh{\mc{Y}}$  for $N=5, M=5, T=100, \zeta=\frac{\pi}{7}$. The numbers are given with a 6 digit accuracy.
The underlined roots are the members of the set  $\wh{\mc{Y}}$.}  \label{tab:N10M5T100}
\end{center}
\end{table}

Among these three states, the 12-th state gives rise to sets  $\wh{\mf{X}}$ and $\wh{\mc{Y}}$ belonging to a class
$\mc{C}^{\eps}_{\a, \varrho}$ \footnote{The class $\mc{C}^{\eps}_{\a, \varrho}$ is defined in Definition \ref{Definition C eps alpha rho class}}: the condition \eqref{condition eloignement racines de 0} is violated for any $\varrho>0$
for the 2nd excited state,   while the 83-th excited state contains singular roots\footnote{To be precise,  they are slightly away from $\pm \i$; they are at 
$\pm 1.00000000629\cdots \i$ }, \textit{viz}. $\wh{\mc{Y}}_{\e{sg}} \not= \emptyset$.

The examples for the same parameters but  with $M=4$ are given in Table \ref{tab:N10M4T100}.

\begin{table}[!h] 
\begin{center}
\begin{tabular}{|l|c|c|c|c|c|}
\hline      &&&\\
    &                  $\wh{\Lambda}_a$&      
      $  \Big\{ \frac{\lambda_a}{\zeta} \Big\}$ &      
       $ \Big\{ \frac{\wh{x}_a}{\zeta} \Big\} $     \\
       
        &&&\\
\hline
$\text{ 10-th }$&       $ 2.10835\times 10^{-4}$ &       
 $ \{\underline{\frac{\pi}{2\zeta} \i},  0, 
 \pm 7.019646\times 10^{-4}   \} $ &      
  $\{\pm 2.96294 \times 10^{-3}\}$                     \\
\hline
$\text{ 17-th}$&       
$1.99318\times 10^{-4}$&      
  $\{ \underline{\pm 7.15137 \times 10^{-2}},    \pm3.14057 \times 10^{-4}\}$&        
     $\{\pm 1.32993  \times 10^{-3}\}$ 
                 \\
\hline
$\text{ 123-th}$&       
$-4.54205\times 10^{-8}$&      
  $\{  \underline{\pm \i}, \pm 1.88769 \times 10^{-8} \} $&        

   \begin{tabular}{c}  
    $\{\pm 2.96082 \times 10^{-3},$        \\
    $ \pm 7.01891 \times 10^{-4} \}$ 
    \end{tabular} 
                 \\
\hline
\end{tabular}
\caption{Examples of eigenstates: $N=5, M=4, T=100, \zeta=\frac{\pi}{7}$.}\label{tab:N10M4T100}
\end{center}
\end{table}

In this sector, the 10-th excited state belongs to $\mc{C}^{\eps}_{\a, \varrho}$, 
$\varrho=0$ for the 17-th excited state and the 123-th excited state contains
$\wh{\mc{Y}}_{\e{sg}}$.

\subsection{\boldmath Optimal choice of parameters  $\epsilon, \alpha$ and $\varrho$ }

The $\mc{C}^{\eps}_{\a, \varrho}$ classes provide a convenient subset in the parameter space of particle roots which allows one to 
easily describe any particle parameters, subject to the constraints \eqref{equation quantification particles et tous cas Trotter fini}. 
Ideally, the parameters $\epsilon, \alpha$ and $\varrho$ should be taken as small as possible. 
However, the analysis developed in Section \ref{Section caracterisation VP a haute tempe} is not refined enough so as to be able to handle rigorously the presumably most optimal case
when these parameters go to zero with some power of T.  Also, it is clear that there might be solution sets $\wh{\mf{X}}$, $\wh{\mc{Y}}$ to 
the constraints \eqref{equation quantification particles et tous cas Trotter fini} such that the parameters $\wh{\mc{Y}}$ do not belong to any class $\mc{C}^{\eps}_{\a, \varrho}$. 
We have investigated these cases  numerically.

 We set 
\begin{equation}
\delta_a:=
(-1)^{n_x-n_y + 1}  
\prod_{ b=1}^{n_y} \bigg\{ \frac{ \sinh( \i \zeta  + y_{b} - y_{a}) }{ \sinh( \i\zeta  +y_{a} - y_{b} ) } \bigg\} \cdot 
 \bigg( \frac{\sinh( \i \zeta + y_{a}) }{ \sinh( \i\zeta  - y_{a}) } \bigg)^{n_x}-1
\label{BAE_spin1d}
\end{equation}
and regard $\{y_a\}$ as a set of solutions if  $\max_a |\delta_a|< \delta$ for a fixed small $\delta$.

With this convention we classify all Eigenstates into  five cases:
\begin{enumerate}
\item The states with $\mc{Y} = \emptyset$.
\item The states  containing singular roots $\wh{\mc{Y}}_{\e{sg}}$.
\item The states   which satisfy 
\[
\Big|(-1)^s  \prod_{y \in \wh{\mc{Y}} }  \frac{ \sinh( \i \zeta + y) }{\sinh( \i \zeta -y)} + 1 \Big| \,< \, \varrho \;.
\]
\item  The states  which belong to  $\mc{C}^{\eps}_{\a, \varrho}$  and satisfy  (\ref{BAE_spin1d}) with a fixed $\delta$.
\item The states  which belong to  $\mc{C}^{\eps}_{\a, \varrho}$  but do not satisfy  (\ref{BAE_spin1d}) with a fixed $\delta$.
\end{enumerate}

The first three cases do not belong to  $\mc{C}^{\eps}_{\a, \varrho}$. \\

The classification depends on the actual choice of parameters 
$\epsilon, \alpha, \varrho$ and $\delta$.
We performed the numerical estimation for various  $N, M$ and $T$ and found that a
consistent choice is
\[
\epsilon \propto \frac{1}{\sqrt{T}}, \qquad \varrho \propto \frac{1}{\sqrt{T}}
, \qquad \delta  \propto \frac{1}{T}.
\] 
They are thus expected to be infinitesimally small in the high-temperature limit.
On the other hand, the choice of $\alpha$ seems almost independent of $T$: it does not change the classification
for $ 10^{-4}\le \alpha \le  10^{-2}$ for $100 \le T \le 1000$.  

A rigorous justification of these observations is left as a future problem.

\subsection{Conclusion from numerics }

Taken the above choice for granted,
  we  have verified that each state is classified into either
(A) not a member of $\mc{C}^{\eps}_{\a, \varrho}$ or
(B) a member of $\mc{C}^{\eps}_{\a, \varrho}$and satisfying  (\ref{BAE_spin1d}) with  $\delta$.
We tabulate the number of  states in each case for $N = 6$, $M = 5, 6$ and $\zeta=\pi/7$ in Table~\ref{tab:classification}.

\begin{table}[!h]
\begin{center}
\begin{tabular}{|l|r|r|r|r|r|r|}
\hline     
    &                                        case 1       &         case 2 &     case 3 &   case 4  \\
\hline
$N=6, M=5, T=100$&                   11&                          26&               178 &                             577\\
$N=6, M=5, T=1000$&                 11&                           26&               178&                        577 \\
$N=6, M=6, T=100$&                    1&                           53&               211&                            659 \\
$N=6, M=6, T=1000$&                  1&                           75&              189&                                  659 \\
\hline
\end{tabular}
\caption{ The distribution of numbers of states into different cases. Here we set $\alpha=0.01$,    $\epsilon=\varrho= 0.6/\sqrt{T}, \delta=10/T$.} \label{tab:classification}  
\end{center}
\end{table}
One notices that the majority of states belongs to case 4, namely to class  $\mc{C}^{\eps}_{\a, \varrho}$,
which satisfies the higher-level Bethe ansatz.
The distribution of numbers of states in each case seems stable against a change in temperature
although it depends on the sector $M$.
It also depends on the Trotter number $N$.  
Although we have tried it only for $N = 4, 5, 6$, we find that the 
relative number of members in class $\mc{C}^{\eps}_{\a, \varrho}$ increases with $N$.
The percentage of ${\cal C}^{\epsilon}_{\alpha,\varrho}$ in the total number of eigenstates 
for each $N, M$ is given in Table~\ref{tab:case4_percentage}.
It suggests that most of the states will belong to class $\mc{C}^{\eps}_{\a, \varrho}$ 
in the Trotter limit $N \rightarrow + \infty$.

\begin{table}[!h]
\begin{center}
\begin{tabular}{|l|c|c|c|c|c|c|}
\hline     
    &                                $N=4$&       $N=5$ &     $N=6$  \\
\hline
$M=N$&                    0.586&                 0.655&             0.713 \\
$M=N-1$&                 0.589&                0.671&              0.729 \\
\hline
\end{tabular}
\caption{The percentage of ${\cal C}^{\epsilon}_{\alpha,\varrho}$ for several $N, M$.}\label{tab:case4_percentage}
\end{center}
\end{table}



\subsection{Higher level Bethe Ansatz equation}

Alternatively we can solve the higher level 
Bethe Ansatz equation directly for the members in  ${\cal C}^{\epsilon}_{\alpha,\varrho}$.
 
Recall that the higher level Bethe Ansatz equations governing the positions of the particle roots take the form 
\begin{equation}
(-1)^{n_x-n_y + 1}  
\prod_{ b=1}^{n_y} \bigg\{ \frac{ \sinh( i \zeta  + y_{b} - y_{a}) }{ \sinh(i\zeta  +y_{a} - y_{b} ) } \bigg\} \cdot 
 \bigg( \frac{\sinh( i \zeta + y_{a}) }{ \sinh( i\zeta  - y_{a}) } \bigg)^{n_x} = 1 \; .
 \label{eq:hlBAE1}
\end{equation}

However, these are obtained upon simplifying the original form of the equations in the case where the hole roots are small. 
In the case where the $x_a$s are not sufficiently small in magnitude,  one should rather consider 

\begin{equation}
(-1)^{n_x-n_y + 1}  
\prod_{ b=1}^{n_y} \bigg\{ \frac{ \sinh( i \zeta  + y_{b} - y_{a}) }{ \sinh(i\zeta  +y_{a} - y_{b} ) } \bigg\} \cdot 
 \prod_{{\ell}=1}^{n_x}  \frac{\sinh( i \zeta + y_{a}-x_{\ell}) }{ \sinh( i\zeta  - y_{a}+x_{\ell}) }   = 1 \;.
\label{eq:hlBAE2}
\end{equation}

While \eqref{eq:hlBAE1} can be solved directly for the $y_a$s, In order to solve  (\ref{eq:hlBAE2}) w.r.t. $\{y_a\}$,  one needs the input data $\{x_{\ell}\}$. The hole roots are 
substituted from the values obtained by the algorithm proposed in \cite{AlbertiniDasmahapatraMcCoySpectrumAndCompleteness3StatePottsComputerStudy}.
 
First we choose the same parameters with  Table \ref{tab:N10M5T100}
and present some examples including the 12-th excited state
in Table \ref{tab:HLBAEN5M5}

\begin{table}[!h]
\begin{center}
\begin{tabular}{|c|c|c|c|c|}
 
\hline     
                               &                    
      $ \{y_a/\zeta\}$ &           
       $ \{\wh{x}_a/\zeta\}$ &    
       $ \{y_a/\zeta\}$ from (\ref{eq:hlBAE1}) &  
        $ \{y_a/\zeta\}$ from (\ref{eq:hlBAE2})       \\
\hline

$\text{ 12-th}$&       
 \begin{tabular}{c}
  $\{ 
   -1.140806 \times 10^{-3}$\\
   $ \pm0.575835 \i \}$
  \end{tabular}&        
   \begin{tabular}{c}  
    $\{-2.979061  \times 10^{-3},$        \\
    $ 7.029353 \times 10^{-4} \}$ 
    \end{tabular} &
  $\{\pm 0.577223 \i \}$&
   \begin{tabular}{c}  
   $\{
  -1.138063\times 10^{-3}$ \\
  $\pm 0.577224 \i\}$
      \end{tabular} 
   \\
\hline
\text{ 27-th}&       
 \begin{tabular}{c}
  $\{ 
 \pm 1.379263 
  + \frac{\pi}{2\zeta} \i \}$
  \end{tabular}&        
   \begin{tabular}{c}  
    $\{ \pm 2.962552  \times 10^{-3} \}$ 
    \end{tabular} &
   \begin{tabular}{c}  
   $\{
 \pm1.378826
  + \frac{\pi}{2\zeta} \i \}$
      \end{tabular} &
   \begin{tabular}{c}  
   $\{
 \pm1.378828
  + \frac{\pi}{2\zeta} \i \}$
      \end{tabular} 
   \\
\hline
\text{ 41-th}&       
 \begin{tabular}{c}
  $\{ 
 -0.381523  \pm$ \\
 $ 0.547636  \i , $ \\
  $ 0.764017  + \frac{\pi}{2\zeta} \i      \}$
  \end{tabular}&        
   \begin{tabular}{c}  
    $\{ -4.518492 \times 10^{-3},$\\
    $5.582956 \times 10^{-4}$,\\
    $  2.169682 \times 10^{-3}   \}$ 
    \end{tabular} &
   \begin{tabular}{c}  
   $\{
  -0.381245\pm $\\
  $ 0.548153 \i , $\\
  $  0.762489 + \frac{\pi}{2\zeta} \i   \}$
      \end{tabular} &
    \begin{tabular}{c}  
     $\{
  -0.381847\pm $\\
  $ 0.548154 \i , $\\
  $  0.761903 + \frac{\pi}{2\zeta} \i   \}$
      \end{tabular} 
   \\
\hline
\text{ 120-th}&       
 \begin{tabular}{c}
  $\{ -0.295790\pm $ \\
 $0.550527  \i , $ \\
  $ 0.2961415\pm$ \\
  $0.550527 \i     \}$
  \end{tabular}&        
   \begin{tabular}{c}  
    $\{ -2.977582 \times 10^{-3},$\\
    $1.183573   \times 10^{-8}, $ \\
    $7.024496 \times 10^{-4}, $\\
    $  2.977334 \times 10^{-3 }     \}$ 
    \end{tabular} &
   \begin{tabular}{c}  
   $\{
 \pm  0.295564 \pm $\\
  $0.551187  \i     \}$
      \end{tabular} &
    \begin{tabular}{c}  
     $\{
 -0.295393 \pm $\\
  $ 0.551188\i   $,\\
  $  0.295745 \pm$ \\
$   0.551188  \i   \}$
      \end{tabular} 
   \\
\hline

\end{tabular}
\caption{The comparison of the result by the algorithm of \cite{AlbertiniDasmahapatraMcCoySpectrumAndCompleteness3StatePottsComputerStudy}, 
and the solutions to higher level Bethe ansatz equations
(\ref{eq:hlBAE1}) and (\ref{eq:hlBAE2})
: $N=5, M=5, T=100, \zeta=\frac{\pi}{7}$.  States are selected ``randomly".}  \label{tab:HLBAEN5M5}
\end{center}
\end{table}

Second we choose the same parameters with  Table \ref{tab:N10M4T100}
and present  examples including the 10-th excited state
in Table \ref{tab:HLBAEN5M4}

\begin{table}[!h]
\begin{center}
\begin{tabular}{|c|c|c|c|c|}
 
\hline     
                               &                    
      $ \{y_a/\zeta\}$ &           
       $ \{\wh{x}_a/\zeta\}$ &    
       $ \{y_a/\zeta\}$ from (\ref{eq:hlBAE1}) &  
        $ \{y_a/\zeta\}$ from (\ref{eq:hlBAE2})       \\
\hline

$\text{ 10-th}$&       
 \begin{tabular}{c}
  $\{ 
   \frac{\pi}{2\zeta} \i \}$
  \end{tabular}&        
   \begin{tabular}{c}  
    $\{\pm 2.962940  \times 10^{-3} \}$ 
    \end{tabular} &
  $\{\frac{\pi}{2\zeta}\i \}$&
  $\{\frac{\pi}{2\zeta}\i \}$
   \\
\hline
\text{ 24-th}&       
 \begin{tabular}{c}
  $\{ 
 3.496022  \times 10^{-4}  +$\\
 $ \frac{\pi}{2\zeta} \i \}$  \end{tabular}&        
   \begin{tabular}{c}  
    $\{ 7.021791 \times 10^{-4}, $\\
    $  1.404799 \times 10^{-7} \}$ 
    \end{tabular} &
   \begin{tabular}{c}  
   $\{
  \frac{\pi}{2\zeta} \i \}$
      \end{tabular} &
   \begin{tabular}{c}  
   $\{
3.511598 \times 10^{-4} $\\
 $ + \frac{\pi}{2\zeta} \i \}$
      \end{tabular} 
   \\
\hline
\text{ 32 -th}&       
 \begin{tabular}{c}
  $\{ 
 -0.466727  \pm$ \\
 $ 0.566923  \i      \}$
  \end{tabular}&        
   \begin{tabular}{c}  
    $\{ -4.520602 \times 10^{-3},$\\
    $ -8.699320 \times 10^{-4}$,\\
    $2.168664 \times 10^{-3}   \}$ 
    \end{tabular} &
   \begin{tabular}{c}  
   $\{
 -0.465800 \pm $\\
  $ 0.567469 \i  \}$
      \end{tabular} &
    \begin{tabular}{c}  
     $\{
  -0.466880 \pm $\\
  $ 0.567469 \i  \}$
      \end{tabular} 
   \\
\hline
\text{ 200-th}&       
 \begin{tabular}{c}
  $\{ -1.537113 + \frac{\pi}{2\zeta} \i, $ \\
 $0.240380+ \frac{\pi}{2\zeta} \i,  $ \\
  $ 1.084164 \pm$ \\
  $0.582417 \i     \}$
  \end{tabular}&        
   \begin{tabular}{c}  
    $\{ -3.715887 \times 10^{-3},$\\
    $-7.987464 \times 10^{-4},$\\
    $-6.083815  \times 10^{-5}, $ \\
    $6.127055 \times 10^{-4}, $\\
    $  2.426153 \times 10^{-3 }     \}$ 
    \end{tabular} &
   \begin{tabular}{c}  
   $\{
-1.536740  + \frac{\pi}{2\zeta} \i,  $\\
 $ 0.240603  + \frac{\pi}{2\zeta} \i,  $\\
  $1.083662  \pm $ \\
  $ 0.582535 \i    \}$
      \end{tabular} &
    \begin{tabular}{c}  
   $\{
-1.537050+ \frac{\pi}{2\zeta} \i,  $\\
 $ 0.240295  + \frac{\pi}{2\zeta} \i,  $\\
  $1.083357\pm $ \\
  $0.582535 \i    \}$
      \end{tabular} 
   \\
\hline

\end{tabular}
\caption{The comparison of the result by the algorithm of \cite{AlbertiniDasmahapatraMcCoySpectrumAndCompleteness3StatePottsComputerStudy}, 
and the solutions to higher level Bethe Ansatz equations
(\ref{eq:hlBAE1}) and (\ref{eq:hlBAE2})
: $N=5, M=4, T=100, \zeta=\frac{\pi}{7}$.  States are selected ``randomly".}  \label{tab:HLBAEN5M4}
\end{center}
\end{table}
  
\section{Conclusion}

This work sets the quantum transfer matrix approach to
the thermodynamics of spin chains into a rigorous framework which
is applicable for sufficiently high temperatures. For temperatures
high enough we have proven all those conjectures raised in the
literature, whose validity is necessary for a rigorous use of the
approach: the exchangeability of the infinite volume and infinite
Trotter number limits, the existence of a maximal in modulus Eigenvalue
of the quantum transfer matrix which, furthermore, is real and non-degenerate,
the well-definiteness of the class of non-linear integral equations
describing the Eigenvalues of the quantum transfer matrix along with
the existence and uniqueness of their solutions and, finally, the
rigorous identification of the non-linear integral equation describing
the dominant Eigenvalue of the quantum transfer matrix. In this way,
we have rigorously established the integral representation for the
\textit{per}-site free energy of the spin-$1/2$ XXZ chain which was
argued earlier on on the basis of heuristic arguments 
\cite{KlumperNLIEfromQTMDescrThermoXYZOneUnknownFcton,DestriDeVegaAsymptoticAnalysisCountingFunctionAndFiniteSizeCorrectionsinTBAFirstpaper}. 

We stress that the uniqueness of solutions to the non-linear integral
equation established in this work is a rather non-trivial property. 
While desirable and potentially expected on the basis of physical
arguments, examples of non-linear integral equations, seemingly similar
to those considered above, are known, for which, on the basis of numerical
investigations, uniqueness has been observed to fail \cite{DoreyTateoExcitedStatesFromAnalyticContinuationTBA}.
Moreover, another type of nonlinear integral equations directly for
the eigenvalues of the quantum transfer matrix \cite{TakahashiSimplifiedFormOfTBAEqns} is
known to have many inequivalent solutions.

While all the exposition of this paper was focused on the spin-$1/2$ XXZ
chain, the reasoning employed in sections \ref{Section Comportement Grand T traces de QTM},
\ref{Section VP dominante et echange limites} is quite general and needs
only very weak assumptions. In particular, upon minor modifications, our proof
of Theorem \ref{Theorem echange limite thermo et Trotter} is  applicable to
all quantum integrable models associated with a fundamental $R$-matrix. 

In this work we have also obtained a characterisation of a subset of
subdominant Eigenvalues of the quantum transfer matrix in the
high-temperature regime, this after taking the infinite Trotter number
limit. As we have shown, the calculation of these Eigenvalues reduces,
to leading order in $T^{-1}$, to finding solutions to the Bethe Ansatz
equations associated with the ordinary transfer matrix of the spin-$1$
XXZ chain. The main tool used in this description are the classes
$\mc{C}^{\eps}_{\a,\varrho}$. The solutions to the Bethe Ansatz equations
have been thoroughly studied numerically, and we have found that, for
Trotter numbers up to $N = 6$, the classes $\mc{C}^{\eps}_{\a,\varrho}$
are non-empty. In fact, it appears that, in this range of $N$, they
capture around 70\% of all subdominant Eigenvalues for large $T$. For
the time being we do not have any further mathematical interpretation
of our numerical data. It will be interesting to attempt to interpret
them on rigorous grounds.

\vspace{3mm}

\noindent {\bf Acknowledgements.}
The authors would like to thank Patrick Dorey, Andreas Kl\"umper and Eric Vernier
for stimulating discussions. The work of FG was supported by the
Deutsche Forschungsgemeinschaft within the framework of the research
unit FOR 2316. The work of SG and KKK was supported by the CNRS, Projet
international de coopération scientifique No. PICS07877:
\textit{Fonctions  de  corrélations  dynamiques  dans  la  chaîne  XXZ  à  température finie}, Allemagne, 2018-2020. 
JS was supported by JSPS KAKENHI Grants, numbers 15K05208, 18K03452 and 18H01141.

\end{document}